\documentclass[12pt]{article}

\usepackage[T1]{fontenc}
\usepackage[utf8]{inputenc}
\usepackage{ytableau}
\usepackage[vcentermath]{youngtab}
\usepackage{yfonts}
\usepackage{hyperref}
\usepackage{graphicx}
\usepackage{amsfonts} 
\usepackage{amsmath}
\usepackage{amssymb}
\usepackage{color} 
\usepackage{pstool}
\usepackage[all]{xy}
\usepackage{mathtools}
\usepackage{xspace}
\usepackage{braket}
\usepackage{multirow}

\usepackage[percent]{overpic}

\usepackage{subfig}
\usepackage{caption}

 \usepackage[usenames,dvipsnames]{pstricks}
 \usepackage{epsfig}
 \usepackage{pst-grad} % For gradients
 \usepackage{pst-plot} % For axes
 \usepackage[space]{grffile} % For spaces in paths
 \usepackage{etoolbox} % For spaces in paths
 \makeatletter % For spaces in paths
 \patchcmd\Gread@eps{\@inputcheck#1 }{\@inputcheck"#1"\relax}{}{}
 \makeatother

\usepackage[numbers,sort&compress]{natbib}

\graphicspath{ {./fig/} }

\newcommand{\pf}{\mathfrak{p}}
\newcommand{\qf}{\mathfrak{q}}

\newcommand{\hj}{{\hat \jmath}}

\newcommand{\ii}{\mathrm{i}}

\newcommand{\Vp}[1]{V^{\pdot}_{#1}}
\newcommand{\Vt}[1]{V^{\tdot}_{#1}}
\newcommand{\Vs}[1]{V^{\star}_{#1}}
\newcommand{\Hp}[1]{H^{\pdot}_{#1}}
\newcommand{\Ht}[1]{H^{\tdot}_{#1}}
\newcommand{\Hs}[1]{H^{\star}_{#1}}

\newcommand{\hOcal}{{\hat \Ocal}}

\newcommand{\hD}{{\hat \D}}

\newcommand{\hl}{{\hat l}}
\newcommand{\hr}{{\hat r}}
\newcommand{\heta}{{\hat \eta}}
\newcommand{\hG}{{\hat G}}
\newcommand{\hg}{{\hat g}}

\newcommand{\hGcal}{{\hat \Gcal}}

\newcommand{\pdot}{{\, \bullet \,}}
\newcommand{\tdot}{{\, \circ\, }}
\newcommand{\Pp}{\Pi_{\bullet }}
\newcommand{\Ps}{\Pi_{\star }}

\newcommand{\Pt}{\Pi_{\circ }}
\newcommand{\pp}{\pi_{\bullet }}
\newcommand{\pt}{\pi_{\circ}}

\newcommand{\gbullet}{{\bullet}}
\newcommand{\wbullet}{{\circ}}
\newcommand{\dd }{\mathtt{d}}
\newcommand{\hO }{{\hat \Ocal}}

\def\beq{\begin{equation}} 
\def\eeq{\end{equation}} 
\newcommand{\be}{\begin{equation}}
\newcommand{\ee}{\end{equation}} 
\newcommand{\ba}{\begin{eqnarray}}
\newcommand{\ea}{\end{eqnarray}}

\newcommand{\Acal}{{\mathcal A}}

\newcommand{\Ccal}{{\mathcal C}}

\newcommand{\Dcal}{{\mathcal D}}

\newcommand{\Fcal}{{\mathcal F}}
\newcommand{\Gcal}{{\mathcal G}}

\newcommand{\Ncal}{{\mathcal N}}
\newcommand{\Ocal}{{\mathcal O}}

\newcommand{\Qcal}{{\mathcal Q}}

\newcommand{\Scal}{{\mathcal S}}

\newcommand{\Vcal}{{\mathcal V}}
\newcommand{\Wcal}{{\mathcal W}}

\let\a=\alpha \let\b=\beta \let\g=\gamma \let\d=\delta 
    
\let\l=\lambda \let\m=\mu \let\n=\nu  \let\r=\rho
\let\s=\sigma    

  \let\D=\Delta

\newcommand{\I}{\textrm{I}\xspace}
\newcommand{\II}{\textrm{II}\xspace}
\newcommand{\III}{\textrm{III}\xspace}
\newcommand{\IV}{\textrm{IV}\xspace}

\DeclareFontFamily{OMX}{MnSymbolE}{}
\DeclareSymbolFont{largesymbolsX}{OMX}{MnSymbolE}{m}{n}
\DeclareFontShape{OMX}{MnSymbolE}{m}{n}{
    <-6>  MnSymbolE5
   <6-7>  MnSymbolE6
   <7-8>  MnSymbolE7
   <8-9>  MnSymbolE8
   <9-10> MnSymbolE9
  <10-12> MnSymbolE10
  <12->   MnSymbolE12}{}

\DeclareMathSymbol{\downbrace}    {\mathord}{largesymbolsX}{'251}
\DeclareMathSymbol{\downbraceg}   {\mathord}{largesymbolsX}{'252}
\DeclareMathSymbol{\downbracegg}  {\mathord}{largesymbolsX}{'253}
\DeclareMathSymbol{\downbraceggg} {\mathord}{largesymbolsX}{'254}
\DeclareMathSymbol{\downbracegggg}{\mathord}{largesymbolsX}{'255}
\DeclareMathSymbol{\upbrace}      {\mathord}{largesymbolsX}{'256}
\DeclareMathSymbol{\upbraceg}     {\mathord}{largesymbolsX}{'257}
\DeclareMathSymbol{\upbracegg}    {\mathord}{largesymbolsX}{'260}
\DeclareMathSymbol{\upbraceggg}   {\mathord}{largesymbolsX}{'261}
\DeclareMathSymbol{\upbracegggg}  {\mathord}{largesymbolsX}{'262}
\DeclareMathSymbol{\braceld}      {\mathord}{largesymbolsX}{'263}
\DeclareMathSymbol{\bracelu}      {\mathord}{largesymbolsX}{'264}
\DeclareMathSymbol{\bracerd}      {\mathord}{largesymbolsX}{'265}
\DeclareMathSymbol{\braceru}      {\mathord}{largesymbolsX}{'266}
\DeclareMathSymbol{\bracemd}      {\mathord}{largesymbolsX}{'267}
\DeclareMathSymbol{\bracemu}      {\mathord}{largesymbolsX}{'270}
\DeclareMathSymbol{\bracemid}     {\mathord}{largesymbolsX}{'271}

\makeatletter
\def\horiz@expandable#1#2#3#4#5#6#7#8{%
  \@mathmeasure\z@#7{#8}%
  \@tempdima=\wd\z@
  \@mathmeasure\z@#7{#1}%
  \ifdim\noexpand\wd\z@>\@tempdima
    $\m@th#7#1$%
  \else
    \@mathmeasure\z@#7{#2}%
    \ifdim\noexpand\wd\z@>\@tempdima
      $\m@th#7#2$%
    \else
      \@mathmeasure\z@#7{#3}%
      \ifdim\noexpand\wd\z@>\@tempdima
        $\m@th#7#3$%
      \else
        \@mathmeasure\z@#7{#4}%
        \ifdim\noexpand\wd\z@>\@tempdima
          $\m@th#7#4$%
        \else
          \@mathmeasure\z@#7{#5}%
          \ifdim\noexpand\wd\z@>\@tempdima
            $\m@th#7#5$%
          \else
           #6#7%
          \fi
        \fi
      \fi
    \fi
  \fi}

\def\overbrace@expandable#1#2#3{\vbox{\m@th\ialign{##\crcr
  #1#2{#3}\crcr\noalign{\kern2\p@\nointerlineskip}%
  $\m@th\hfil#2#3\hfil$\crcr}}}
\def\underbrace@expandable#1#2#3{\vtop{\m@th\ialign{##\crcr
  $\m@th\hfil#2#3\hfil$\crcr
  \noalign{\kern2\p@\nointerlineskip}%
  #1#2{#3}\crcr}}}

\def\overbrace@#1#2#3{\vbox{\m@th\ialign{##\crcr
  #1#2\crcr\noalign{\kern2\p@\nointerlineskip}%
  $\m@th\hfil#2#3\hfil$\crcr}}}
\def\underbrace@#1#2#3{\vtop{\m@th\ialign{##\crcr
  $\m@th\hfil#2#3\hfil$\crcr
  \noalign{\kern2\p@\nointerlineskip}%
  #1#2\crcr}}}

\def\bracefill@#1#2#3#4#5{$\m@th#5#1\leaders\hbox{$#4$}\hfill#2\leaders\hbox{$#4$}\hfill#3$}

\def\downbracefill@{\bracefill@\braceld\bracemd\bracerd\bracemid}
\def\upbracefill@{\bracefill@\bracelu\bracemu\braceru\bracemid}

\def\upbrace@expandable{%
  \horiz@expandable
    \upbrace
    \upbraceg
    \upbracegg
    \upbraceggg
    \upbracegggg
    \upbracefill@}
\def\downbrace@expandable{%
  \horiz@expandable
    \downbrace
    \downbraceg
    \downbracegg
    \downbraceggg
    \downbracegggg
    \downbracefill@}

\DeclareRobustCommand{\overbrace}[1]{\mathop{\mathpalette{\overbrace@expandable\downbrace@expandable}{#1}}\limits}
\DeclareRobustCommand{\underbrace}[1]{\mathop{\mathpalette{\underbrace@expandable\upbrace@expandable}{#1}}\limits}
\makeatother

\begin{document}

\begin{titlepage}

\begin{center}
\vspace{1.5cm}

{\Large \bf Spinning operators and defects in conformal field theory}

\vspace{0.8cm}

{\bf Edoardo Lauria$^{1}$, Marco Meineri$^{2}$,  Emilio Trevisani$^{3}$}

\vspace{.5cm}
{
\small
{\it $^{1}$ Instituut voor Theoretische Fysica, KU Leuven, \\Celestijnenlaan 200D, B-3001 Leuven, Belgium \\
\it $^{2}$Institute of Physics, École Polytechnique Fédérale de Lausanne (EPFL), \\
 CH-1015 Lausanne, Switzerland \\
\it  $^3$
Laboratoire de Physique Th\'eorique, \'Ecole Normale Sup\'erieure \& PSL Research University, \\
24 rue Lhomond, 75231 Paris Cedex 05, France
\\
Institut des Hautes \'Etudes Scientifiques, Bures-sur-Yvette, France
}}
\end{center}

\vspace{1cm}

\begin{abstract}
We study the kinematics of correlation functions of local and extended operators in a conformal field theory. We present a new method for constructing the tensor structures associated to primary operators in an arbitrary bosonic representation of the Lorentz group. The recipe yields the explicit structures in embedding space, and can be applied to any correlator of local operators, with or without a defect. 
 We then focus on the two-point function of traceless symmetric primaries in the presence of a conformal defect, and explain how to compute the conformal blocks. In particular, we illustrate various techniques to generate the bulk channel blocks either from a radial expansion or  by acting with differential operators on simpler \emph{seed} blocks.  For the defect channel, we detail a method to compute the blocks in closed form, in terms of projectors into mixed symmetry representations of the orthogonal group.

\end{abstract}

\bigskip

\end{titlepage}

%%%%%%%%%%%%%%%%%%%%%%%%%%%%%%%%%%%%%%%%%%%%%%%%%%
\tableofcontents
%%%%%%%%%%%%%%%%%%%%%%%%%%%%%%%%%%%%%%%%%%%%%%%%%%

\section{Introduction}

The most natural observables in a conformal field theory (CFT) are correlation functions. In this paper, we target the correlators which involve one extended operator and multiple local insertions. In fact, conformal invariant extended operators, or conformal defects for short, have been studied extensively at least since the early days of two dimensional CFTs, starting from the seminal work of Cardy \cite{Cardy:1984bb} on boundary conditions in the minimal models. Rather than attempting a review of the relevance of conformal defects in both low and high energy physics, let us only mention the most recent motivation for the present work. 

On one hand, both numerical \cite{arXiv:1210.4258,Gaiotto:2013nva,arXiv:1502.07217,Gliozzi:2016cmg,Liendo:2018ukf} and analytical \cite{Hogervorst:2017kbj,Lemos:2017vnx} conformal bootstrap techniques\footnote{See \cite{Poland:2018epd} for a comprehensive review of the numerical bootstrap.} have been applied to the study of conformal defects. The main targets have been so far the two-point function of scalar primaries in the presence of a flat defect, and the four-point function of local operators living on the defect itself. Boundaries and interfaces provide an exception: there, external stress-tensors were considered in \cite{arXiv:1210.4258}. A natural generalization of this setup is the bootstrap of the correlators of two bulk local operators with spin and a defect. Conserved currents and the stress-tensor are of course the main candidates. As we shall demonstrate in this paper, the more complicated kinematics offers a considerably smaller challenge with respect to the case of a four-point function of local operators with spin \cite{Dymarsky:2017xzb, Dymarsky:2013wla, Dymarsky:2017yzx}. 

On the other hand, control over the kinematics involved in correlators of spinning operators with a defect should be useful also when tackling specific examples with techniques different from the bootstrap. For instance, the technology of defect CFT played a crucial role in proving the Quantum Null Energy Condition (QNEC) \cite{Balakrishnan:2017bjg} via the replica trick. In particular, the Operator Product Expansion (OPE) of the stress tensor with the so-called replica defect \cite{Hung:2014npa,Bianchi:2015liz} contains the non trivial information about the matrix elements of the modular Hamiltonian, which not only lies at the heart of the proof of the QNEC, but is also an important quantity in its own right. Therefore, the two-point function of the stress tensor is again a natural observable to focus on in this context. Another example is provided by the class of line defects, \emph{e.g.} Wilson and 't Hooft lines, which correspond to massive probes. When these external objects surf the vacuum on a generic worldline, they emit radiation. This real time process, so relevant in the case of a gauge theory, is again captured by correlation functions of the stress tensor with the line defect. In a supersymmetric setup, a recent proof of a series of conjectures concerning the energy emitted by an accelerated quark \cite{Lewkowycz:2013laa,Fiol:2015spa} has been obtained by studying the coupling of the stress tensor with a line defect \cite{Bianchi:2018zpb}.

Let us begin by recalling a few definitions. The conformal defect will always be taken either flat or spherical, and the following convention is adopted:
\beq
p=\textup{dimension of the defect,}\quad q=\textup{codimension of the defect,}\quad d=\textup{dimension of spacetime,}
\eeq
so that $p+q=d$. We call bulk OPE the fusion of local operators away from the defect,
\beq
\Ocal_1(x_1) \Ocal_2(x_2) \sim \sum_{\Ocal} c_{12\Ocal} \Ocal(x_2)~.
\label{bOPEschem}
\eeq
A conformal defect can be excited locally by a set of defect operators, which appear in the OPE of a bulk operator with the defect, or defect OPE for short:
\beq
\mathcal{O}(x) \sim \sum_{\hat{\mathcal{O}}} b_{\Ocal\hOcal} \hOcal(x_{\parallel})~.
\label{dOPEschem}
\eeq
Here, the presence of a flat defect is understood, defect operators are denoted with a hat, and $x_{\parallel}$ denotes the projection of $x$ onto the defect.

Let us recapitulate the status of the art in the analysis of the symmetry constraints on correlation functions of local operators with a defect. The case of a boundary in higher dimensional CFT was first studied in \cite{McAvity:1995zd}. In the case of a defect of generic codimension, one-point functions\footnote{We use a terminology which leaves the present presence of the defect as understood. For instance, a one-point function is the correlator of a bulk operator with the defect. In section \ref{sec:mixedCFT}, though, we discuss correlation functions of local operators without defects: we hope that this creates no confusion.} of bulk operators, correlators of a symmetric traceless bulk operator and a defect operator, and two-point functions of bulk symmetric traceless operators were analyzed in \cite{Billo:2016cpy}. The tensor structures which appear in the correlation function of mixed symmetry bulk operators were recently studied in \cite{Guha:2018snh}. The bulk OPE was considered from a different point of view in \cite{Fukuda:2017cup}: this paper studies the expansion of a spherical defect in a sum over local operators, and describes the OPE-blocks for this kind of fusion. In \cite{Liendo:2016ymz}, the additional constraints implied by $\mathcal{N}=4$ superconformal symmetry were tackled. Finally, Mellin space for defect CFT was considered in \cite{Rastelli:2017ecj,Goncalves:2018fwx}.

The minimal correlator which admits an expansion both in the bulk and the defect channels is the two-point function of local operators in the presence of the defect. Results for conformal blocks are available in the literature, in the case of scalar external primaries \cite{McAvity:1995zd,Billo:2016cpy,Lauria:2017wav}. In particular, in \cite{Lauria:2017wav} a convenient set of cross-ratios was defined, the so-called radial coordinates, which we shall also adopt here. 

Finally, in the recent paper \cite{Isachenkov:2018pef}, the conformal blocks for pairs of defects were studied. The authors map the problem of finding the blocks into the problem of finding eigenfunctions of a Calogero-Sutherland Hamiltonian. The approach allows to extend a set of known dualities between blocks \cite{Billo:2016cpy,Gadde:2016fbj,Liendo:2016ymz}, and as a special case applies to the bulk channel blocks for the two-point function of external scalars with a single defect.

The content of the paper is two-fold. Sections \ref{sec:mixedCFT} and \ref{SO(d+1,1)SO(q)} are dedicated to the tensor structures appearing in correlation functions of local operators in arbitrary representations of the rotation group. We describe a way of explicitly building the structures in embedding space, which we apply both in the ordinary CFT setup, and in the presence of a defect. In section \ref{sec:spinning}, we turn to the computation of the conformal blocks for the two-point function of traceless symmetric primaries. In the bulk channel, we extend the results of \cite{Lauria:2017wav} and explain how to efficiently generate the blocks in an expansion in radial coordinates and by mean of the spinning  differential operators of  \cite{Costa2011}.  In the defect channel, the full set of conformal blocks can be computed in closed form, and we describe the general solution. Finally, in section \ref{sec:example} we illustrate the results in the simple context of a free defect CFT.

%%%%%%%%%%%%%%%%%%%%%%%
\section{Mixed symmetry representations and CFTs}
\label{sec:mixedCFT}
%%%%%%%%%%%%%%%%%%%%%%%

In this section we introduce new tensor structures for mixed symmetry representations which generalize the $H_{ij}$ and $V_{i,jk}$ introduced in \cite{SpinningCC}. 
We will find a minimal choice of polynomials which are in $1-1$ correspondence with the conformal invariant tensor structures in a correlation function. Our structures differ therefore from the ones introduced in \cite{MixedSym}, which are not minimal and cannot be used for the counting of tensor structures.

%%%%%%%%%%%%%%%%%%%%%%%
\subsection{Tensor structures for $SO(n)$ mixed symmetry representations}\label{ss:SOn_mixed_rep}
%%%%%%%%%%%%%%%%%%%%%%%

We begin in the context of the orthogonal group. This allows us to review some background material and set up a technology that, with minor modifications, will be applied to correlation functions constrained by the full conformal group. Furthermore, defect operators enjoy a global $SO(q)$ symmetry, and the content of this subsection can be used verbatim to take care of the associated representation theory.

\subsubsection{Mixed symmetry tensors as polynomials}
\label{subsec:so(n)}
A tensor $t_{l}$  in a irreducible representation $l=l_1,\dots,l_{[\frac{n}{2}]}$ can be labelled by a Young tableau which has indices in each box. The indices in the rows are symmetrized, while antisymmetrization is performed on the indices in each column. Finally, all the traces are removed. In order to make the symmetrization manifest we can contract all the indices of the $i-$th row with the same polarization vector $z^{(i)}$,
  \ytableausetup{centertableaux,boxsize=1.7 em}
\be
t_l(z) \equiv t_{l}
\left(
{ \scriptsize
\begin{ytableau}
z^{(1)}&\, _{\cdots}&\, _{\cdots}&\, _{\cdots}&\, _{\cdots}&\, _{\cdots}&\,z^{(1)}\\
z^{(2)}&\, _{\cdots}&\, _{\cdots}&\, _{\cdots}&z^{(2)}\\
\resizebox{!}{0.3 cm}{\bf \vdots}&\resizebox{!}{0.3 cm}{\bf \vdots}& \resizebox{!}{0.3 cm}{\bf \vdots}& \raisebox{0.1 cm}{\resizebox{!}{0.12 cm}{\rotatebox[origin=c]{45}{\dots}}} \\
z^{(k)}&\, _{\cdots}&z^{(k)}
\end{ytableau}
}\
\right)
=
t_{l}
\left(
{ \scriptsize
\begin{ytableau}
\m^{1}_{\!1}&\, _{\cdots}&\, _{\cdots}&\, _{\cdots}&\, _{\cdots}&\, _{\cdots}&\,\m^{1}_{l_{1}}\\
\m^{2}_{\!1}&\, _{\cdots}&\, _{\cdots}&\, _{\cdots}&\m^{2}_{l_{2}}\\
\resizebox{!}{0.3 cm}{\bf \vdots}&\resizebox{!}{0.3 cm}{\bf \vdots}& \resizebox{!}{0.3 cm}{\bf \vdots}& \raisebox{0.1 cm}{\resizebox{!}{0.12 cm}{\rotatebox[origin=c]{45}{\dots}}} \\
\m^{k}_{\!1}&\, _{\cdots}& \m^{k}_{l_{ k}}
\end{ytableau}
}\
\right)
\prod_{i=1}^k z^{(i)}_{\m^{i}_{\!1}} \cdots z^{(i)}_{\m^{i}_{l_{i}}} \ ,
\label{tlgeneral}
\ee
where $k\leq [\frac{n}{2}]$.
A vector (\emph{e.g.} $z^{(i)}$) inside a box means that the index of the vector is contracted with the index of the box. 
The result is a polynomial $ t_l(z)$ which has homogeneity $l_i$ for all the $z^{(i)}$,
\be
\label{scalingSOn}
 z^{(i)} \cdot \partial_{z^{(i)}} \ t_l(z)= l_i \ t_l(z) \ .
\ee
Antisymmetry of the columns (or better mixed symmetry of the Young tableau) is the statement that it is not possible to symmetrize an index of a row $j$ with all the indices of a given a row $i$, with $i<j$. In terms of the polynomials this 
condition can be imposed asking that 
\be
\label{antisym}
 z^{(i)} \cdot \partial_{z^{(j)}}  \ t_l(z)= 0 \, , \qquad \forall j>i \ .
\ee
Alternatively, we can say that $t_l(z)$ is invariant under the map $z^{(j)}\rightarrow z^{(j)}+ \a z^{(i)}$ for $j>i$ and for any $\a \in \mathbb{R}$.
Finally, tracelessness implies
\be
\label{tracelessness}
 \partial_{z^{(i)}}  \cdot \partial_{z^{(j)}}  \ t_l(z)= 0 \, , \qquad \forall i,j \ .
\ee
Rotations act naturally on $t_l(z)$, the generators being $L^{\m \n}=\sum_{i=1}^{[\frac{n}{2}]} (z^{(i) \, \m} \, \partial^{\n}_{z^{(i)}}-z^{(i) \, \n} \, \partial^{\m}_{z^{(i)}} )$. It follows that $t_l(z)$ are eigenfunctions of the Casimir operator $ C\equiv -\frac{1}{2}L^{\m \n}L_{\m \n}$:
\be
\label{casimirtensor}
C t_l(z) = c_{l} t_l(z) \ ,
\qquad 
c_{l}=\sum_{i=1}^{[\frac{n}{2}]}  l_{i} (l_{i}+n-2i) 
\ .
\ee

So far the vectors $z^{(j)}$ are unconstrained. However, there is a cheaper way to encode a tensor in terms a polynomial $t_l(z)$, by asking that it is defined in the following subspace
\be
\Ncal \equiv \left\{z^{(1)} , \dots z^{([\frac{n}{2}])} \in \mathbb{R}^n: (z^{(i)} \cdot z^{(j)})=0, \mbox{for all } i,j \right\}.
\ee
Indeed, the tensor can be uniquely recovered from the polynomial restricted to the subspace $\Ncal$. 
It is important that the tensor should be transverse, namely
\be
\label{spaceS}
t_l(z) \in \Scal \equiv \{ f:\Ncal \rightarrow \mathbb{R} \mbox{ such that } z^{(i)} \cdot \partial_{z^{(j)}}  f= 0\} \ .
\ee
Transversality has useful consequences. For instance, the action of the Casimir operator on functions  $t_l(z) \in \Scal$ reduces to
\be
\label{CasimirTransverse}
Ct_l(z)=\sum_{i=1}^{[\frac{n}{2}]}(z^{(i)} \cdot \partial_{z^{(i)}}+n-2i) (z^{(i)} \cdot \partial_{z^{(i)}})t_l(z) \ , \qquad  \forall\ t_l(z) \in \Scal\ .
\ee
Therefore any function $t_l(z) \in \Scal $ with the correct homogeneity in $z^{(i)}$ automatically satisfies the Casimir equation. 

Equations \eqref{spaceS} and \eqref{CasimirTransverse} make sense because the vectors $L_{\mu\nu}$ and $z^{(i)} \cdot \partial_{z^{(j)}}$ are tangent to the manifold $\Ncal$. This is not true of the partial derivative $\partial_{z^(i)}$. It follows that indices cannot simply be opened while staying inside $\Ncal$, nor can the tracelessness condition \eqref{tracelessness} be verified. Still, a recipe exists to recover the tensor from the polynomial restricted to $\Ncal$. 
The recipe \cite{MixedSym, projectors} is to derive all the $z^{(i)}$ of $t_l(z)$ and contract the resulting open indices with a projector $\pi^{SO(n)}_{l}$ into the representation $l$ of $SO(n)$.  
This prescription defines a (generalization of) the differential operator introduced in \cite{PhysRevD.13.887}. 
Namely, given  $t_l(z)\in \Scal$ we have
 \be
\!
 \label{recovertensor}
  \resizebox{0.93 \textwidth}{!}{$
    \ytableausetup{centertableaux,boxsize=2.1 em}
 t_{l}
\left(
{ \scriptsize
\begin{ytableau}
\m^{1}_{\!1}&\, _{\cdots}&\, _{\cdots}&\, _{\cdots}&\, _{\cdots}&\, _{\cdots}&\,\m^{1}_{l_{1}}\\
\m^{2}_{\!1}&\, _{\cdots}&\, _{\cdots}&\, _{\cdots}&\m^{2}_{l_{2}}\\
\resizebox{!}{0.3 cm}{\bf \vdots}&\resizebox{!}{0.3 cm}{\bf \vdots}& \resizebox{!}{0.3 cm}{\bf \vdots}& \raisebox{0.1 cm}{\resizebox{!}{0.12 cm}{\rotatebox[origin=c]{45}{\dots}}} \\
\m^{k}_{\!1}&\, _{\cdots}& \m^{k}_{l_{ k}}
\end{ytableau}
}\
\right)= c \, \pi^{SO(n)}_{l}
\left(
{ \scriptsize
\begin{ytableau}
\m^{1}_{\!1}&\, _{\cdots}&\, _{\cdots}&\, _{\cdots}&\, _{\cdots}&\, _{\cdots}&\,\m^{1}_{l_{1}}\\
\m^{2}_{\!1}&\, _{\cdots}&\, _{\cdots}&\, _{\cdots}&\m^{2}_{l_{2}}\\
\resizebox{!}{0.3 cm}{\bf \vdots}&\resizebox{!}{0.3 cm}{\bf \vdots}& \resizebox{!}{0.3 cm}{\bf \vdots}& \raisebox{0.1 cm}{\resizebox{!}{0.12 cm}{\rotatebox[origin=c]{45}{\dots}}} \\
\m^{k}_{\!1}&\, _{\cdots}& \m^{k}_{l_{ k}}
\end{ytableau}
}\ ,
{ \scriptsize
\begin{ytableau}
\, \partial_{\!z^{(1)}}&\, _{\cdots}&\, _{\cdots}&\, _{\cdots}&\, _{\cdots}&\, _{\cdots}&\, \partial_{\!z^{(1)}}\\
\, \partial_{\!z^{(2)}}&\, _{\cdots}&\, _{\cdots}&\, _{\cdots}&\, \partial_{\!z^{(2)}}\\
\resizebox{!}{0.3 cm}{\bf \vdots}&\resizebox{!}{0.3 cm}{\bf \vdots}& \resizebox{!}{0.3 cm}{\bf \vdots}& \raisebox{0.1 cm}{\resizebox{!}{0.12 cm}{\rotatebox[origin=c]{45}{\dots}}} \\
\, \partial_{\!z^{(k)}}&\, _{\cdots}&\, \partial_{\!z^{(k)}}
\end{ytableau}
}\
\right)
 t_{l}(z),
 $ }
 \ee
 where $c=1/(l_1! \dots l_k!)$ comes from the derivatives. We will comment on the definition of $ \pi^{SO(n)}_{l}$ in the next section. 
 Let us briefly explain why \eqref{recovertensor}  works. On one hand  the operation \eqref{recovertensor}  recovers the original tensor if applied to  $t_l(z)$ with unconstrained polarization vectors. On the other hand any   $t_l(z)$ defined in $\Ncal$ differs from the unconstrained  $t_l(z)$ only by terms proportional to $z^{(i)}\cdot z^{(j)}$. However these terms are automatically  annihilated by the operation  \eqref{recovertensor}, because the projector is traceless. The result follows.

Let us summarize. A tensor in a representation $l=(l_1,\dots l_{[n/2]})$ of $SO(n)$ is encoded in a polynomial $t_l(z)$ with the following three properties
 \begin{itemize}
\item $t_l(z)$ is defined on the subspace  $(z^{(i)} \cdot z^{(j)})=0$,
\item $t_l(z)$ has  homogeneity $l_i$ in $z^{(i)}$, \emph{i.e.} it satisfies \eqref{scalingSOn},
\item $t_l(z)$ is transverse: it is invariant under $z^{(j)}\rightarrow z^{(j)}+ \a z^{(i)}$ for $j>i$, \emph{i.e.} it satisfies \eqref{antisym}.
\end{itemize}
We recover the initial tensor from the polynomial $t_l(z)$ by performing the operation \eqref{recovertensor}.

Given any tensor $t$, there is a simple way to project its indices onto a representation $l$. 
One can just construct the associated polynomial $t_l(z)$ as follows. For each column with $m$ boxes in the Young tableau of the representation, contract $m$ of the indices of the tensor $t$ with the following antisymmetric tensor
  \ytableausetup{centertableaux,boxsize=1.7 em}
\be
\label{buildingC}
 c^{(m)}_{z \; \; \; \m_1\dots \m_m} \equiv z^{(1)}_{\; \; \;  [\m_1 }\ \cdots \ z^{(m) }_{\; \; \; \m_m]}  \ , \qquad m=1,\dots, \left[\frac{n}{2}\right] \ .
\ee
The tensors $c_z^{(m)}$ are automatically  transverse in all the $z^{(i)}$ such that $1\leq i \leq [n/2]$. For instance, given a tensor $t$ with $8$ indices, we obtain the polynomial $t_l(z)$ associated to the representation $l=(4,3,1)$ as follow:
\be
t_l(z) = t
\left(
{ \scriptsize
\begin{ytableau}
\m_1&\m_2 &\m_3 &\m_4\\
\n_1&\n_2&\n_3\\
\r_1
\end{ytableau}
}\
\right) 
\ c^{(3) \m_1 \n_1 \r_1}_{z} \ c^{(2) \m_2 \n_2}_{z} \ c^{(2) \m_3 \n_3}_{z} \ c^{(1) \m_4}_{z}~.
\ee
Notice that the polynomial $t_l(z)$ now scales correctly in $z^{(i)}$ and it is automatically transverse. From now on we can therefore think that any $t_l(z)$ is just a tensor contracted opportunely with a set of $c_z^{(m)}$.
%%%%%%%%%%%%%%%%%%%%%
\subsubsection{Projectors onto representations of $SO(n)$}
\label{subsec:proj_son}
%%%%%%%%%%%%%%%%%%%%%

As explained, the projectors $ \pi^{SO(n)}_{l}$ onto a representation $l$ of $SO(n)$ are useful objects. In order to make the paper self contained we review here their definition and the state of the art on the subject. For more details, see for instance  \cite{projectors}.

A projector $\pi^{SO(n)}_{l}(a,b)$ depends on two sets of tensor indices $a, b$ ($a=\a_1, \dots \a_{|l|}$, $b=\b_1, \dots \b_{|l|}$, where $|l|=l_1+\dots+l_{[n/2]}$). Both sets of indices have the symmetries of the Young tableaux of the representation $l$ of $SO(n)$. The projector is invariant under conjugation by an element of $SO(n)$. As a consequence, when the projector is contracted on one side with an arbitrary tensor $t_a$, the result is a new tensor $t'_b$ which transforms in the irreducible representation  $l$ of $SO(n)$:
\be
t'_b=\pi^{SO(n)}_{l}(a,b)\, t_a \ .
\ee
The projector is also idempotent,
\be
\pi^{SO(n)}_{l}(a,b)\pi^{SO(n)}_{l}(b,c)=\pi^{SO(n)}_{l}(a,c).
\ee
When we contract with vectors the two sets of tensor indices in a projector we find a polynomial\footnote{In the following, we will use these polynomials both in physical and in embedding space. The coefficients of the polynomials do not depend on the signature of the metric, while the variables, which are scalar products in $X_i$ and $Y_i$, do.
With abuse of notation, we will intend the polynomial ${\bf P}^{d}$ as a function of scalar products built out of the metric of the physical space $\mathbb{R}^{d}$, while ${\bf P}^{d+2}$ will depend on scalar products built out of the metric of the embedding space $\mathbb{R}^{1,d+1}$.
\label{footnotePd}
}
\be
\label{Pnl}
{\bf P}^{n}_{l_1,\dots, l_k}(X_1,\dots, X_k; Y_1, \dots , Y_k )
\equiv 
{\pi}^{SO(n)}_{l_1,\dots, l_k} \left (
 \
  { \scriptsize
  \ytableausetup{centertableaux,boxsize=1.35 em}
\begin{ytableau}
\, X_{\!1}&\, _{\cdots}&\, _{\cdots}&\, _{\cdots}&\, _{\cdots}&\, _{\cdots}&\,X_{\!1}\\
\, X_{\!2}&\, _{\cdots}&\, _{\cdots}&\, _{\cdots}&\,X_{\!2}\\
\resizebox{!}{0.3 cm}{\bf \vdots}&\resizebox{!}{0.3 cm}{\bf \vdots}& \resizebox{!}{0.3 cm}{\bf \vdots}& \raisebox{0.1 cm}{\resizebox{!}{0.12 cm}{\rotatebox[origin=c]{45}{\dots}}} \\
\,X_{\!k}&\, _{\cdots}& \,X_{\!k}
\end{ytableau}
}
\, ,\,
{ \scriptsize
\begin{ytableau}
Y_{\!1}&\, _{\cdots}&\, _{\cdots}&\, _{\cdots}&\, _{\cdots}&\, _{\cdots}&\,Y_{\!1}\\
Y_{\!2}&\, _{\cdots}&\, _{\cdots}&\, _{\cdots}&Y_{\!2}\\
\resizebox{!}{0.3 cm}{\bf \vdots}&\resizebox{!}{0.3 cm}{\bf \vdots}& \resizebox{!}{0.3 cm}{\bf \vdots}& \raisebox{0.1 cm}{\resizebox{!}{0.12 cm}{\rotatebox[origin=c]{45}{\dots}}} \\
Y_{\!k}&\, _{\cdots}& Y_{\!k}
\end{ytableau}
}\
\right) \ ,
\ee
where $k\leq [n/2]$.
The vectors $X_i$ and $Y_i$ are meant to be unconstrained ($X_i \cdot X_j \neq 0$ and similarly for $Y_i$). It follows from the discussion of the previous section that these polynomials need to satisfy scaling \eqref{scalingSOn}, transversality \eqref{antisym}, tracelessness \eqref{tracelessness} and the Casimir equation \eqref{casimirtensor} both in the $X_i$ and in the $Y_i$. One can use these requirements to bootstrap the form of the polynomials \eqref{Pnl}. This approach was used in \cite{projectors} in order to obtain a vast class of such polynomials for generic $l_1$, and many choices of small integer values of $l_2, l_3$.

The simplest polynomial is the symmetric and traceless one,
\be
\label{Pnl_TST}
\frac{{\bf P}^{n}_{l}(X_1; Y_1)}{
(X_1 \cdot X_1)^{l/2} (Y_1 \cdot Y_1)^{l/2}
}= \frac{l!}{2^l (\frac{n}{2}-1)_l}  \ C^{\frac{n}{2}-1}_{l}\left(x \right) \ ,
\qquad x\equiv \frac{X_1 \cdot Y_1}{\sqrt{(X_1 \cdot X_1) (Y_1 \cdot Y_1)}}.
\ee
A less trivial example is the polynomial ${\bf P}^{n}_{l,1}$ which can be obtained in a closed form for any $l$ \cite{projectors, Rejon-Barrera:2015bpa} in terms of operations performed on the projector \eqref{Pnl_TST},
\ba
\!\!\!\!
\label{Pnl1}
\begin{array}{ll}
\dfrac{{\bf P}^{n}_{l,1}(X_1,X_2; Y_1,Y_2)}{(X_1\cdot X_1)^{\frac{l}{2}} \ (Y_1 \cdot Y_1)^{\frac{l}{2}}}= & c_{l,1}
\Big[ n\left(\frac{(X_1\cdot Y_2) (X_2\cdot Y_1)}{\sqrt{X_1\cdot X_1} \sqrt{Y_1\cdot Y_1}}-x (X_2\! \cdot \! Y_2)\right) \partial_x + \Big(x\frac{\left(X_1\cdot X_2 Y_1\cdot Y_2+X_1\cdot Y_2 X_2\cdot Y_1\right)}{\sqrt{X_1\cdot X_1} \sqrt{Y_1\cdot Y_1}}\\
& \quad \ -\frac{X_1\cdot X_2 X_1\cdot Y_2}{X_1\cdot X_1}-\frac{Y_1\cdot Y_2 X_2\cdot Y_1}{Y_1\cdot Y_1}-\left(x^2-1\right) X_2\! \cdot
\! Y_2\Big) \partial_x^2 
\Big]   C^{\frac{n}{2}-1}_{l}\left(x \right) \ ,
\end{array}
\ea
where $x$ is defined as in \eqref{Pnl_TST} and $c_{l,1}$
 is a normalization coefficient, irrelevant for the purposes of this paper.
Importantly, $l$ appears in \eqref{Pnl1} only through the Gegenbauer polynomial (beside the overall normalization $c_{l,1}$). This makes the formula convenient for generic integer $l$, and even suggests its analytic continuation to real values.
Moreover, in all the known cases the functions ${\bf P}^{n}_{l_1,\dots, l_k}$ take the form \cite{projectors}
\be
\label{P_Generic}
\frac{
{\bf P}^{n}_{l_1,\dots, l_k}(X_1,\dots, X_k; Y_1, \dots , Y_k )
}{
(X_1 \cdot X_1)^{l_1/2} (Y_1 \cdot Y_1)^{l_1/2}}
= \# \ {\bf D}^{n}_{l_2,\dots, l_k}(X_1,\dots, X_k; Y_1, \dots , Y_k, \partial_x)
C^{\frac{n}{2}-1}_{l_1}\left(x\right) 
\ee
where ${\bf D}^n_{l_2,\dots, l_k}$ are some explicit differential operators which can be found in \cite{projectors}. For example, ${\bf D}^n_{1}$ is the one defined in the square brackets in  \eqref{Pnl1}. 
Again, notice that in \eqref{P_Generic} the full $l_1$ dependence is carried by the Gegenbauer polynomial. 
In \cite{AdSWeightShifting} it was also found that  one can generate all the operators ${\bf D}^n_{l_2}$ (for any $l_2$) by acting successively with some weight shifting differential operators \cite{Karateev:2017jgd}.

%%%%%%%%%%%%%%%%%%%%%%%
\subsection{Tensor structures for $SO(d+1,1)$ mixed symmetry representations}
\label{SO(d+1,1)}
%%%%%%%%%%%%%%%%%%%%%%%
In this section we construct tensor structures for mixed symmetry representations of $SO(d+1,1)$, exploiting the fact that they can be seen, roughly speaking, as analytic continuations of representations of $SO(d+2)$.

It is convenient to lift CFT operators to the embedding space \cite{SpinningCC}. Given a primary $\Ocal^{\a_1,\dots \a_{|l|}}(x)$, defined on $x\in \mathbb{R}^{d}$, with $\a_i=1\dots d$, conformal dimension $\D$ and $SO(d)$ spin $l=(l_1\dots,l_{[d/2]})$ we can lift it to the embedding space as an operator whose indices have the same symmetries:
\be
\Ocal^{\a_1,\dots \a_{|l|}}(x) \rightarrow \Ocal^{A_1,\dots A_{|l|}}(P) \, 
\ee
where $|l|=l_1+\dots +l_{[d/2]}$ and  $P \in \mathbb{R}^{d+1,1}$. 
The tensor can be as usual encoded in a polynomial:
  \ytableausetup{centertableaux,boxsize=1.9 em}
\be
\label{SO(d+2)poly}
 \Ocal(P,Z^{(i)}) \equiv
 \Ocal(P)
\left(\,
{ \scriptsize
\begin{ytableau}
A^{1}_{\!1}&\, _{\cdots}&\, _{\cdots}&\, _{\cdots}& A^{1}_{l_{1}}\\
\resizebox{!}{0.3 cm}{\bf \vdots}&\resizebox{!}{0.3 cm}{\bf \vdots}& \resizebox{!}{0.3 cm}{\bf \vdots}& \raisebox{0.1 cm}{\resizebox{!}{0.12 cm}{\rotatebox[origin=c]{45}{\dots}}} \\
A^{k}_{\!1}&\, _{\cdots}& A^{k}_{l_{k}}
\end{ytableau}
}\
\right)
 \  \prod_{i=1}^k \ Z^{(i)}_{\; \,A^{i}_{1}} \cdots Z^{(i)}_{\; \,A^{i}_{l_{i}}} \ ,
\ee
where $k\leq [d/2]$.
The operator \eqref{SO(d+2)poly} is required to satisfy the following scaling and transversality relations
\begin{eqnarray}
&\displaystyle{\Ocal(\a P, \b_i Z^{(i)}) = \Ocal( P,  Z^{(i)}) \  \a^{-\D} \ \prod_{i=1}^{[\frac{d}{2}]}\ \b_i^{l_i}} \, , \label{ScalingConditions1}\\ 
&P \cdot \partial_{Z^{(j)}} \Ocal(P,Z^{(i)}) =0 \ , \qquad\qquad \ \  Z^{(k)}\cdot \partial_{Z^{(j)}} \Ocal(P,Z^{(i)}) =0 \, ,  \ (j<k). \label{ScalingConditions2}
\end{eqnarray}
We recognize the scaling and transversality conditions which we imposed for tensors of $SO(n)$. The only difference from what discussed in section \ref{ss:SOn_mixed_rep} is that the embedding space operator does not scale in $P$ as a polynomial. 
However, when $-\D \in \mathbb{N}$ we can think of $ \Ocal(P,Z^{(i)})$ as a tensor of $SO(d+2)$ associated to the Young tableau $(-\D,l_1,\dots,l_{[d/2]})$ with an extra line of $-\D$ symmetric indices contracted with ``polarization vectors'' $P$. In this case one can write the operator $ \Ocal$ as a polynomial following the recipe of section \ref{ss:SOn_mixed_rep},
  \ytableausetup{centertableaux,boxsize=1.9 em}
\be
\label{SO(d+2)operators}
 \Ocal(P,Z^{(i)}) 
 \sim
 \Ocal
\left(\,
{ \scriptsize
\begin{ytableau}
 *(yellow) A^{0}_{\!1}&  *(yellow) \, _{\cdots}&  *(yellow) \, _{\cdots}&  *(yellow) \, _{\cdots}&  *(yellow) \, _{\cdots}& *(yellow) \, _{\cdots}& *(yellow) \ A^{0}_{\!\mbox{-}\!\D}\\
A^{1}_{\!1}&\, _{\cdots}&\, _{\cdots}&\, _{\cdots}& A^{1}_{l_{1}}\\
\resizebox{!}{0.3 cm}{\bf \vdots}&\resizebox{!}{0.3 cm}{\bf \vdots}& \resizebox{!}{0.3 cm}{\bf \vdots}& \raisebox{0.1 cm}{\resizebox{!}{0.12 cm}{\rotatebox[origin=c]{45}{\dots}}} \\
A^{k}_{\!1}&\, _{\cdots}& A^{k}_{l_{k}}
\end{ytableau}
}\
\right)
 P_{A^{0}_{\!1}} \cdots P_{A^{0}_{- \D}} \;  \prod_{i=1}^k Z^{(i)}_{A^{i}_{\!1}} \cdots Z^{(i)}_{A^{i}_{l_{i}}} \ .
\ee
Of course, the indices in the first row (highlighted in yellow) cannot be defined for generic $\D\in \mathbb{R}$. Nevertheless, we keep in mind the picture \eqref{SO(d+2)operators} to motivate the following prescriptions, in analogy with the discussion on $SO(n)$ presented in section \ref{ss:SOn_mixed_rep}.
First, following section \ref{subsec:so(n)}, we consider vectors $P$ and $Z^{(i)}$  satisfying the conditions
\be\label{bulk_pol}
P\cdot P=0 \ , \qquad P \cdot Z^{(j)} =0 \ , \qquad Z^{(k)} \cdot Z^{(j)} =0 \ .
\ee
These conditions match the ones derived in \cite{SpinningCC} and \cite{MixedSym}. Notice, however, that the first two conditions have a different status here: they are forced on us by the projection onto physical space, as we review in subsection \ref{ss:real_rad}.

According to section \ref{subsec:so(n)}, we can think of the operator $ \Ocal(P,Z^{(i)})$ in \eqref{SO(d+2)operators} as contracted with antisymmetric tensors of the form
  \ytableausetup{centertableaux,boxsize=1.7 em}
\be
\label{Cbuilding}
C^{(m) }_{P \; \; A_1 \dots A_m} \equiv
P_{\; \; [A_1 } \ \ Z^{(1)}_{ \; \; A_2} \ \ \cdots \ \ Z^{(m-1)}_{ \; \; \; A_m]}  \ , \ \ \ \qquad m=1,\dots,\left[\frac{d}{2}\right]+1 \ .
\ee
To avoid cluttering, we do not explicitly denote the dependence of $C^{(m) A_1 \dots A_m}_{P}$ on the polarization vectors $Z^{(j)}$.

Using \eqref{Cbuilding} one can write a correlation function of generic operators $\Ocal_i(P_i,Z_i^{(j)})$ in terms of scalar contractions of the associated antisymmetric tensors $C^{(m) A_1 \dots A_m}_{P_i}$. 
As an example we define the following class of scalar contractions,
\be
\label{Tso(d+1,1)}
T^{n,m_1 m_2}_{i, j k} \equiv  C^{(n) \; A_1\dots A_n}_{P_i} \ \  \ C^{(m_1)}_{P_j \; \;A_1\dots A_{m_1}} \ \ \ C^{(m_2)}_{P_k \; \; A_{m_1+1} \dots A_{n}} \ ,
\ee
where $n=m_1+m_2$. Although more contractions are in general possible among the tensors \eqref{Cbuilding}, the \eqref{Tso(d+1,1)} are sufficient for the purposes of this work.
The structures \eqref{Tso(d+1,1)} satisfy the properties 
\be
T^{n,n 0}_{i, j k}\equiv T^{n,n}_{i, j}=T^{n,n}_{j, i}\ , 
\qquad
T^{n, m_1 m_2}_{i, j k}=(-1)^{m_1 m_2} \,  T^{n, m_1 m_2}_{i, k j} \ ,
\qquad
T^{n, m_1 m_2}_{i, j j}=0 \ ,
\ee
which easily descend from the symmetries of the tensors \eqref{Cbuilding}.
 The simplest instance is the scalar product
\be
T^{1,1}_{i, j} = P_i \cdot P_j \ .
\ee
The well known  $V_{i,jk}$ and $H_{ij}$ introduced in  \cite{SpinningCC} are in correspondence with the \eqref{Tso(d+1,1)} as well:
\be\label{three_point_BBB0}
\begin{array}{cclll}
T^{2,1 1}_{i, j k} &\leftrightarrow& V_{i,jk} 
&
\equiv \dfrac{(P_i\cdot P_j) (Z_i\cdot P_{k}) - (P_i\cdot P_{k}) (Z_i\cdot P_j)}{\sqrt{- 2 (P_i\cdot P_j)( P_j\cdot P_k)( P_k\cdot P_i)}} \ ,\\
T^{2,2}_{i, j} &\leftrightarrow& H_{ij} 
&
\equiv \dfrac{\left(P_i\cdot P_j) (Z_i\cdot Z_j)-(P_i\cdot Z_j) (P_j\cdot Z_i\right)}{\left(P_i \cdot P_j\right)   } 
\ .
\\
\end{array}
\ee
The structures $T^{2,1 1}_{i, j k}$ and $T^{2,2}_{i, j}$ scale with the $P_i$, while  $V_{i,jk}$ and $H_{ij}$ were chosen to be scale invariant. This is convenient, and we shall also often define structures with degree zero in the $P_i$, by an appropriate choice of factors $ P_i \cdot P_j$.
Let us now use the formalism to characterize two and three-point functions.

Before we proceed, we want to comment on one important difference between $SO(n)$ tensors and embedding operators seen as \eqref{SO(d+2)operators}. For $SO(n)$ tensors we chose polarization vectors $z^{(k)}$ which satisfy $z^{(k)} \cdot z^{(j)} = 0 $ as a trick, but in the end we needed to restore their dependence on unconstrained $z^{(k)} \in \mathbb{R}^n$ by using the prescription \eqref{recovertensor}. For embedding operators instead we will never want to restore their dependence on unconstrained $P\in \mathbb{R}^{d+1,1}$, since the subspace described by the null cone $P^2=0$ (and similarly $P\cdot Z^{(k)}=0$) is still redundant (it is a $d+1$ dimensional space, while the physical space is $\mathbb{R}^d$). In subsection \ref{ss:real_rad} we review how to  recover the physical space operators form the operators defined on the null cone by further restricting the vectors $P$ to a $d$-dimensional subspace of the null cone. On the other hand, one may want to lift the results to the case of embedding space polarization vectors which are not constrained to be transverse to each other, $Z^{(k)}\cdot Z^{(j)} \neq 0$. If the operator and the polarizations are transverse to $P$  -- \emph{i.e.} $\Ocal$ obeys the first of the \eqref{ScalingConditions2} and $P\cdot Z^{(i)}=0$ -- this operation can be performed by using the prescription of section  \eqref{recovertensor}, using $SO(d)$ projectors \cite{SpinningCC}.

%%%%%%%%%%%%%%%%%%%%%%%%%%%%%%%%%%%%%%
\subsection{Examples of correlation functions}
\label{correlatorsCFT}
%%%%%%%%%%%%%%%%%%%%%%%%%%%%%%%%%%%%%%
\paragraph{Two-point functions\\}
Using the structures \eqref{Tso(d+1,1)} it is trivial to see that all the two-point functions of operators $\Ocal_i(P_i,Z_i^{(j)})$, transforming in a representation $ \ell_i \in SO(d)$, are fixed in terms of a unique combination of structures. This combination is only allowed when $\D_1=\D_2\equiv \D$ and $\ell_1=\ell_2\equiv l=(l_1,\dots, l_{[\frac{d}{2}]})$ and it reads 
\be
\label{2ptBULK}
\langle 
\Ocal(P_1,Z_1^{(j)})
\Ocal(P_2,Z_2^{(j)})
\rangle
\propto \prod_{i=0}^{[\frac{d}{2}]} (T^{i+1,i+1}_{1, 2} )^{n_i}~.
\ee
In \eqref{2ptBULK} we wrote $\D,l$ in terms of the associated Dynkin labels $n=[n_0,n_1,\dots n_{[\frac{d}{2}]}]$ such that $n_0=-\D-l_1$, $n_{[\frac{d}{2}]}=l_{[\frac{d}{2}]}$ and $n_{i}=l_i-l_{i+1}$ (for $i=1\dots [\frac{d}{2}]-1$).
We stress that \eqref{2ptBULK} is the only possible combination of structures which satisfies equations (\ref{ScalingConditions1} - \ref{ScalingConditions2}).  As an example, the two-point function of traceless and symmetric operators of spin $l$  reduces to the usual expression,
\be
\langle 
\Ocal(P_1,Z_1)
\Ocal(P_2,Z_2)
\rangle =
\frac{H_{12}^{l} }{(-2 P_1\cdot P_2)^{\D}} \ .
\ee
Notice that a two-point function for $Z_i^{(j)}$ such that $Z_{i}^{(j)} \cdot Z_{i}^{(k)}\neq 0$  is fixed in terms of the polynomial ${\bf P}^{n}$ defined in \eqref{Pnl},\footnote{Actually a more elegant equivalent way to define the two point function is in terms of the analytically continued projector ${\bf P}^{d+2}_{-\D l_1,\dots, l_{[d/2]}}(P_1, Z_1^{(1)},\dots, Z_1^{([d/2])};P_2, Z_2^{(1)},\dots, Z_2^{([d/2])})$ which will be used in section \ref{Seed_Projectors}. It is easy to check that this analytically continued projector reduces to \eqref{2ptBULKprojector} when $P\cdot P=0$ and $P \cdot Z^{(k)}=0$. }
\be
\label{2ptBULKprojector}
\langle 
\Ocal(P_1,Z_1^{(i)})
\Ocal(P_2,Z_2^{(j)})
\rangle
\propto \frac{{\bf P}^{d+2}_{l_1 l_1,\dots, l_{[d/2]}}(P_1, Z_1^{(1)},\dots, Z_1^{([d/2])};P_2, Z_2^{(1)},\dots, Z_2^{([d/2])})}{(P_1 \cdot P_2)^{\D+l_1} }\ ,
\ee
It is possible to check that, setting $Z_{i}^{(j)} \cdot Z_{i}^{(k)}=0$ in  \eqref{2ptBULKprojector},  one does recover exactly \eqref{2ptBULK}. 
One can also check that \eqref{2ptBULKprojector} reduces to the two-point function in physical space once we write it in the Poincar\'e section described in subsection \ref{ss:real_rad}, with generic polarizations $z_{i}^{(j)}$.

%%%%%%%%%%%%%%%%%%%%%%%%
\paragraph{Three-point functions\\}
Here we classify the tensor structures in the OPE of two traceless and symmetric operators. 
We claim that any three point-function of operators $\Ocal_1,\Ocal_2$ traceless and symmetric with spin $l_1,l_2$ and an operator $\Ocal_3$ in a  representation with generic spin $(l_3^{(1)},l_3^{(2)},l_3^{(3)})$ can be written as follows
 \be
 \resizebox{0.93 \textwidth}{!}{$
 \langle 
\Ocal_1(P_1,Z_1)\Ocal_2(P_2,Z_2) \Ocal_3(P_3,Z^{(i)}_3)
\rangle
= 
\displaystyle{\sum_{\pf}} 
\!\!\!
\raisebox{1.7em}{
$
\xymatrix@=0.01pt{\Ocal_1 \ar@{-}[rdd]& &&&  \\  
&&&&\\
& *+[o][F]{\mbox{\tiny $\pf$}}  \ar@{-}[rrr] && &\Ocal_3  \\
&&&&\\
\Ocal_2 \ar@{-}[ruu]&&&& }
$
}
\!\!\!
=
\frac{\sum_{\pf} c^{(\pf)}_{123} Q^{(\pf)}(P_i,Z_i,Z^{(j)}_3) }{P_{12}^{\frac{\D_1+\D_2-\D_3}{2}}P_{13}^{\frac{\D_1+\D_3-\D_2}{2}}P_{23}^{\frac{\D_2+\D_3-\D_1}{2}}} \ .
$
}
\label{three_point_BBB}
\ee
where $c^{(\pf)}_{123}$ are the OPE coefficients and $P_{ij}\equiv - 2 (P_i\cdot P_j)$. Each OPE coefficient in \eqref{three_point_BBB} is multiplied by a conformal invariant structure of the form\footnote{The structures \eqref{3ptallstructures} are scaleless in all the $P_i$. The three point function \eqref{three_point_BBB} could have also been written in a compact way in terms of the structures $T^{n,m_1 m_2}_{i, j k}$ alone, but we decided for a form which may be more familiar to the reader.}
\be
\label{3ptallstructures}
Q^{(\pf)}(P_i,Z_i,Z^{(i)}_3)=\frac{\prod_{i=1}^3(V_i)^{n_i}  \prod_{i<j}(H_{ij})^{n_{ij}}  (T^{3,2 1}_{3, 1 2})^{k_{1}} (T^{3,2 1}_{3, 2 1})^{k_{2}} (T^{4, 2 2}_{3,1 2})^{k}}{[- 2 (P_1\cdot P_2)( P_2\cdot P_3)( P_3\cdot P_1)]^{\frac{k_1+k_2+k}{2}}} \, ,
\ee
where $V_1\equiv V_{1,23},  V_2\equiv V_{2,31},  V_3\equiv V_{3,12}$. 
The values of $\pf$ in \eqref{3ptallstructures} label the choices of exponents in the right hand side of \eqref{3ptallstructures}, which are non-negative integers subject to the conditions
\ba
\label{3ptconditions}
\begin{split}
k=l_3^{(3)} \ , \\
k_1+k_2+k=l_3^{(2)} \ , \\
n_3+ n_{13}+ n_{23}+k_1+k_2+k=l^{(1)}_3 \ , \\
n_1+ n_{12}+ n_{13}+k_1+k=l_1 \ , \\
n_2+ n_{12}+ n_{23}+k_2+k=l_2 \  .
\end{split}
\ea
As an example we write the number of structures $Q^{(\pf)}$ in some three point functions for fixed spin $l_1,l_2$ and generic $l^{(1)}_3$:
\ba
\label{table3pt}
\begin{split}
\!\!\!\!\!\!\!\!\!\!\!\!&
l_1,l_2=1,\ l^{(1)}_3 \geq 2\ ,  \quad
&&
\begin{array}{| c | c c c c |}
\hline 
(l^{(2)}_3,l^{(3)}_3) & (0,0) & (1,0) & (1,1) & (2,0)
\\
\hline
\# &  5 & 4  & 1 & 1  
\\
\hline
\end{array}
\\
\!\!\!\!\!\!\!\!\!\!\!\!&l_1,l_2=2,\ l^{(1)}_3 \geq 4 \ ,  \quad &&
\begin{array}{| c | ccccc ccc c|}
\hline 
(l^{(2)}_3,l^{(3)}_3) & (0,0) & (1,0) & (2,0)&(3,0)&(4,0)&(1,1)&(2,1)&(3,1)&(2,2)
\\
\hline
\# &  14 & 16 & 11&4&1&5&4&1&1
\\
\hline
\end{array}
\end{split}
\ea
It follows from the relations \eqref{3ptconditions}, as one can check from the table \eqref{table3pt}, that the non-vanishing three-point functions need to satisfy the conditions
\be
\label{existence3pt}
l_1+l_2 \geq l_3^{(2)}+l_3^{(3)} ~, \qquad l_1 \geq l_3^{(3)}~,
\qquad l_2 \geq l_3^{(3)}~.
\ee
In the table we also recognize some \emph{seed} three point functions, which are defined as the three point functions with only one tensor structure for generic $l_3^{(1)}$ (in this case $\pf$ only takes the value $1$, so we will drop it). We treat $\Ocal_3$ differently, asking for the length of its first row to be generic, because we think of it as the exchanged operator in the OPE of $\Ocal_1$ and $\Ocal_2$, see section \ref{sec:spinning}. 
From the conditions \eqref{3ptconditions} it is easy to see  that  a seed three point function is generated by exhausting all the polarization vectors $Z_{1}$ and $Z_2$ while building the tensor structures $T^{3,2 1}_{3, 1 2},T^{3,2 1}_{3, 2 1},T^{4, 2 2}_{3,1 2}$. In other words, seed three point functions can be obtained by looking for solutions of \eqref{3ptconditions} with $n_{ij}= n_1=n_2=0$. These are uniquely obtained as
\be
k=l_3^{(3)}~, 
\quad
n_3=l_3^{(1)}-l_3^{(2)}~,
\quad
k_1=l_1-l_3^{(3)}
~,
\quad
k_2=l_2-l_3^{(3)} ~,
\ee
provided that the external operators satisfy the following seed condition\footnote{Actually, also when $l_1\,(\mbox{or }l_2)=l_3^{(3)},\ l_3^{(1)}=l_3^{(2)}$ and $l_1+l_2 > l_3^{(2)}+l_3^{(3)}$ there is only one structure. This additional case, where $l_3^{(1)}$ is bounded for fixed $l_1,\,l_2$, is not counted among the seeds. Indeed, the same $\Ocal_3$ is exchanged by external primaries with lower spin, which saturate the first of the \eqref{existence3pt}. The single structure of the additional case can then be obtained by applying the spinning operators of subsection \ref{Spinning_Differential_Operators_Bulk}.}
\be
\label{3ptseed}
l_1+l_2=l_3^{(2)}+l_3^{(3)} \quad  \Leftrightarrow\quad \mbox{seed} 
\ .
\ee
The seed three point functions saturate the first of the three conditions \eqref{existence3pt}.
 The requirement \eqref{3ptseed} matches the one obtained in \cite{projectors}.
 
 The prominence of the seeds stems from the fact that all other three-point functions are obtained by acting on them with a set of differential operators \cite{Costa2011} which increase the spins $l_1$ or $l_2$. In fact, as we explain in subsection \ref{Spinning_Differential_Operators_Bulk} and in appendix \ref{recrelBULK},  the minimal set is even smaller. Out of the $l^{(2)}_3-l^{(3)}_3+1$ pairs $(l_1,l_2)$ which exchange a given $\Ocal_3$ as a seed, only one is necessary. The others can be in fact obtained by acting with differential operators \cite{projectors} which map seeds into seeds (see appendix \ref{recrelBULK}).
It is therefore convenient to choose a representative seed three point function for each $\Ocal_3$ exchanged. A natural choice is to consider seeds that also saturate another of the \eqref{existence3pt}, say $l_2=l_3^{(3)}$,
 \be
\label{3ptseedrepresentative}
 \mbox{seed representative} \equiv \!\!
 \raisebox{2.1em}{
$
\xymatrix@=0.01pt{
{\Ocal_{\D_1 \,  l_1= l_3^{(2)}}}
\ar@{-}[rdd]    \\  
\\
& *+[o][F]{}  \ar@{-}[rrr] && &\Ocal_{\D_3 \, l^{(1)}_3  \, l^{(2)}_3 \, l_3^{(3)} }  \\
{\Ocal_{\D_2 \,  l_2= l_3^{(3)}}}
\ar@{-}[ru] 
}
$
}
\ .
\ee
In appendix \ref{recrelBULK} it is detailed how to obtain all the three-point functions by acting with a set of differential operators on the representative seeds \eqref{3ptseedrepresentative}.

As a last remark, we would like to discuss conservation of seed correlation functions. Let us consider the case of a seed three-point function $\langle \Ocal_1 \Ocal_2 \Ocal_3\rangle$ in which one symmetric and traceless operator (let say $\Ocal_1^{\m_1\dots \m_{l_1}}$) is conserved, namely it satisfies $\partial_{\m_1} \Ocal_1^{\m_1\dots \m_{l_1}}=0$. It is trivial to see that the seed three-point function is automatically conserved. Indeed, since the three-point function with the operator $\Ocal_1$ saturates the condition \eqref{existence3pt}, the three-point function with $(\partial_{\m_1} \Ocal_1^{\m_1\dots \m_{l_1}})$ violates it, thus it vanishes.
The same argument holds for more generic seed correlation functions, because seed correlations functions saturate a condition of existence of the kind \eqref{existence3pt}. In subsection \ref{subsec:corr_defect}, we shall see another example of conservation of seed correlation functions in the case of the bulk-defect two-point functions.

%%%%%%%%%%%%%%%%%%%%%%%
\section{Mixed symmetry representations and defect CFTs}
\label{SO(d+1,1)SO(q)}
%%%%%%%%%%%%%%%%%%%%%%%
In a defect CFT, a $p$ dimensional defect breaks the $SO(d+1,1)$ symmetry to a $SO(p+1,1)\times SO(q)$ (with $p+q=d$) subgroup of the original conformal group. As in the pure CFT case, the non linear realization of the stability group of the vacuum makes it hard in general to implement the symmetry constraints on correlation functions. The uplift to the embedding space for defect CFTs of general codimension was worked out in \cite{Billo:2016cpy}. In the present section, we will extend the analysis of \cite{Billo:2016cpy} to operators transforming in mixed symmetry representations of $SO(d)$. This problem was addressed in the recent paper \cite{Guha:2018snh}, using the formalism of \cite{MixedSym}. Our solution, as in section \ref{sec:mixedCFT}, employs commuting polarization vectors to build a minimal set of structures with no redundancy, thus facilitating the task of enumerating them.

Before presenting the results, let us set our conventions up. Following \cite{Billo:2016cpy}, in the embedding space it is convenient to split the $(d+2)$-dimensional scalar product $P\cdot Q \equiv \sum_{M} P^M Q_M$ into its counterparts parallel and transverse to the defect, which is always lifted to a $(p+2)$-dimensional \emph{time-like} plane. Following the convention of \cite{Lauria:2017wav}, we implement the splitting by defining projectors $\Pp,\Pt$:
\ba
\label{Pdot}
P{\pdot} Q & \displaystyle\equiv P \cdot \Pp \cdot Q 
& \qquad \mbox{(parallel)}~, \\
\label{Tdot}
P {\tdot } Q  &\displaystyle \equiv P \cdot \Pt \cdot Q
&\qquad \mbox{(orthogonal)}~,
\ea
with $\Pp+\Pt=\text{diag}(-1,1,\dots,1)$. The shape of the defect in physical space  can be chosen by specifying the form of the projectors (\ref{Pdot} - \ref{Tdot}). With the usual conventions for the projection onto the Poincaré section, to define a flat defect it is sufficient to take the axis $P^{-}$ to lie on the parallel subspace, while in general the defect will be spherical \cite{Lauria:2017wav}. Of course, equations (\ref{Pdot} - \ref{Tdot}) define a splitting of the physical space scalar product $x \cdot y\equiv x^\mu y^\nu \delta_{\mu \nu}$ as well
\ba
\label{Prdot}
x{\pdot} y & \displaystyle\equiv x \cdot \pp \cdot y 
 & \qquad \mbox{(parallel)}~, \\
\label{Trdot}
x{\tdot} y & \displaystyle\equiv x \cdot \pt \cdot y 
& \qquad \mbox{(orthogonal)}~,
\ea
with $\pp+\pt=\delta$. When the defect is spherical and centered in the origin, the $p+1$ directions in which the defect is embedded are defined to be parallel.

\subsection{Operators and tensor structures}
There are two classes of operators: bulk operators $\Ocal$ and defect operators $\hOcal$. Bulk insertions are the same local operators of a $d$ dimensional CFT. We discussed them in section \ref{SO(d+1,1)}.
Defect operators $\hOcal$ deserve a separate treatment. Since they live on the defect, they can be thought of as operators of a $p$ dimensional CFT, with quantum numbers under symmetries acting \emph{parallel} to the defect. They also transform under an $SO(q)$ global symmetry, the rotations \emph{trasverse} to the defect. 
The defect operators $\hOcal$  are therefore labeled by  $SO(p+1,1)\times SO(q)$ quantum numbers, in particular
\be
\begin{array}{l}
\left.
\begin{array}{l l}
\hD & \mbox{ conformal dimension,} \\
\hl=\big(\hl_1,\dots \hl_{[p/2]}\big) & \mbox{ \emph{parallel} $SO(p)$ spin,} \\
\end{array}
\right\}  SO(p+1,1)
\\
\\
s=\big(s_1,\dots s_{[q/2]}\big) \mbox{ \emph{transverse} $SO(q)$ spin.} \\
\end{array}
\ee
As we did in \eqref{SO(d+2)poly}, we consider operators in embedding space, $\hOcal(P)$ with $P$ living on the defect ($P_M \Pp^{MN}=P^N$), and contract the $SO(p)$ indices with polarization vectors $Z^{(1)}\dots Z^{([p/2])}$ and the $SO(q)$ indices with new polarization vectors $W^{(1)}, \dots W^{([q/2])}$. 
In order to make analogies with  \ref{subsec:so(n)}, we also repeat the construction \eqref{SO(d+2)operators} for defect operators,
\ytableausetup{centertableaux,boxsize=2. em}
\be
 \hOcal \big(P,Z^{(j)},W^{(j)}\big)
\sim
 \hOcal
\left(
{ \scriptsize
\begin{ytableau}
 *(yellow) P&  *(yellow) \, _{\cdots}&  *(yellow) \, _{\cdots}&  *(yellow) \, _{\cdots}&  *(yellow) \, _{\cdots}& *(yellow) \, _{\cdots}& *(yellow) \ P\\
\, Z^{\resizebox{.28 cm}{!}{$(1)$} }&\, _{\cdots}&\, _{\cdots}&\, _{\cdots}& \, Z^{\resizebox{.28 cm}{!}{$(1)$} }\\
\resizebox{!}{0.3 cm}{\bf \vdots}&\resizebox{!}{0.3 cm}{\bf \vdots}& \resizebox{!}{0.3 cm}{\bf \vdots}& \raisebox{0.1 cm}{\resizebox{!}{0.12 cm}{\rotatebox[origin=c]{45}{\dots}}} \\
Z^{\resizebox{.28 cm}{!}{$(k)$} }
&\, _{\cdots}& 
Z^{\resizebox{.28 cm}{!}{$(k)$} }
\end{ytableau}
}\,
,
\ytableausetup{centertableaux,boxsize=2.2 em}
{ \scriptsize
\begin{ytableau}
\, W^{\resizebox{.28 cm}{!}{$(1)$} }&\, _{\cdots}&\, _{\cdots}&\, _{\cdots}&\, W^{\resizebox{.28 cm}{!}{$(1)$} }\\
\resizebox{!}{0.3 cm}{\bf \vdots}&\resizebox{!}{0.3 cm}{\bf \vdots}& \resizebox{!}{0.3 cm}{\bf \vdots}& \raisebox{0.1 cm}{\resizebox{!}{0.12 cm}{\rotatebox[origin=c]{45}{\dots}}} \\
\,W^{\resizebox{.28 cm}{!}{$(h)$} }
&\, _{\cdots}& 
\,W^{\resizebox{.28 cm}{!}{$(h)$} }
\end{ytableau}
}\
\right) \ ,
\ee
where $k \leq [\frac{p}{2}]$ and $h\leq [\frac{q}{2}]$.
As we did in \eqref{SO(d+2)operators}, we coloured in yellow the line of the tableau which makes sense for $-\hD \in \mathbb{N}$. We think of $P_M,\, Z^{(j)}_{\; M} ,\, W_{\; M}^{(j)}$ as vectors in $\mathbb{R}^{d+1,1}$ (therefore $M=0,\dots d+1$), but  $P,\,Z^{(j)}$ only have non zero components parallel to the defect, while the only non zero components of  $W^{(j)}$ are transverse to the defect, namely $Z^{(j)}_{\; M} \ \Pp^{MN}=Z^{(j)\; N}$  and $W^{(j)}_{\; M} \ \Pt^{MN}=W^{(j) \; N}$.

We associate to a defect operator two sets of antisymmetric tensors: the $C^{(n) M_1 \dots M_n}_{P}$ as defined in \eqref{Cbuilding}, with $n=1,\dots,\left[\frac{p}{2}\right]+1$, and the following
\be
\label{CWbuilding}
C^{(n) }_{W \; \; \; M_1 \dots M_n} \equiv
 W^{(1)}_{ \; \; \;  [M_1} \ \cdots \ W^{(n)}_{ \; \; \; M_n]}  \ , \qquad n=1,\dots,\left[\frac{q}{2}\right] \ ,
\ee
 which are of the form \eqref{buildingC}. 
We can build all the conformal invariant structures appearing in a correlation function of bulk and defect operators by contracting the tensors $C^{(n)}_{P_i}$ and  $C^{(n)}_{W_j}$. The only extra ingredient are the projectors (\ref{Pdot} - \ref{Tdot}), namely the indices of the $C^{(m)}$ can be contracted either with  $\Pp$ or with $\Pt$.
For our purposes, the following class of structures will suffice:
\be
\label{defTso(d+1,1)}
T^{\star \, n,m_1 m_2}_{I, J K} \equiv  C^{(n)}_{I \; \; \; M_1\dots M_n} \ \ C^{(m_1)}_{J \; \; \; N_1\dots N_{m_1}} \ \  C^{(m_2)}_{K \; \; \; N_{m_1+1} \dots N_{n}} \  \ \Ps^{M_1 N_1} \cdots  \Ps^{M_n N_n}  \ ,
\ee
where $\star=\pdot,\tdot$ labels the two projectors and the capital letters $I,J,K=P_i,W_i$ are introduced  to distinguish between $C_{P_i}$ and $C_{W_i}$. For simplicity we denote $T^{\star \, n,n0}_{I, J K}\equiv T^{\star \, n,n}_{I, J}$.

Explicit examples are the building blocks analogous to the $H_{ij}$ and $V_{i,jk}$ of \eqref{three_point_BBB0}
\be
\begin{array}{cclll}
\vspace{0.2 cm}
T^{\star \, 2,1 1}_{P_i, P_j P_k}& \leftrightarrow& \Vs{i,jk} 
&
\equiv \dfrac{(P_i\star P_j) (Z_i\star P_{k}) - (P_i\star P_{k}) (Z_i\star P_j)}{\sqrt{- (P_i\pdot P_j)( P_j\pdot P_k)( P_k\pdot P_i)}} \ ,\\
T^{\star \, 2,2}_{P_i, P_j }& \leftrightarrow&  \Hs{ij} 
&
\equiv \dfrac{\left(P_i\star P_j) (Z_i\star Z_j)-(P_i\star Z_j) (P_j\star Z_i\right)}{\left(P_i \star P_j\right)   } 
\ ,
\\
\end{array}
\label{HVdef}
\ee
where $\star= \pdot, \tdot $. By taking an opportune set of linearly independent $ \Vs{i,jk} $ and $\Hs{ij} $ one can write any correlation function of bulk traceless and symmetric operators. In the following, we will give a set of linearly independent structures for the two-point function.

Correlation functions of defect operators only can be also written in terms of structures of the kind \eqref{HVdef}, but in this case the label $\star=\pdot$, since the $P$ and $Z^{(j)}$ live in the parallel space. In order to take into account the global symmetry $SO(q)$ of the operators, we need to add to the mix similar structures obtained by transverse contractions of $C^{(n)}_{W_i}$, like $T^{\tdot \, 1,1}_{W_i ,W_j }=W^{(1)}_i\tdot W^{(1)}_j$.

A more involved set of structures appears for correlation functions involving both bulk and defect operators. Indeed, it is possible to contract  $C^{(n)}_{W_i}$ of a defect operator with $C^{(n)}_{P_j}$ of a bulk operator, in this case using the transverse product, since the $W_i$ are orthogonal to the defect. The simplest structures are
\be
\label{WStructures}
\begin{array}{cclll}
 \vspace{0.1cm}
T^{\tdot \, 1,1}_{W_i, P_j } &\leftrightarrow & K^i_{j}&\equiv  \dfrac{ W^{(1)}_{i} \tdot P_j}{(P_j \tdot P_j)^{1/2}} \ , 
  \\
   \vspace{0.1cm}
T^{\tdot \, 2,1 1}_{P_i, W_j P_k } &\leftrightarrow &Y^j_{i, k} &\equiv \dfrac{ (P_i\tdot P_k) (Z_i\tdot W^{(1)}_j) - (P_i\tdot W^{(1)}_j) (P_k\tdot Z_i) }{P_i \tdot P_k}\ , \\
 \vspace{0.1cm}
T^{\tdot \, 2,2}_{W_i, P_j } &\leftrightarrow &S^{i}_{j}&\equiv \dfrac{(W^{(1)}_i \tdot P_j) (W^{(2)}_i \tdot Z_j)-(W^{(2)}_i \tdot P_j) (W^{(1)}_i \tdot Z_j)}{\left(P_j \tdot P_j\right)^{1/2}   } \ .
\end{array}
\ee
Again we normalized the structures so that they have degree zero in $P_i$.
In the next subsection, we exemplify the formalism.

\subsection{Examples of correlation functions}
\label{subsec:corr_defect}
%%%%%%%%%%%%%%%%%%%%%%%%%%%%%%
\paragraph{One-point functions\\}
%%%%%%%%%%%%%%%%%%%%%%%%%%%%%%
The one-point function of a bulk operator $\Ocal(P_1,Z_1^{(j)})$ can be constructed with the structure $T^{\pdot \, n, n}_{P,P}$. Notice that, since $P^2=P\pdot P+P\tdot P=0$, the structure defined with $\tdot$ is not linearly independent, $T^{\tdot \, n, n}_{P,P}=(-1)^n  T^{\pdot \, n, n}_{P,P}$. We obtain, up to a normalization constant,
\be
\label{1pt}
\langle 
\Ocal(P,Z^{(j)})
\rangle
\propto \prod_{i=0}^{i_\textup{max}} (T^{\pdot i+1,i+1}_{P, P} )^{n_i/2} ~,\qquad i_\textup{max} = \min\left(p+1,q-1,\left[d/2\right]\right)~.
\ee

In \eqref{1pt}, as in \eqref{2ptBULK}, we wrote $\D,l$ in terms of the associated Dynkin labels $n=[n_0,n_1,\dots n_{[\frac{d}{2}]}]$ such that $n_0=-\D-l_1$, $n_{[\frac{d}{2}]}=l_{[\frac{d}{2}]}$ and $n_{i}=l_i-l_{i+1}$ (for $i=1\dots [\frac{d}{2}]-1$). 
For example, the one-point function of  a traceless and symmetric  primary of spin $l$ is fixed as follows
 \be
 \langle 
\Ocal(P_1,Z_1)
\rangle
= a_{\Ocal}
\frac{(\Hp{11})^{l/2} }{(-P_1 \pdot P_1)^{\frac{\D}{2}}} \ .
\label{onepoint}
\ee
The function \eqref{1pt} must be a polynomial in the $Z_1^{(j)}$, and this only happens if all the $l_{i}$ are even. We conclude that the operators with non vanishing one-point function are\footnote{This rule applies to parity even primaries. The one-point function of parity odd operators is constructed by contracting structures $C^{(n)}$ with the epsilon tensor of the $p+2$ or $q$ dimensional space. Using this recipe, it is not hard to classify the parity odd operators which can acquire a one-point function.}
\be\label{nonzero1pt}
\Ocal_{\D \, 2 l_1 \dots 2 l_{i_\textup{max}}} ~.
\ee
The maximum number of rows $i_\textup{max}$ cannot exceed $[d/2]$, of course, nor can it be larger than  $\min(p+1,q-1)$. The latter bound follows from the fact that it is impossible to antisymmetrize a larger number of vectors in the parallel or in the orthogonal subspaces, which makes the structures \eqref{Cbuilding} vanish identically if they have too many indices.
As in the case of the two-point functions \eqref{2ptBULK}, the one-point function can be written in terms of a projector when the $Z^{(j)}$ obey $Z^{(j)}\cdot P=0$, but $Z^{(i)}\cdot Z^{(j)}$ is unconstrained.
In this case, the projector is contracted on one side with the $Z^{(j)}$ and on the other side with $\Pp^{M N}$:
\be
\label{1ptproj}
\langle 
\Ocal_{\D l}(P_1,Z_1^{(j)})
\rangle
\propto \frac{{\bf P}^{\pdot d+2}_{l_1 l_1,\dots, l_{[d/2]}}(P_1,Z_1^{(1)},\dots, Z_1^{([d/2])})}{(P_1 \pdot P_1)^{\D+l_1} }\ .
\ee
Here we introduced the notation ${\bf P}^{\pdot d+2}$ for the following polynomial 
\be
\begin{array}{ll}
\label{Ptnl}
{\bf P}^{\pdot d+2}_{l_1,\dots, l_k}(X_1,\dots, X_k)
\equiv 
&
{\pi}^{SO(d+2)}_{l_1,\dots, l_k} \left (
 \
  { \scriptsize
  \ytableausetup{centertableaux,boxsize=1.7 em}
\begin{ytableau}
\, X_{\!1}&\, _{\cdots}&\, _{\cdots}&\, _{\cdots}&\, _{\cdots}&\, _{\cdots}&\,X_{\! 1}\\
\, X_{\!2}&\, _{\cdots}&\, _{\cdots}&\, _{\cdots}&\,X_{\!2}\\
\resizebox{!}{0.3 cm}{\bf \vdots}&\resizebox{!}{0.3 cm}{\bf \vdots}& \resizebox{!}{0.3 cm}{\bf \vdots}& \raisebox{0.1 cm}{\resizebox{!}{0.12 cm}{\rotatebox[origin=c]{45}{\dots}}} \\
\,X_{\!k}&\, _{\cdots}& \,X_{\!k}
\end{ytableau}
}
\, ,\,
{ \scriptsize
\begin{ytableau}
M^{1}_{\!1}&\, _{\cdots}&\, _{\cdots}&\, _{\cdots}&\, _{\cdots}&\, _{\cdots}&\,M^{1}_{l_{1}}\\
M^{2}_{\!1}&\, _{\cdots}&\, _{\cdots}&\, _{\cdots}&M^{2}_{l_{2}}\\
\resizebox{!}{0.3 cm}{\bf \vdots}&\resizebox{!}{0.3 cm}{\bf \vdots}& \resizebox{!}{0.3 cm}{\bf \vdots}& \raisebox{0.1 cm}{\resizebox{!}{0.12 cm}{\rotatebox[origin=c]{45}{\dots}}} \\
M^{k}_{\!1}&\, _{\cdots}& M^{k}_{l_{ k}}
\end{ytableau}
}\
\right) 
\vspace{0.2 cm}
\\
& \times 
\prod_{j=1}^{k} \Pp^{M^{j}_{\!1} M^{j}_{\!2}} \cdots \Pp^{M^{j}_{\! l_{j-1}} M^{j}_{\! l_{j}}} \ ,
\end{array}
\ee
where $k\leq [d/2]+1$.
As in footnote \ref{footnotePd}, with abuse of notation we will use the functions \eqref{Ptnl} both in physical and in embedding space. The metric for the scalar products is intended as the metric of these spaces. Moreover when we use \eqref{Ptnl} in physical space we contract the $SO(d)$ projector to  $\pp^{\m \n}$ -- defined in \eqref{Prdot} -- instead of its embedding space counterpart $\Pp^{M N}$.
By construction, these polynomials are non zero only when $l_1,\dots,l_k$ are all even numbers. Also, it is easy to check that \eqref{1ptproj} reduces to \eqref{1pt} when  $Z^{(j)} \cdot Z^{(k)}=0$. 
We can give some examples of these polynomials 
$$
{\bf P}^{\pdot d}_{j}(X_1)=\frac{\left(\frac{-j-q+3}{2} \right)_{\frac{j}{2}} }{\left(\frac{d+j-2}{2} \right)_{\frac{j}{2}}}
(X_1 \cdot X_1)^{j/2} \, _2F_1\left(-\frac{j}{2},\frac{d+j-2}{2};\frac{p+1}{2};\frac{X_1 \pdot X_1}{X_1 \cdot X_1}\right)
$$
and similarly in appendix \ref{poly_Proj_j2} we define the function ${\bf P}^{\pdot d}_{j,2}$. 

In the next section, the polynomials ${\bf P}^{\pdot n}$ will play an important role in the computation of the bulk channel conformal blocks.

%%%%%%%%%%%%%%%%%%%%%%%%%%%%%%%%%%%%%%%%%
\paragraph{Two-point functions of bulk traceless and symmetric operators\\}
%%%%%%%%%%%%%%%%%%%%%%%%%%%%%%%%%%%%%%%%%
In this case  the set of linearly independent building blocks is 
\be
\label{2pt_Structures}
\begin{array}{l l l l}
\Vp{1 } \ , & \quad \Vp{2}\ , & \quad \Vt{1}\ , &\quad \Vt{2} \ , \\
\Hp{12}\  ,& \quad \Ht{12}\ , &\quad \Hp{11} \  , &\quad \Hp{22} \ ,
\end{array}
\ee
where $\Vs{i}$ is defined in terms of the structures $\Vs{1,12}$ and $\Vs{2,21}$ of \eqref{HVdef}:
\be
\Vs{i} \equiv \Vs{i,i(3-i)} \left[ \frac{P_1\pdot P_2}{P_1\star P_2}\sqrt{ \frac{P_1\pdot P_2}{P_1\cdot P_2}} \right] 
\ .
\label{NormalizationHV}
\ee
Terms in square brackets are cross ratios, thus are not essential. They have been chosen so that the tensor structures remain finite and linearly independent in both the bulk and the defect OPE limits (see appendix \ref{tensor structures to radial frame}). In order to prove completeness of the structures \eqref{2pt_Structures}, it is convenient to use the radial coordinates defined in subsection \ref{ss:real_rad}. Using the bulk radial coordinates to fix ideas, the elementary building blocks are the bilinears in the physical space polarizations $z_1,\ z_2$ and in the angle $n$. It is easy to see that said bilinears are in one-to-one correspondence with the \eqref{2pt_Structures}, thus proving the completeness of the latter.\footnote{This basis can be also written in terms of the structures defined in \cite{Billo:2016cpy}.}

In sum, a two-point function of bulk operators $\Ocal_i$ of dimension $\D_i$ and spin $l_i$ can be written as follows
\be\label{newStructEmi}
\langle \Ocal_1(P_1,Z_1) \Ocal_2(P_2,Z_2) \rangle = \frac{1}{(P_1 \tdot P_1)^{\frac{\D_1}{2}}(P_2 \tdot P_2)^{\frac{\D_2}{2}}} 
\sum_{k} f^k(\{u_a\}) Q_k(P_1,P_2,Z_1,Z_2) \ .
\ee
The structures  $Q_k$ are given in terms of the building blocks
\be
\label{BB_tensor_structures}
Q_k=(\Hp{12})^{m^{\pdot}_{12}} (\Ht{12})^{m^{\tdot}_{12}}  \prod_{i=1}^2 \ (\Vp{i})^{n^\pdot_{i}}  (\Vt{i})^{n^\tdot_{i}} (\Hp{ii})^{m^\pdot_{ii}} \ ,
\ee
where the index $k$ labels the choice of non-negative integers $n^\star_{i}$ and $m^\star_{ij}$  which satisfy the relation
\be
l_i= m^\pdot_{12}+m^\tdot_{12}+n^\pdot_{i}+n^\tdot_{i}+2 m^\pdot_{ii} \ .
\ee
The functions $f^k(\{u_a\})$ depend on two cross ratios \cite{Lauria:2017wav}.

%%%%%%%%%%%%%%%%%%%%%%%%%%%%%%%%%%%%%%%%%
\paragraph{Two-point functions of generic defect operators\\}
%%%%%%%%%%%%%%%%%%%%%%%%%%%%%%%%%%%%%%%%%
The two-point function of a defect  operator $\hOcal$ with transverse spin $s$ and parallel spin $\hl$ is  fixed  in a combination of tensor structures 
 \be
 \langle 
\hat \Ocal(P_1,Z^{(j)}_1,W^{(j)}_1)
\hat \Ocal(P_2,Z^{(j)}_2,W^{(j)}_2)
\rangle
\propto  \prod_{i=0}^{[\frac{p}{2}]} (T^{\pdot i+1,i+1}_{P_1, P_2} )^{n_i}  \times \prod_{i=1}^{[\frac{q}{2}]} (T^{\tdot i,i}_{W_1, W_2} )^{m_i} \ .
\ee
As before  $n=[n_0,n_1,\dots n_{[\frac{p}{2}]}]$ are defined as the Dynkin labels associated to $(-\D,l_1,\dots l_{[p/2]})$ such that $n_0=-\D-l_1$, $n_{[\frac{p}{2}]}=l_{[\frac{p}{2}]}$ and $n_{i}=l_i-l_{i+1}$ (for $i=1\dots [\frac{p}{2}]-1$). The extra labels $m=[m_1,\dots m_{[\frac{q}{2}]}]$ are instead the Dynkin labels associated to $s=(s_1,\dots s_{[q/2]})$.

As an explicit example, the two-point function of a defect  operator $\hOcal$ with symmetric and traceless parallel and  transverse spin  $\hl$ and   $s$, is fixed as follows: 
 \be
 \langle 
\hat \Ocal(P_1,Z_1,W_1)
\hat \Ocal(P_2,Z_2,W_2)
\rangle
= 
\frac{(\Hp{{1}{2}})^{\hl} \ (W_1 \tdot W_2)^s  }{(-2 P_1 \pdot P_2)^{\frac{\hD}{2}}} \ .
\ee

It is easy to generalize this result to higher point functions of defect operators, since they coincide with those of a $p$-dimensional CFT with a global symmetry. Vice versa one can also use this formalism in order to write correlation functions of CFTs with global symmetry. 
%%%%%%%%%%%%%%%%%%%%%%%%%%
\paragraph{Bulk-defect two-point functions\\}
\label{ss:bulk-to-defect}
%%%%%%%%%%%%%%%%%%%%%%%%%%
We consider now a correlation function of a bulk symmetric traceless primary $ \Ocal$, with dimension $\D$ and spin $l$, and a generic defect primary $\hOcal$, with dimension $\hD$ parallel spin $\hl$ and transverse spin $s$. Using the structures \eqref{defTso(d+1,1)}, it is easy to see that the defect operator is fixed to be a traceless and symmetric parallel tensor of spin $\hl$ and a tensor of orthogonal spin $s=(s_1,s_2)$, with $s_i=0$ for $i>2$. Therefore
\begin{equation}\label{btodefect2pt}
\langle \Ocal(P_1,Z_1)  \hO(P_2,Z_2,W^{(j)}_2) \rangle = 
\sum_{\pf} 
 \raisebox{1.6em}{
$
\xymatrix@=0.3pt{
&&& \ar@{=}[dd] & \\
&&& &\\
{\Ocal}\ar@{-}[rrr]& &&*+[o][F]{\mbox{\tiny$\pf$}} \ar@{=}[dd]&  \hO\\  
&  &&& \\
&&&&\\
}
$
} 
=\frac{\sum_{\pf} b^{(\pf)}_{\Ocal \hO} Q^{(\pf)}(P_1,Z_1,P_2,Z_2,W^{(j)}_2)}{(-2 P_1 \pdot P_2)^{\hD} (P_1 \tdot P_1)^{\frac{\D-\hD}{2}}} 
 \ ,
\end{equation}
where\footnote{The building blocks map to the ones defined in \cite{Billo:2016cpy} as follows:
\begin{align}
\Hp{1{2}}=Q_{BD}^0, \quad K^{ 2}_{1}=Q_{BD}^1, \quad \Vp{1,1{2}}= -Q_{BD}^2, \quad Y^{{2}}_{1,1}=Q_{BD}^3, \quad \Hp{11}=Q_{BD}^4.
\end{align}
We introduced an extra structure in order to take into account operators with transverse spin $s_2>0$.
}
\be\label{BuildBlocksBulkToDef}
 Q^{(\pf)}=\left(\Hp{1{2}}\right)^{\hl}\ (S^{2}_{1})^{s_2} \  (\Hp{11})^{m_{11}}  \ (\Vp{1,1{2}})^{n_{1}} \  (K^{ 2}_{1})^{n_{2}} \ (Y^{ 2}_{1 ,1})^{m_{12}} \, .
\ee
The structures involved are defined in \eqref{HVdef} and \eqref{WStructures}.
The index $\pf$ labels the choices of non-negative integers $m_{ij}$ and $n_i$ which satisfy the constraints
\ba
\label{condition_BD_integers}
\begin{split}
l-\hl-s_2=&2 m_{11 }+n_{1}+m_{1 2} \ ,
\\
s_1-s_2=&n_{ 2}+m_{1  2} \ .
\end{split}
\ea
Eq. \eqref{condition_BD_integers} implies the  requirement
\be
l \geq \hl +s_2 \ .
\ee
Let us now exemplify the counting of tensor structures for fixed  spin of the bulk operator. Analogously to the discussion in subsection \ref{SO(d+1,1)}, we are mainly interested in defect operators with label $s_1$ generic, which in this case means $s_1\geq l-\hl$ :
\ba
\!\!\!\!\!\!\!\!\!\!\!\!&
l=1~,\ s_1 \geq 1-\hl~,  \quad
&
\begin{array}{| c | c c c |}
\hline 
(\hl,s_2) & (0,0) & (1,0) &(0,1)
\\
\hline
\# &  2 & 1& 1
\\
\hline
\end{array}
\\
\!\!\!\!\!\!\!\!\!\!\!\!&l=2~,\ s_1 \geq 2-\hl~,  \quad &
\begin{array}{| c | ccccc c|}
\hline 
(\hl,s_2) & (0,0) & (1,0) & (0,1)&(2,0)&(1,1)&(0,2)
\\
\hline
\# &  4 & 2 & 2 &1&1&1
\\
\hline
\end{array}
\ea
From formula \eqref{condition_BD_integers} and from the table it is clear that, for large enough $s_1$, the number of structures only depends on the difference $N \equiv l-\hl-s_2$ and in particular it is equal to $\#=(1+N/2)^2$ for even $N$ and $\#= (1+N) (3+N)/4$ for odd $N$.
We are led to a simple characterization of the seed two-point functions, which are again defined as the correlation functions which appear with a single structure $\#=1$ (for large enough $s_1$). These are forced to have $N=0$ or equivalently
\be
l=\hl+s_2\quad \Leftrightarrow\quad \mbox{seed} \equiv
\!\!\!
 \raisebox{1.7em}{
$
\xymatrix@=0.3pt{
&&& \ar@{=}[dd] & \\
&&& &\\
{\Ocal_{\D l}}\ar@{-}[rrr]& &&*+[o][F]{} \ar@{=}[dd]&  \hO_{\hD,\hl=l-s_2,(s_1,s_2)}\\  
&  &&& \\
&&&&\\
}
$
} \ ,
\ee
in which case $m_{ij}=0=n_1$ (of course $s_1\geq s_2$). 

Using more generic structures \eqref{defTso(d+1,1)} one can characterize the bulk-defect two-point function also when the bulk operator transforms in a generic representation $l=(l_1,\dots,l_{[d/2]})$ and the defect operator has generic parallel  $\hl=(\hl_1,\dots,\hl_{[p/2]})$  and transverse $s=(s_1,\dots,s_{[q/2]})$ spin.
In particular it is easy to see that all the non zero two-point functions satisfy $l_i \geq \hl_i+s_{i+1}$. Moreover, this condition is saturated by the bulk-defect seeds, which obey $l_i=\hl_i+s_{i+1}$.
As a last comment, we would like stress that, as we already explained at the end of subsection \ref{SO(d+1,1)} for bulk seed structures, the seed bulk-defect two-point function of a conserved bulk operator is automatically conserved.

%%%%%%%%%%%%%%%%
\subsection{Correlation functions in physical space}
\label{ss:real_rad}
%%%%%%%%%%%%%%%%%

This section is dedicated to the projection of the embedding space expressions to physical space. We will simply gather the relevant formulae, referring the reader to the literature for a complete discussion \cite{SpinningCC,Billo:2016cpy,Lauria:2017wav}. 

As mentioned in subsection \ref{SO(d+1,1)}, the CFT lives on the null cone of  $\mathbb{R}^{1,d+1}$. 
The usual flat Euclidean space with Cartesian coordinates $x^\mu$ is obtained by restricting the operators to lie on the Poincaré section:
\be
\label{PoincareSec}
P_{\mbox{\scriptsize Poincar\'e}}^M=\left(\frac{1+x^2}{2} , x^\m , \frac{1-x^2}{2} \right) \ ,
\ee
where the first coordinate is the time-like one. The embedding space indices of the operators are projected onto the physical ones via the Jacobian of the immersion \eqref{PoincareSec}. The polarization vectors in embedding and in physical space are related by requiring that such a projection, applied to the tensor structures in embedding space, yields the polynomial which encodes the tensor structures in physical space. Since in the following the only external operators will be bulk primaries, we only explicitly consider their polarizations. A possible choice is
\be
Z^{(k) \,M}=z^{(k) \, \mu} \ \frac{\partial }{\partial x^\mu}P_{\mbox{\scriptsize Poincar\'e}}^M=(z^{(k)}\cdot x, z^{(k) \, \m}, -z^{(k)}\cdot x)~.
\label{Zz}
\ee
The polarizations defined this way obey \eqref{bulk_pol} if $z^{(i)}\cdot z^{(j)}=0$, but the choice \eqref{Zz} is non unique, since a shift $Z^{(k)}_M \to Z^{(k) }_M+ \alpha P_M$ leaves the correlation functions invariant.

In the next section, we will be concerned with the two-point function of bulk primaries. In \cite{Lauria:2017wav}, two convenient configurations for this correlator were discussed, which we call the \emph{bulk radial frame} and the \emph{defect radial frame}. 

\begin{figure}[h]
\centering
\includegraphics[scale=0.70]{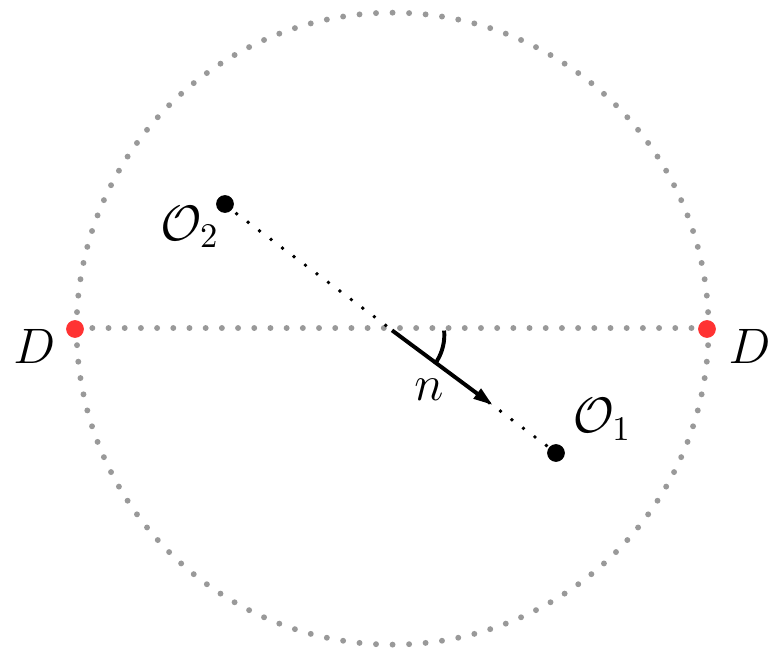}\hspace{0.1\columnwidth}
\includegraphics[scale=0.70]{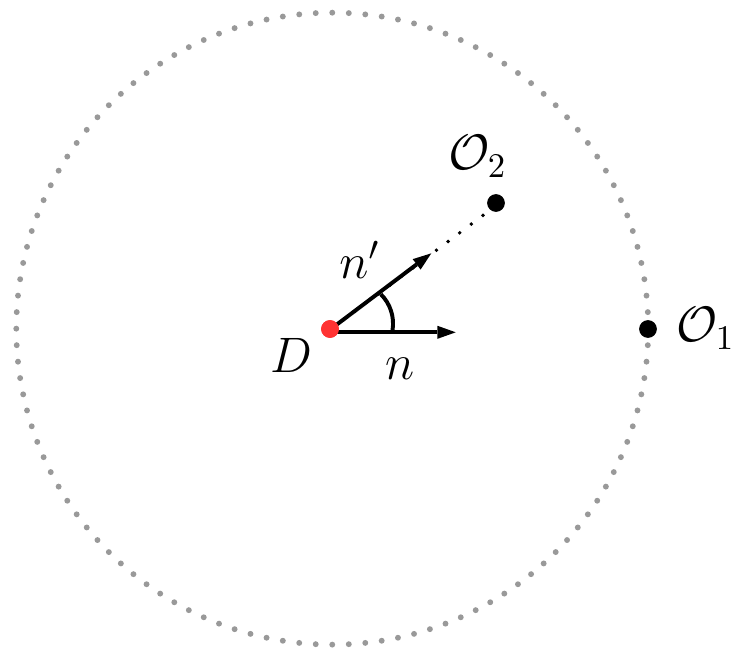}
\caption{The bulk and defect radial frames, corresponding to equations \eqref{B_Poin_Conf} and \eqref{D_Poin_Conf}.  \emph{Bulk radial frame} (left): the defect is spherical and orthogonal to the plane drawn in the figure, and crosses it at the position marked by the red dots. The operators $\mathcal{O}_1$ and $\mathcal{O}_2$ sit at the same radius $r$. \emph{Defect radial frame} (right): the defect is flat and orthogonal to the plane drawn in the figure, and crosses it at the position marked by the red dot. The operator $\mathcal{O}_1$ sits at unit radius, while $\mathcal{O}_2$ lies at radius $\hr$.}
\label{fig:rho}
\end{figure}

\paragraph{Bulk radial frame.} In the bulk radial frame, the defect is a $p$-sphere of unit radius centered in the origin. The operators are inserted in $P_1$ and $P_2$, with
\be
\label{B_Poin_Conf}
P_1=\left(\frac{1+r^2}{2}, r n^\mu, \frac{1-r^2}{2}\right)~,
\qquad
P_2=\left(\frac{1+r^2}{2},-r n^\mu,\frac{1-r^2}{2}\right)~,
\ee
where $n$ is a unit vector in $\mathbb{R}^d$ and $0<r<1$. The configuration, which is depicted on the left in fig. \ref{fig:rho}, naturally defines the two cross-ratios
\be
 r~, \qquad \eta^2\equiv n \pdot n~.
 \label{bulkreta}
\ee
The polarization vectors are
\beq
Z_1^{(k)}=( r\,z_1^{(k)}\cdot n, z_1^{(k)\, \mu}, - r\, z^{(k)}_1\cdot n)\,,\qquad 
Z_2=(-r \,z^{(k)}_2\cdot n, z_2^{(k)\, \mu}, r \,z_2^{(k)}\cdot n)\,,
\label{B_Poin_Z}
\eeq

\paragraph{Defect radial frame.}
On the other hand, in the defect radial frame the defect is taken to be flat and the operators are located at
\be
\label{D_Poin_Conf}
P_1=(1, n^\mu, 0)~,
\qquad
P_2=\left(\frac{1+\hr^2}{2}, \hr n'^\mu,\frac{1-\hr^2}{2}\right)~,
\ee
where $0<\hr<1$, and $n,$ $n'$ are now unit vectors in the transverse space $\mathbb{R}^q$, \emph{i.e.} $\pp^{\mu\nu} n_\nu=\pp^{\mu\nu} n'_\nu=0$. The coordinates of $P_2$ can be taken as cross ratios:
\be
\hr~, \qquad \heta\equiv  n \tdot  n' 
\ ,
\label{defectreta}
\ee
and the configuration is depicted on the right in fig. \ref{fig:rho}. The polarization vectors are
\beq
Z_1^{(k)}=( z_1^{(k)}\tdot n, z_1^{(k)\,\mu}, - z_1^{(k)}\tdot n)\,,\qquad 
Z_2^{(k)}=(\hr \,z_2^{(k)}\tdot n', z_2^{(k)\, \mu}, -\hr \,z_2^{(k)}\tdot n')\,.
\label{D_Poin_Z}
\eeq

%%%%%%%%%%%%%%%%%%%%%%%%%%%%
\section{Spinning conformal blocks}
\label{sec:spinning}
%%%%%%%%%%%%%%%%%%%%%%%%%%%%
We  would like to study the two-point function of symmetric and traceless bulk operators $\Ocal_i$, with dimension $\D_i$ and spin $l_i$, in the presence of a defect:
\be
\Fcal_2(\{ P_i,Z_i\}) \equiv \langle \Ocal_1(P_1,Z_1) \Ocal_2(P_2,Z_2)\rangle~.
\ee
We are going to consider the conformal partial wave decomposition of the two-point function both in the bulk and the defect channel. 
 In the bulk channel one has
\ba
\label{Spin_CB_Decomposition_Bulk}
\begin{split}
\Fcal_2(\{ P_i,Z_i\})
&
=    \sum_{\Ocal}  \sum_{\pf} \, a_\Ocal \, c^{(\pf)}_{12 \Ocal} \,  G^{(\pf)}_{\Ocal}(\{ P_i,Z_i\}) 
 \\
 &
 =
 \sum_{\Ocal}   \sum_{\pf}  \, a_\Ocal \, c^{(\pf)}_{12\Ocal}
  \raisebox{1.8em}{
 $
\xymatrix@=1.3pt{
{\Ocal_{1}}\ar@{-}[rdd]& &&&{ \ar@{=}[dd]  }\\  
&&&&\\
& *+[o][F]{\mbox{\tiny{$\pf$}}}  \ar@{.}[rrr]^{\Ocal }& & & *+[o][F]{} \ar@{=}[dd]  \\
&&&\\
\Ocal_{2} \ar@{-}[ruu]&&&& }
$}
\  \ .
\end{split}
\ea
Here $a_{\Ocal}$ is the one-point function coefficient, which is implicit in \eqref{1pt} and appears explicitly in \eqref{onepoint}. The $c^{(\pf)}_{12\Ocal}$ are the three-point function OPE coefficients defined in \eqref{three_point_BBB}. 
The exchanged operator $\Ocal$ with conformal dimension $\D$ and $SO(d)$ spin $l$ needs to have a non-vanishing one-point function in order to appear in the bulk OPE.
From the discussion of subsection \ref{subsec:corr_defect} we therefore conclude that all the Cartan labels $l=(\ell_1, \dots, \ell_{[d/2]})$ of $\Ocal$ are forced to be even numbers.

The defect channel expansion is written as follows:
\ba
\label{Spin_CB_Decomposition_Defect}
\begin{split}
\Fcal_2(\{ P_i,Z_i\})
&
= 
  \sum_{\hOcal}   \sum_{\pf,\qf}\, b^{(\pf)}_{1 \hOcal} \, b^{(\qf)}_{2 \hOcal} \,  \hG^{(\pf,\qf)}_{\hOcal}(\{ P_i,Z_i\}) 
  \\
  &
=
 \sum_{\hOcal}   \sum_{\pf,\qf} \, b^{(\pf)}_{1 \hOcal} \, b^{(\qf)}_{2 \hOcal} \,  
  \raisebox{2.6em}{
$
\xymatrix@=1.6pt{
&&& \ar@{=}[dd]  \\
&&& \\
{\Ocal_{1}}\ar@{-}[rrr]& &&*+[o][F]{\mbox{\tiny $\pf$}} \ar@{:}[dd]  \\  
&  && % { \mbox{       \scriptsize $\quad \ \ \ \hO$}} 
\\
\Ocal_{2} \ar@{-}[rrr]&&&*+[o][F]{\mbox{\tiny $\qf$}} \ar@{=}[dd] \\
&&& \\
&&& }
$
}
\! \mbox{\scriptsize$\hO$}
   \ \,
\end{split}
\ea
where bulk-defect OPE coefficients $b^{(\pf)}_{1 \hOcal}$, $b^{(\qf)}_{2 \hOcal} $ were defined in \eqref{btodefect2pt}.
The exchanged operator $\hO$ is labelled by its conformal dimension $\hD$, the $SO(p)$ representation $\hl$ and the $SO(q)$ representation $s$.
As explained in subsection \ref{ss:bulk-to-defect}, the parallel spin $\hl$ is always traceless and symmetric, while the transverse spin $s$ may be in a mixed symmetric representation $s=(s_1,s_2)$.

It is often convenient to study conformal blocks in the radial coordinates defined in \ref{ss:real_rad}. Following the conventions of \cite{Lauria:2017wav}, we expand the partial waves in terms of a sum of conformal blocks, which depend on two cross ratios. In the bulk channel, from \eqref{Spin_CB_Decomposition_Bulk} we define 
\be
\label{DEF_CB_Bulk_Spin_Generic}
G^{(\pf)}_{\Ocal}(\{ P_i,Z_i\})=\frac{\Acal(r,\eta)}{(P_1 \wbullet P_1 )^{\frac{\D_1}{2}} (P_2 \wbullet P_2 )^{\frac{\D_2}{2}}} 
 \sum_{k=1}^{k_{max}} g^{(\pf),k}_{\Ocal}(r,\eta) Q_k(\{ P_i,Z_i\}) \ ,
\ee
where the structures $Q_k$ are defined in \eqref{newStructEmi}
and $\Acal$ is the following function
\begin{align}
\label{S_prefactor}
   \Acal(r,\eta) \equiv (2 r)^{-\Delta _1-\Delta _2} \left(r^4-4 \eta ^2 r^2+2 r^2+1\right)^{\frac{1}{2} \left(\Delta _1+\Delta _2\right)} \ .
\end{align}
A similar expansion holds for the defect channel partial waves  \eqref{Spin_CB_Decomposition_Defect}, where we define $\hr,\heta$ as in \ref{ss:real_rad}:
\be
\label{DEF_CB_Defect_Spin_Generic}
\hG^{(\pf,\qf)}_{\hOcal}(\{ P_i,Z_i\})=\frac{1}{(P_1 \wbullet P_1 )^{\frac{\D_1}{2}} (P_2 \wbullet P_2 )^{\frac{\D_2}{2}}}  \sum_{k=1}^{k_{max}}  \hg^{(\pf,\qf),k}_{\hOcal}(\hr,\heta) Q_k(\{ P_i,Z_i\}) \ .
\ee

In tables (\ref{TAB_2pt_Spin_Example_bulk}-\ref{TAB_2pt_Spin_Example_def}) we give some examples of conformal blocks 
which appear in various two-point functions of bulk operators with spin $l_1$ and $l_2$. The notation is as follows. When the number of OPE tensor structures (which we recall are labeled by $\pf$ in the bulk channel and $(\pf,\qf)$ in the defect channel) is $1$, we call the associated conformal block a seed block.
The label $k_{max}$ is the number of tensor structures in the two-point function, according to \eqref{DEF_CB_Bulk_Spin_Generic} and \eqref{DEF_CB_Defect_Spin_Generic}.
\arraycolsep=1.4pt\def\arraystretch{2.0}
\be
\label{TAB_2pt_Spin_Example_bulk}
\begin{array}{| cc || c | l | }
\hline
\ l_1 \ &\ l_2 \ &\ k_{max}\ &\ \ \mbox{Bulk OPE}  \ \\
%%%%%%%%%%%%%%%
\hline
0&0&1&
\
\,
 \raisebox{1.4em}{
 $
\xymatrix@=1.3pt{
\ar@{-}[rdd]&& &&&{ \ar@{=}[dd]  }\\  
&&&&&\\
& *+[o][F]{}  \ar@{.}[rrrr]^{\D,l }& & && *+[o][F]{} \ar@{=}[dd]  \\
&&&&\\
\ar@{-}[ruu]&&&&& }
$}
\\
%%%%%%%%%%%%%%%
\hline
1&0&2&
\
 \raisebox{-0.5em}{
 $
\xymatrix@=1.3pt{
\ar@{-}[rdd]& &&&&{ \ar@{=}[dd]  }\\  
&&&&&\\
& *+[o][F]{\mbox{\tiny{$\pf$}}}  \ar@{.}[rrrr]^{\D, l }& && & *+[o][F]{} \ar@{=}[dd]  \\
&&&&\\
\ar@{-}[ruu]&&&&& }
$}
\atop 
\ \ \ 
\raisebox{0.4 em}{\scriptsize $\pf=1,2$}
\\
%%%%%%%%%%%%%%%%
\hline
1&1&6&
\
{
\raisebox{-0.5em}{
 $
\xymatrix@=1.3pt{
\ar@{-}[rdd]& &&&&{ \ar@{=}[dd]  }\\  
&&&&&\\
& *+[o][F]{\mbox{\tiny{$\pf$}}}  \ar@{.}[rrrr]^{\D, l }& && & *+[o][F]{} \ar@{=}[dd]  \\
&&&&\\
\ar@{-}[ruu]&&&&& }
$}
\ \ \
\atop 
\raisebox{0.4 em}{\scriptsize $\pf=1,\dots,5$}
}
\quad \ \
{
 \raisebox{-0.5em}{
 $
\xymatrix@=1.3pt{
\ar@{-}[rdd]&& &&&{ \ar@{=}[dd]  }\\  
&&&&&\\
& *+[o][F]{}  \ar@{.}[rrrr]^{\D,(l,2) }& & && *+[o][F]{} \ar@{=}[dd]  \\
&&&&\\
\ar@{-}[ruu]&&&&& }
$}
\atop
\raisebox{0.4 em}{\ \ \scriptsize seed CB}
}
\\
%%%%%%%%%%%%%%%%
\hline
2&2& 27 &
\
{
\raisebox{-0.5em}{
 $
\xymatrix@=2.5pt{
\ar@{-}[rdd]& &&&&{ \ar@{=}[dd]  }\\  
&&&&&\\
& *+[o][F]{\mbox{\tiny{$\pf$}}}  \ar@{.}[rrrr]^{\D, l }& && & *+[o][F]{} \ar@{=}[dd]  \\
&&&&\\
\ar@{-}[ruu]&&&&& }
$}
\ \ 
\atop 
\raisebox{0.4 em}{\scriptsize $\pf=1,\dots 14$}
}
\quad
{
\raisebox{-0.5em}{
 $
\xymatrix@=2.5pt{
\ar@{-}[rdd]& &&&&{ \ar@{=}[dd]  }\\  
&&&&&\\
& *+[o][F]{\mbox{\tiny{$\pf$}}}  \ar@{.}[rrrr]^{\D, (l,2) }& && & *+[o][F]{} \ar@{=}[dd]  \\
&&&&\\
\ar@{-}[ruu]&&&&& }
$}
\ \ 
\atop 
\raisebox{0.4 em}{\scriptsize $\pf=1,\dots 11$}
}
\quad
{
\raisebox{-0.5em}{
 $
\xymatrix@=2.5pt{
\ar@{-}[rdd]& &&&&{ \ar@{=}[dd]  }\\  
&&&&&\\
& *+[o][F]{}  \ar@{.}[rrrr]^{\D, (l,4) }& && & *+[o][F]{} \ar@{=}[dd]  \\
&&&&\\
\ar@{-}[ruu]&&&&& }
$}
\ \ \
\atop
\raisebox{0.4 em}{\!\!\!\!\! \scriptsize seed CB}
}
\quad
{
\raisebox{-0.5em}{
 $
\xymatrix@=2.9pt{
\ar@{-}[rdd]& &&&&{ \ar@{=}[dd]  }\\  
&&&&&\\
& *+[o][F]{}  \ar@{.}[rrrr]^{\D, (l,2,2) }& && & *+[o][F]{} \ar@{=}[dd]  \\
&&&&\\
\ar@{-}[ruu]&&&&& }
$}
\ \ \
\atop
\raisebox{0.4 em}{\!\!\!\!\! \scriptsize seed CB}
}
\\
\hline
%%%%%%%%%%%%
\end{array}
\ee

%%%%%%%%%%%%%%%%%%%%%%%%%%%%%%%%%%%%%%%%%%%%%
From tables (\ref{TAB_2pt_Spin_Example_bulk}) and (\ref{TAB_2pt_Spin_Example_def}), we see that the total number of partial waves in the bulk and defect channels is equal. They are also equal to the number of two-point function tensor structures  $k_{max}$. For instance, the two-point function of spin one operators can be decomposed in $5+1=6$ bulk partial waves, or in  $2^2+1+1=6$ defect partial waves and has $k_{max}=6$. Similarly, the two-point function of spin two operators is decomposed in $14+11+1+1=27$ bulk CBs and in $4^2+2 \times 2^2+3 \times 1=27$ defect ones and again $k_{max}=27$. 
In fact one can check that this match continues also for external operators with higher spin. 
It would be interesting to justify this match by using representation theory as it was done in  \cite{Kravchuk:2016qvl} for correlation functions in CFTs without defects.
 %%%%%%%%%%%%%%%%%%%%%%%%%%%%%%%%%%%%%%%%%%%%%
\arraycolsep=1.4pt\def\arraystretch{2.0}
\be
\label{TAB_2pt_Spin_Example_def}
\begin{array}{|cc || c | l | l |}
\hline
\ l_1 \ &\ l_2 \ &\ k_{max}\ &\  \ \mbox{Defect OPE} \ \\
\hline
0&0&1&
 \raisebox{2.4em}{
$
\xymatrix@=0.3pt{
&&& &\ar@{=}[dd]  &\\
&&&& &\\
\ar@{-}[rrrr]&& &&*+[o][F]{} \ar@{:}[dd] & \\  
&  &&& &{ \mbox{       \scriptsize $\hD, \hl=0, s $}} \\
 \ar@{-}[rrrr]&&&&*+[o][F]{} \ar@{=}[dd] \\
&&& &&\\
&&&&& }
$
}
\\
%%%%%%%%%%%%%%%
\hline
1&0&2&
 \raisebox{0. em}{
$
\xymatrix@=0.3pt{
&&& &\ar@{=}[dd]  &\\
&&&& &\\
\ar@{-}[rrrr]&& &&*+[o][F]{\mbox{\tiny $\pf$} } \ar@{:}[dd] & \\  
&  &&& &{ \mbox{       \scriptsize $\hD, \hl=0, s $}} \\
 \ar@{-}[rrrr]&&&&*+[o][F]{} \ar@{=}[dd] \\
&&& &&\\
&&&&& }
$
}
\atop
\raisebox{0.4 em}{
\!\!\!\!\!\!\!\!
 \scriptsize$\pf=1,2$
}
\\
%%%%%%%%%%%%%%%%
\hline
1&1&6&
{
 \raisebox{0.em}{
$
\xymatrix@=0.3pt{
&&& &\ar@{=}[dd]  &\\
&&&& &\\
\ar@{-}[rrrr]&& &&*+[o][F]{\mbox{\tiny $\pf$} } \ar@{:}[dd] & \\  
&  &&& &{ \mbox{       \scriptsize $\hD, \hl=0, s $}} \\
 \ar@{-}[rrrr]&&&&*+[o][F]{\mbox{\tiny $\qf$} } \ar@{=}[dd] \\
&&& &&\\
&&&&& }
$
}
\atop
\raisebox{0.4 em}{\scriptsize$\pf,\qf=1,2$}
}
\
{
 \raisebox{0.em}{
$
\xymatrix@=0.3pt{
&&& &\ar@{=}[dd]  &\\
&&&& &\\
\ar@{-}[rrrr]&& &&*+[o][F]{} \ar@{:}[dd] & \\  
&  &&& &{ \mbox{       \scriptsize $\hD, \hl=0, (s,1) $}} \\
 \ar@{-}[rrrr]&&&&*+[o][F]{} \ar@{=}[dd] \\
&&& &&\\
&&&&& }
$
}
\atop
%\raisebox{0.4 em}{\phantom{\scriptsize$\pf,\qf=1,2$}}
\raisebox{0.4 em}{\!\!\!\!\!\!\!\!\!\!\!\!\!\!\!\! \scriptsize seed CB}
}
\
{
 \raisebox{0.em}{
$
\xymatrix@=0.3pt{
&&& &\ar@{=}[dd]  &\\
&&&& &\\
\ar@{-}[rrrr]&& &&*+[o][F]{} \ar@{:}[dd] & \\  
&  &&& &{ \mbox{       \scriptsize $\hD, \hl=1, s $}} \\
 \ar@{-}[rrrr]&&&&*+[o][F]{} \ar@{=}[dd] \\
&&& &&\\
&&&&& }
$
}
\atop
%\phantom{\pf=1,2}
\raisebox{0.4 em}{\!\!\!\!\!\!\!\!\!\!\!\!  \scriptsize seed CB}
}
\\
%%%%%%%%%%%%%%%%
\hline
2&2& 27 &
{
 \raisebox{0.em}{
$
\xymatrix@=0.3pt{
&&& &\ar@{=}[dd]  &\\
&&&& &\\
\ar@{-}[rrrr]&& &&*+[o][F]{\mbox{\tiny $\pf$} } \ar@{:}[dd] & \\  
&  &&& &{ \mbox{       \scriptsize $\hD, \hl=0, s $}} \\
 \ar@{-}[rrrr]&&&&*+[o][F]{\mbox{\tiny $\qf$} } \ar@{=}[dd] \\
&&& &&\\
&&&&& }
$
}
\atop
\raisebox{0.4 em}{\scriptsize$\pf,\qf=1,\dots,4$}
}
\
{
 \raisebox{0.em}{
$
\xymatrix@=0.3pt{
&&& &\ar@{=}[dd]  &\\
&&&& &\\
\ar@{-}[rrrr]&& &&*+[o][F]{\mbox{\tiny $\pf$} } \ar@{:}[dd] & \\  
&  &&& &{\phantom{\hD} } \\
 \ar@{-}[rrrr]&&&&*+[o][F]{\mbox{\tiny $\qf$} } \ar@{=}[dd] \\
&&& &&\\
&&&&& }
$
}
\atop
\raisebox{0.4 em}{\ \ \ \ \scriptsize$\pf,\qf=1,2$}
}
\!\!\!\!\!\!\!\!
{
\raisebox{0.6 em}{$
\scriptsize
\begin{array}{l}
\vspace{-0.15cm}
\hD,\hl =1, s \\
\hD,\hl =0, (s,1) 
\end{array}
$
}
}
\
{
 \raisebox{0.em}{
$
\xymatrix@=0.3pt{
&&& &\ar@{=}[dd]  &\\
&&&& &\\
\ar@{-}[rrrr]&& &&*+[o][F]{} \ar@{:}[dd] & \\  
&  &&& &{\phantom{\hD}} \\
 \ar@{-}[rrrr]&&&&*+[o][F]{} \ar@{=}[dd] \\
&&& &&\\
&&&&& }
$
}
\atop
\raisebox{0.4 em}{\ \ \  \scriptsize seed CBs}
}
\!\!\!\!\!\!\!\!
{
\raisebox{0.6 em}{$
\scriptsize
\begin{array}{l}
\vspace{-0.15cm}
\hD,\hl =2, s \\
\vspace{-0.15cm}
\hD,\hl =1, (s,1) \\
\hD,\hl =0, (s,2) 
\end{array}
$
}
}
\\
\hline
%%%%%%%%%%%%
\end{array}
\ee
\arraycolsep=1.4pt\def\arraystretch{1.0}

In the following, we describe  some techniques to determine the conformal blocks $g$ and $\hg$ in formulae \eqref{DEF_CB_Bulk_Spin_Generic} and \eqref{DEF_CB_Defect_Spin_Generic}. In general, the bulk CBs are going to be computable only as an expansion in radial coordinates: roughly speaking the bulk CBs are as hard as the CBs for the four-point function, with which they share the same bulk OPE.
On the other hand we will be able to determine a closed form formula for any defect channel CB.

In appendix \ref{SpinCB_bulk_examples}, \ref{RadialCoordForCB}, \ref{SpinningZam} and \ref{App_Defect_Spinning_Diff_Ops} we will exemplify the techniques by computing CBs with $l_1=1,l_2=0$ and $l_1,l_2=1$.

%%%%%%%%%%%%%%%%%%%%%%%%%%%%
\subsection{Bulk channel }
%%%%%%%%%%%%%%%%%%%%%%%%%%%%
The bulk channel partial waves \eqref{Spin_CB_Decomposition_Defect} are eigenfunctions of the quadratic Casimir operator: 
\be
-\frac{1}{2}(J_1+J_2)^2 \; G^{(\pf)}_{\Ocal}(\{ P_i,Z_i\})=c_{\D l} \; G^{(\pf)}_{\Ocal}(\{ P_i,Z_i\}) \ .
\label{Casimir_spinning_B}
\ee
The eigenvalue is 
$c_{\D l} =\D(\D-d) + l (l+d-2)$ 
and  the generators are $J_i^{M N} \equiv 2 P_i^{[M} \partial_{P_i}^{N]}+2 Z_i^{[M} \partial_{Z_i}^{N]}$. Different partial waves associated to the same operator are distinguished by the asymptotic behavior in the OPE limit. 
Equation \eqref{Casimir_spinning_B} can be cast into a set of second order partial differential equations which couple the functions $g^{(\pf),k}_{\Ocal}(r,\eta) $ defined in \eqref{DEF_CB_Bulk_Spin_Generic} for $k=1,\dots,k_{max}$. We schematically write
\be
\sum_{k'=1}^{k_{max}} M_{k k'}(\partial_r,\partial_\eta) g^{(\pf),k'}_{\Ocal}(r,\eta) = 0 \ .
\ee
Here the matrix $M$ depends on the cross ratios $r$, $\eta$ and the derivatives $\partial_r,\partial_\eta$.
A closed form solution  for generic dimension and codimension is not known. 

The goal of this section is to compute spinning conformal blocks by generalizing  different methods that were used in \cite{Lauria:2017wav} to compute the scalar blocks. First, we  explain how to write CBs as a series expansion in the radial coordinates. The coefficients of the expansion can be computed in various ways. In particular, we comment on an efficient way to generate them through a recurrence relation of the kind introduced by Zamolodchikov in \cite{Zamolodchikov:1985ie}. Finally, we explain how to obtain the spinning conformal blocks by acting with differential operators on seed blocks, following the idea of \cite{Costa2011}.

%%%%%%%%%%%%%%%%%%%%%%%%%
\subsubsection{Radial expansion}
\label{sec:Spinning_Casimir_Bulk}
%%%%%%%%%%%%%%%%%%%%%%%%%
The existence of the bulk OPE implies that the bulk CBs can be written as a power expansion in the radial coordinates of section \ref{ss:real_rad} (see \cite{Hogervorst:2013sma, radial_expansion, Lauria:2017wav}). 
In fact, by writing the two-point function in the cylinder frame \eqref{B_Poin_Conf}, it becomes clear that the powers of $r$ measure the cylinder energy of the operators exchanged in the OPE. The dependence on the unit vector $n$ is fixed by the $SO(d)$ representations of the descendants, and encoded in the polynomials $ {\bf P}^{\pdot d}$ of \eqref{Ptnl}. Finally, the expansion can be conveniently repackaged in a finite number of functions $W(r,\eta,n,z_i)$ in one-to-one correspondence with the three-point function tensor structures given in subsection \ref{correlatorsCFT}.

The object of interest is the following matrix element in radial quantization:
\be
\label{Gcalpq_spin}
\Gcal_{\Ocal}
=
 \langle \hat 0 | 
 r^{H_{cyl}} \mathcal{P}_{\Ocal} \
\mathcal{O}_1(n,z_1) \mathcal{O}_2(-n,z_2)
| 0
\rangle
 \ ,
\ee
where  $\mathcal{P}_{\Ocal}$ is the projector onto the conformal family with highest weight labelled by $\D$ and $SO(d)$ spin $l$ and $H_{cyl}$ is the Hamiltonian conjugate to the cylinder time $\tau = \log r$.
The function $\Gcal_{\Ocal}$ is equivalently obtained by writing the conformal partial waves into the bulk radial frame\footnote{ 
In practice, the freedom contained in the coefficients $c^{(\pf)}$ is translated in the freedom of choosing the coefficients $w^{(\pf)}(0,l)$ in eq. \eqref{Wk_Bulk_Spin} below.
}
\be
\label{GToGcal_brf_Spin}
 \sum_{\pf} 
 %a_\Ocal \, c^{(\pf)}_\Ocal  
 c^{(\pf)}
 G^{(\pf)}_{\Ocal}(\{ P_i,Z_i\}) \underset{b.r.f.}{\longrightarrow} r^{-\D_1-\D_2}\Gcal_{\Ocal}(r, \eta , n\cdot z_k, n \pdot z_k ,z_1\cdot z_2, z_k  \pdot z_j) \ ,
\ee
as explained in appendix \ref{tensor structures to radial frame}.
In this section we define $a\pdot b \equiv a\cdot   \pp \cdot b $ with
 $\pp$ the parallel projector for spherical defects, namely the diagonal matrix with $p+1$ ones followed by $q-1$ zeros. 
Eq. \eqref{GToGcal_brf_Spin} makes it manifest that in the radial frame the functions $\Gcal_{\Ocal}$ can be expanded in tensor structures generated by the building blocks $n\cdot z_k,\, n \pdot z_k ,\,z_1\cdot z_2,\, z_k  \pdot z_j$.
It is natural to rewrite the projector  $\mathcal{P}_{\Ocal}$  in \eqref{Gcalpq_spin} as a sum over a complete basis of bulk states:
\begin{align} \label{GcalSumStates_spin}
&\Gcal_{\Ocal} =  \sum_{m=0}^\infty r^{\Delta+m}
\sum_{j}
\sum_{\dd} 
   \langle \hat 0 |  m, j ,\dd 
\rangle
\langle
m, j,
\dd | 
\mathcal{O}_1(n,z_1) \mathcal{O}_2(-n,z_2)
| 0  \rangle
\,,
\end{align}
where we sum over all states at level $m$ of the conformal family, organized in irreducible representations (irreps) with spin $j=(j_1,j_2, \dots, j_{[d/2]})$ of $SO(d)$. $\dd$ labels the degeneracy of such states.

The one-point function $ \langle \hat 0 |  m, j ,\dd \rangle$ is always fixed in terms of a single tensor structure. For example when $j=(k,2)$ we have
%%%%%%%%%%%
\be
  \langle \hat 0 |  m, { \scriptsize
  \ytableausetup{centertableaux,boxsize=1.4 em}
\begin{ytableau}
\m_{1}&\, \m_{2}&\, _{\cdots}&\,\m_{k}\\
\n_{1}&\, \n_{2}\\
\end{ytableau}
},\dd 
\rangle
 = 	
 v(m;j,2;\dd) 
 \pi^{SO(d)}_{k,2} 
 \left(
 { \scriptsize
  \ytableausetup{centertableaux,boxsize=1.4 em}
\begin{ytableau}
\m_{1}&\, \m_{2}&\, _{\cdots}&\,\m_{k}\\
\n_{1}&\, \n_{2}\\
\end{ytableau}
},
{ \scriptsize
  \ytableausetup{centertableaux,boxsize=1.4 em}
\begin{ytableau}
\r_{1}&\, \r_{2}&\, _{\cdots}&\,\r_{k}\\
\s_{1}&\, \s_{2}\\
\end{ytableau}
}\right)
\pp^{\s_{1} \, \s_{2}} \pp^{\r_{1} \, \r_{2}} \cdots \pp^{\r_{k-1} \, \r_{k}} \ .
 \label{left_overlap_Spin_Bulk}
\ee
The case of generic $j$ is just a straightforward generalization.
Equation \eqref{left_overlap_Spin_Bulk} implies that the only allowed one-point functions correspond to states for which all the Cartan labels $j_1,j_2, \dots j_{[d/2]}$ are even integers. While we pointed out in subsection \ref{subsec:corr_defect} that this is true for primary operators, we now see that the same holds for descendants as well, but only when acting on the vacuum at the center of a spherical defect.

The overlaps $\langle
m, j,
\dd | 
\mathcal{O}_1(n,z_1) \mathcal{O}_2(-n,z_2)
| 0  \rangle$ of \eqref{GcalSumStates_spin} were already considered in \cite{radial_expansion}. 
For concreteness  we present here some examples of their form. If $j$ is a symmetric and traceless $SO(d)$ representation, the overlap is fixed by Lorentz invariance up to a few coefficients $u^{(\pf)}(m,j,\dd)$,
\be
 \resizebox{0.92 \textwidth}{!}{$
  \langle
m, { \scriptsize
\ytableausetup{centertableaux,boxsize=1.4 em}
\begin{ytableau}
\, \a_{1}&\, _{\cdots}&\,\a_{j}\\
\end{ytableau}
},
\dd | 
\mathcal{O}_1(n,z_1) \mathcal{O}_2(-n,z_2)
| 0  \rangle=
\displaystyle{ \sum_{\pf}} u^{(\pf)}(m,j,\dd) \, t^{(\pf)}(n,\nabla_n,z_1,z_2) 
 \pi^{SO(d)}_{j} 
 \!
 \left(
 { \scriptsize
  \ytableausetup{centertableaux,boxsize=1.4 em}
\begin{ytableau}
\, \a_{1}&\, _{\cdots}&\,\a_{j}\\
\end{ytableau}
},
{ \scriptsize
  \ytableausetup{centertableaux,boxsize=1.4 em}
\begin{ytableau}
n&\, _{\cdots}&\, n\\
\end{ytableau}
}\right)
,
 $}
 \label{right_overlap_Spin_Bulk_tracelesssym}
\ee
%%%%%%
where we introduced the covariant derivative on the sphere $\nabla_n^\m=\partial^\m_n-n^\m \, (n\cdot \partial_n)$.
The coefficients $u^{(\pf)}(m,j,\dd)$ multiply the tensor structures $t^{(\pf)}$. The tensor structures $t^{(\pf)}$ are homogeneous functions of $z_i$ of degree $l_i$ and are generated as products of the following five building blocks (we take all the derivatives to be ordered on the right of the polynomial):
\be
(n\cdot z_1), \quad  (n\cdot z_2), \quad (z_1\cdot \nabla_n) ,\quad  (z_2\cdot \nabla_n) ,\quad  (z_1 \cdot z_2) \ .
\ee
For example, for one external vector we get two structures
\be
t^{(1)}=(z_1\cdot n)   \ , \qquad t^{(2)}= (z_1\cdot \nabla_n)\ .
\ee
While for two external vectors we get five structures 
\be
\label{tk(11)}
\begin{array}{lll}
t^{(1)}=(z_1\cdot n)(z_2\cdot n)   \ , 
\qquad 
&
t^{(2)}= (z_1\cdot n) (z_2 \cdot \nabla_n)  \ , \qquad
& t^{(3)}=(z_2\cdot n)  (z_1 \cdot \nabla_n)  \ , \\
t^{(4)}= (z_1 \cdot \nabla_n) (z_2 \cdot \nabla_n)  \ , 
\qquad 
& t^{(5)}= (z_1\cdot z_2) 
 \ .
 &
\end{array}
\ee
The exchange of other representations $j$ require appropriate $SO(d)$ projectors in eq. \eqref{right_overlap_Spin_Bulk_tracelesssym}.
For instance, if the primaries  $\mathcal{O}_1$ and $\mathcal{O}_2$ are vectors, they also exchange operators in the representation $j=(\ell,2)$,
\be
  \langle
m, { \scriptsize
\ytableausetup{centertableaux,boxsize=1.4 em}
\begin{ytableau}
\, \a_{1}&\, \a_{2}&\, _{\cdots}&\,\a_{\ell}\\
\b_{1}&\b_{2}\\
\end{ytableau}
},
\dd | 
\mathcal{O}_1(n,z_1) \mathcal{O}_2(-n,z_2)
| 0  \rangle
 = u(m,\ell,2,\dd) \,
 \pi^{SO(d)}_{\ell,2} 
 \left(
 { \scriptsize
  \ytableausetup{centertableaux,boxsize=1.4 em}
\begin{ytableau}
\, \a_{1}&\, \a_{2}&\, _{\cdots}&\,\a_{\ell}\\
\b_{1}&\b_{2}\\
\end{ytableau}
},
{ \scriptsize
  \ytableausetup{centertableaux,boxsize=1.4 em}
\begin{ytableau}
n&\, n&\, _{\cdots}&n \\
z_{1}&z_{2}\\
\end{ytableau}
}\right)
 .
 \label{right_overlap_Spin_Bulk_l2}
\ee
Notice that in this case there are no tensor structures $t^{(\pf)}$ since all the polarization vectors $z_1$, $z_2$ are being contracted with the $SO(d)$ projector.
In fact this is a seed three-point function as defined in \eqref{3ptseed}.

Putting all together, one obtains a general formula for the radial expansion:
\begin{align}
\Gcal_{\D,l_1,\dots,l_{[d/2]}}(r,\eta,n,z_i)  &= 
\sum_{\pf} \sum_{j_2,\dots,j_{[d/2]}} 
  t^{(\pf)}(n,\nabla_n,z_1,z_2)  W^{(\pf)}_{j_2,\dots, j_{[d/2]}}(r,\eta,n,z_i) \ ,
  \label{CB_Expansion_BULK}
\\
W^{(\pf)}_{j_2,\dots, j_{[d/2]}}(r,\eta,n,z_i) &\equiv 
\sum_{m=0}^\infty
\sum_{j_1=\max[l_1-m,j_2]}^{l_1+m}
w^{(\pf)}_{j_2,\dots, j_{[d/2]}}(m,j_1) \ 
 r^{\Delta+m}  \ {\bf P}^{\pdot d }_{j_1,\dots, j_{[d/2]}}(n, z_i)~, \label{Wk_Bulk_Spin}
\end{align}
where $w^{(\pf)}_{j_2,\dots, j_{[d/2]}} (m,j_1)\equiv \sum_{\dd}  u^{(\pf)}(m,j,\dd)v(m,j,\dd)$. 

Notice that the sums in \eqref{CB_Expansion_BULK} over $\pf$ and $j_2,\dots j_{[d/2]}$ span a finite sets of elements: $\pf$ is bounded by the number of tensor structures in the OPE $\Ocal_1 \times \Ocal_2$, while the Cartan labels $j_2,\dots j_{[d/2]}$ are bounded by the possible representations exchanged in the OPE $\Ocal_1 \times \Ocal_2$ and need to be even. This means that to compute the conformal blocks we need to know a finite number of functions $W^{(\pf)}_{j_2,\dots, j_{[d/2]}}$. They are constrained by Lorentz symmetry as shown in equation \eqref{Wk_Bulk_Spin}, where the only unknowns are the coefficients $w^{(\pf)}$. Equation \eqref{Wk_Bulk_Spin} provides a natural expansion in radial coordinates, where at each new level in $r$ there is a finite number of coefficients to be computed.

In \eqref{Wk_Bulk_Spin} we loosely write ${\bf P}^{\pdot d }_{j_1,\dots, j_{[d/2]}}(n, z_i)$, where $z_i$ are the polarization vectors of the external operators. This is a schematic formula to stress that all the complication of a mixed symmetry exchanged representation are encoded in the polynomials  ${\bf P}^{\pdot d }_j$ defined in \eqref{Ptnl}. More precisely, in the definition \eqref{Ptnl} each line of the Young tableau is contracted with the same polarization vector, while in \eqref{CB_Expansion_BULK} different polarization vectors may appear in the same line.
Let us exemplify this construction with the two-point function of vector operators,
\ba
\Gcal_{\D,l_1,l_2}(r,\eta,n,z_i)  &=& 
\sum_{\pf=1}^5 
  t^{(\pf)}(n,\nabla_n,z_1,z_2)  W^{(\pf)}_{0}(r,\eta,n) +
 W_{2}(r,\eta,n,z_i) \ ,
\ea
where the structures $t^{(\pf)}$ are defined in \eqref{tk(11)}.
We dropped the label $\pf$ from $W_{2}$ since it only has one tensor structure and  we omitted  the dependence on $z_i$ from $W^{(\pf)}_{0}$ since it does not depend on them. The definition of the functions $W$ is as follows,
\ba
W^{(\pf)}_{0}(r,\eta,n)&=& \sum_{m=0}^\infty
\,
\sum_{j_1=\max[l_1-m,0]}^{l_1+m}
\,
w^{(\pf)}_0(m,j_1) \;
 r^{\Delta+m} \, {\bf P}^{\pdot d }_{j_1}(n)\\
W_{2}(r,\eta,n,z_i)&=& \sum_{m=0}^\infty
\,
\sum_{j_1=\max[l_1-m,2]}^{l_1+m}
\,
w_2 (m,j_1) \;
 r^{\Delta+m} (z_1\cdot \partial_{z_2}) \, {\bf P}^{\pdot d }_{j_1,2}(n, z_2)
\ea
The purpose of the derivative $(z_1\cdot \partial_{z_2})$ is to insert the polarization vector $z_1$ in the second line of the Young tableau, in accordance with equation \eqref{right_overlap_Spin_Bulk_l2}. 

The final task is to fix the coefficients $w$. A possible strategy uses the Casimir equation, which can be cast as a recurrence relation for the coefficients $w$. This strategy was used for example in \cite{Lauria:2017wav} to compute the bulk channel scalar CBs. 
In the next subsection, we explain instead how to compute them using a recurrence relation akin to the one Zamolodchikov proposed for $2d$ Virasoro CBs  \cite{Zamolodchikov:1985ie}.

%%%%%%%%%%%%%%%%%%%%%%%%%%%%
\subsubsection{Zamolodchikov recurrence relation}
\label{BulkZamolodchikovSpinning}
%%%%%%%%%%%%%%%%%%%%%%%%%%%%
In \cite{Lauria:2017wav} we explained how to write a Zamolodchikov recurrence relation \cite{Zamolodchikov:1985ie} for scalar bulk blocks following the recipe of \cite{arXiv:1406.4858, arXiv:1307.6856, recrel}.
Here, we show that it is easy to generalize the Zamolodchikov recurrence relation to the case of external operators with spin. 
Following the argument of \cite{recrel}, a conformal block for the exchange of an operator $\Ocal$ of conformal dimension $\D$ and $SO(d)$ spin $l=(l_1,l_2\dots,l_{[d/2]})$, has the following pole structure as a function of $\D$:
\be
G^{(\pf)}_{\D l}(P_i,Z_i) = \frac{1}{\D-\D^\star_A}  \sum_{\pf'} (R_A)_{\pf \pf'} G^{(\pf')}_{ \D_A  l_A}(P_i,Z_i) +O((\D-\D^\star_A)^0) \, .
\label{eq:GDeltal1}
\ee
In equation \eqref{eq:GDeltal1} we denoted by $\Delta_A$ and $l_A=(l_{A\, 1},l_{A\, 2},\dots l_{A\, [d/2]})$  the labels of the operator $\Ocal_A$ which is a descendant of $\Ocal$. The descendant operator $\Ocal_A$  becomes primary when we tune the dimension of the primary $\Ocal$ to $\D=\D^\star_A$, in which case $\Delta_A=\D^\star_A+n_A$, where $n_A\in \mathbb{N}$ is the level.
Being both a primary and a descendant, $\Ocal_A$ has a vanishing norm,  which in turn gives rise to a pole in the conformal block \cite{recrel}. We break up the matrix $R_A$ into the following pieces:
 \be
( R_A)_{\pf \pf'}=M_A^{(L)} Q_A (M_A^{(R)})_{\pf \pf'}~,
 \label{QMM_spin_bulk}
 \ee
where the coefficients $M_A$ and $Q_A$ take into account the different normalization of the operator $\Ocal_A$ with respect to a canonically normalized operator $\Ocal$ with the same $SO(d)$ spin as $\Ocal_A$  \cite{Lauria:2017wav, recrel}. In particular $Q_A, M_A^{(L)}$ are defined by
\ba
\langle \Ocal_A\Ocal_A\rangle^{-1}= \frac{Q_A}{\D-\D^\star_A} \langle \Ocal\Ocal\rangle^{-1}+O(\D-\D^\star_A)
\ , \qquad 
\langle \Ocal_A \rangle=  M_A^{(L)} \langle \Ocal\rangle \ ,
\ea
while the matrix $ M_A^{(R)}$ implements the following change of basis, 
\be
\raisebox{1.7em}{
$
\xymatrix@=0.01pt{\Ocal_1 \ar@{-}[rdd]& &&&  \\  
&&&&\\
& *+[o][F]{\mbox{\tiny $\pf$}}  \ar@{-}[rrr] && &\Ocal_A  \\
&&&&\\
\Ocal_2 \ar@{-}[ruu]&&&& }
$
}
=
 \sum_{\pf'}(M_A^{(R)})_{\pf \pf'}  
 \raisebox{1.7em}{
$
\xymatrix@=0.01pt{\Ocal_1 \ar@{-}[rdd]& &&&  \\  
&&&&\\
& *+[o][F]{\mbox{\tiny $\pf \resizebox{0.07cm}{!}{$'$}$}}  \ar@{-}[rrr] && &\Ocal  \\
&&&&\\
\Ocal_2 \ar@{-}[ruu]&&&& }
$
}
\ .
 \ee
 
As we explained in \cite{Lauria:2017wav}, the pole structure matches the one of the conformal blocks for theories without defects \cite{recrel, radial_expansion}. However, in this case the spins $l_i, l_{A \, i}$ are even integers, while the four-point function of local operators also exchanges odd integers \cite{recrel, radial_expansion}.
For completeness, we write the full set of poles $\D^\star_A$ and the quantum numbers of the  associated primary descendants $\Ocal_A$ in the following table
\be \label{AllTheLabels}
\begin{array}{|l | ccc|}
\hline
\phantom{\Big(}	\qquad \quad\quad	A								&\D^\star_A 					&n_A 										&l_{A\, k}	 \\ 
\hline
\phantom{\Big(}  \I_k, \; \; n=2,4, \dots ,l_{k-1}-l_k			& k-l_k-n 			 		& n 	  	   		&\quad l_k + n \quad \\ 
\phantom{\Big(} \II_k,  \;n=2,4, \dots ,l_{k}-l_{k+1}  \;\; 	&      \quad   d+l_k-k-n \quad & \quad n \quad  			&\quad l_k - n \quad  \\
\phantom{\Big(} \III,     \, \,   n=1,2, \dots \infty \;				& \frac{d}{2}-n  						& 2n       			 	 &l_k     \\ 
\phantom{\Big(} \IV,     \, \,   n=1,2, \dots ,l_{[d/2]} \; 				& \frac{d+1}{2}-n  					& 2n       			 	 &l_k     \\ 
\hline
\end{array}
\ee
where $k=1,\dots ,[d/2]$.
 
Also the coefficients $Q_A$ and the matrices $M_A^{(R)}$ are the same as the ones defined in \cite{recrel}. They were computed for the vector-scalar and the vector-vector cases in \cite{recrel, radial_expansion, Dymarsky:2017xzb}. Finally, $M_A^{(L)}$ was computed in Appendix B.2 of  \cite{Lauria:2017wav} for all the operators $\Ocal_A$ in a symmetric and traceless representation.

The conformal blocks  \eqref{DEF_CB_Bulk_Spin_Generic} are obtained by summing over all the poles in $\D$ and the regular part as follows:
\begin{align}
&h^{(\pf),k}_{\D l}(r,\eta) \equiv (4 r)^{-\D} g^{(\pf),k}_{\D l}(r,\eta) \underset{\D \rightarrow \infty }{\longrightarrow} h^{(\pf),k}_{\infty l}(r,\eta)
\\
&h^{(\pf),k}_{\D l}(r,\eta)=h^{(\pf),k}_{\infty l}(r,\eta)+\sum_{A} \sum_{\pf'} \frac{(R_A )_{\pf \pf'}}{\D-\D^\star_A} (4 r)^{n_A}
h^{(\pf'),k}_{\D_A\,l_A}(r,\eta) \,.
\label{eq:rec_Spin}
\end{align}
The functions $h^{(\pf),k}_{\infty l}(r,\eta)$ can be computed by solving the Casimir equation at leading order for $\D\rightarrow \infty$. 
Notice that, since $n_A>0$, we can use this recurrence relation to compute the radial expansion of the conformal blocks.

%%%%%%%%%%%%%%%%%%%%%
\subsubsection{Spinning differential operators}
\label{Spinning_Differential_Operators_Bulk}
%%%%%%%%%%%%%%%%%%%%%
It is possible to obtain spinning conformal partial waves in the presence of defects by acting on  seed conformal partial waves with appropriate differential operators $\Dcal^{(\pf)}$, first defined in  \cite{Costa2011}.
These differential operators act by effectively increasing the spin of operators in a three-point function
\be
\label{DiffOpSchematic}
\Dcal^{(\pf)} \
\raisebox{1.7em}{
$
\xymatrix@=0.01pt{
{\Ocal_{\D_1 \,  l_1= l^{(2)}}}
\ar@{-}[rdd]    \\  
\\
& *+[o][F]{}  \ar@{-}[rrr] && &\Ocal_{\D \, l }  \\
\\
{\Ocal_{\D_2 \,  l_2= l^{(3)}}} \ar@{-}[ruu] 
}
$
}
=
 \raisebox{1.7em}{
$
\xymatrix@=0.01pt{{\Ocal_{\D_1 \, l'_1}}\ar@{-}[rdd]& &&&  \\  
&&&&\\
& *+[o][F]{\mbox{\tiny{$\pf$}}}  \ar@{-}[rrr] && &\Ocal_{\D \, l}  \\
&&&&\\
\Ocal_{\D_2 \, l'_2} \ar@{-}[ruu]&&&& }
$
}
\ ,
\ee
where $l'_i\geq l_i$ and $l \equiv (l^{(1)}, l^{(2)}, l^{(3)})$. The three-point function on the left hand side of \eqref{DiffOpSchematic} is a representative seed of the kind \eqref{3ptseedrepresentative} 
 (here we named $\Ocal_3=\Ocal$), while the one on the right hand side is a generic three-point function, thus it is labeled by a tensor structure index $\pf$.
The result of \cite{Costa2011} is that $\Dcal^{(\pf)}$ can be constructed as compositions of the following elementary operators
\begin{align}
D_{11} & \equiv 
(P_1 \cdot P_2){ } (Z_1 \cdot \partial_{P_2})-
(Z_1 \cdot P_2){ }(P_1 \cdot \partial_{P_2})-
(Z_1 \cdot Z_2){ }(P_1 \cdot \partial_{Z_2})+
(P_1 \cdot Z_2){ }(Z_1 \cdot \partial_{Z_2}),\nonumber\\
D_{12}& \equiv 
(P_1 \cdot P_2){ }(Z_1 \cdot \partial_{P_1})-
(Z_1 \cdot P_2){ }(P_1 \cdot \partial_{P_1})+
(Z_1 \cdot P_2){ }(Z_1 \cdot \partial_{Z_1}) \ , \nonumber
\\
D_{22}& \equiv 
(P_2 \cdot P_1){ }(Z_2 \cdot \partial_{P_1})-
(Z_2 \cdot P_1){ }(P_2 \cdot \partial_{P_1})-
(Z_2 \cdot Z_1){ }(P_2 \cdot \partial_{Z_1})+
(P_2 \cdot Z_1){ }(Z_2 \cdot \partial_{Z_1}),\nonumber \\
D_{21}& \equiv 
(P_2 \cdot P_1){ }(Z_2 \cdot \partial_{P_2})-
(Z_2 \cdot P_1){ }(P_2 \cdot \partial_{P_2})+
(Z_2 \cdot P_1){ }(Z_2 \cdot \partial_{Z_2}),
\label{bulkdiff}
\end{align}
and $H_{12}$ as defined in \eqref{three_point_BBB}.
With the above definitions $D_{ij}$ increases the degree in $Z_i$ and  $P_j$ by one unit, while $H_{12}$ increases both $Z_1$ and $Z_2$.
With respect to \cite{Costa2011}, we want to consider an extra operator which increases the degree in $Z_j$ while deceasing the degree of $Z_i$ by one unit.
 This is the so called \emph{spin transfer operator} $D^{(T)}_{ij}$ of  \cite{projectors} defined in equation \eqref{def:SpinTransfer}  in appendix \ref{SpinCB_bulk_examples}. 
This has the special role of mapping seeds in seeds (see appendix \eqref{recrelBULK}). Therefore it allows us to construct all the  seed three-point functions \eqref{3ptseed} from its action onto the seed representatives \eqref{3ptseedrepresentative}.

The bulk OPE is not affected by the presence of the defect. This means that it is possible to generate the spinning blocks in the bulk channel by  acting with $\Dcal^{(\pf)}$ on the (representative) seed partial waves,
\be
\Dcal^{(\pf)}
 \raisebox{1.8em}{
$
\xymatrix@=1.7pt{
{\Ocal_{\D_1\, l_1= l^{(2)}}}\ar@{-}[rdd]& &&&{ \ar@{=}[dd]  }\\  
&&&&\\
& *+[o][F]{}  \ar@{.}[rrr]^{\Ocal } & && *+[o][F]{} \ar@{=}[dd]  \\
&&&\\
\Ocal_{\D_2 \, l_2= l^{(3)}} \ar@{-}[ruu]&&&& }
$
}
=
 \raisebox{1.8em}{
 $
\xymatrix@=1.3pt{
{\Ocal_{\D_1 \, l'_1}}\ar@{-}[rdd]& &&&{ \ar@{=}[dd]  }\\  
&&&&\\
& *+[o][F]{\mbox{\tiny{$\pf$}}}  \ar@{.}[rrr]^{\Ocal }& & & *+[o][F]{} \ar@{=}[dd]  \\
&&&\\
\Ocal_{\D_2 \, l'_2} \ar@{-}[ruu]&&&& }
$
}
\ .
\ee
In appendix \eqref{recrelBULK} we show that the full set of partial waves (for $\Ocal_i$ in traceless and symmetric representations) is obtained by acting on seed partial waves with the following combinations:
\begin{equation}\label{diffBulk}
\Dcal^{(\bar \pf)}=
H_{12}^{n_{12}}D_{12}^{n_{13}}D_{21}^{n_{23}}D_{11}^{n_{1}}D_{22}^{n_{2}} D^{(T)\, k_2}_{12} \ \Sigma^{n_1+n_{23},n_2+n_{13}} \ .
\end{equation} 
The label $\bar \pf$ counts the choices of non-negative integers $n_{ij},n_{i}$ which satisfy the constraints \eqref{3ptconditions}.
The operator $\Sigma^{x_1,x_2}$ implements the shift on the external dimensions $\Delta_i \rightarrow \Delta_i+x_i$.

We put a bar on the OPE label $\bar \pf$ because the basis of differential operators is different from the OPE basis (labelled by $\pf$) defined in \eqref{BB_tensor_structures}.
There is however a linear map between the two bases $\Dcal^{(\pf)}=\sum_{\bar \pf} (a_{\bar \pf \, \pf})^{-1} \Dcal^{(\bar \pf)}$ which can be easily obtained acting with $\Dcal^{(\bar \pf)}$ on the representative seed and expressing the result in terms of the OPE basis \eqref{BB_tensor_structures}.  This problem was already addressed in the paper \cite{Costa2011} (see equation (3.31)).
The matrix $(a_{\bar \pf \, \pf})^{-1}$ is computed explicitly in few examples in appendix \ref{SpinCB_bulk_examples}.

As in the case of a four-point function of local operators, the differential operators \eqref{diffBulk} are not sufficient to generate all the blocks, since they do not provide a way to compute the seed representatives. In the four-point function case the problem was solved by the introduction of \emph{weight shifting} operators \cite{Karateev:2017jgd} which can be used to generate seed blocks by acting on scalar ones. It would be interesting to generalize this technology to defect CFTs. 

Finally, it is important to stress that there are just three new seed blocks which need to be computed in the case of the two-point function of spin two operators with a defect in generic dimensions. This has to be compared with the eight (non symmetric and traceless) seeds which are needed to tackle the case of the four-point function of stress tensors. In order to compute the missing seeds (in radial expansion)  one can apply the techniques explained in the sections above.  Moreover, in three spacetime dimensions only traceless and symmetric representations are allowed, therefore by acting with  spinning differential operators \cite{Costa2011} on the scalar bulk channel CB one can generate the full set of bulk channel blocks.

%%%%%%

%%%%%%%%%%%%%%%%%%%%%%%%%%%%%%%%%%%%%%%%%%%%%%%%%%%%%%%%%%%%%%% 
%%%%%%%%%%%%%%%%%%%%%%%%%%%%%%%%%%%%%%%%%%%%%%%%%%%%%%%%%%%%%%%
\subsection{Defect channel}
%%%%%%%%%%%%%%%%%%%%%%%%%%%%%%%%%%%%%%%%%%%%%%%%%%%%%%%%%%%%%%% 
%%%%%%%%%%%%%%%%%%%%%%%%%%%%%%%%%%%%%%%%%%%%%%%%%%%%%%%%%%%%%%%

In the following, we show that the full set of defect partial waves can be written in a closed form. In particular, they are simply related to the set of special functions $ {\bf P}^{n}$ introduced in \eqref{Pnl}. The functions  ${\bf P}^{n}$ explicitly computed in \cite{projectors, AdSWeightShifting}  are sufficient for obtaining all the defect CBs ``of interest''.
In order to further illustrate and check the results, in subsections \ref{defect_Spinning_CBs} and \ref{SpinZamDef} we also extend the radial expansion techniques to the defect channel. A list of computed conformal blocks for external vector operators is presented in appendix \ref{FinalRes}.

The factorized form of the defect symmetry group $SO(p+1,1)\times SO(q)$ gives rise to two independent Casimir equations for the parallel and transverse factors. We claim, and check in various cases, that the conformal partial waves \eqref{DEF_CB_Defect_Spin_Generic} can be written in embedding space in a completely factorized form:
\be
\label{Factorization_Spinning_Partial_Wave}
 \hG^{(\pf,\qf)}_{\hO}(P_i,Z_i) \equiv  \frac{ 
 \hG^{\pdot (\pf,\qf)}_{\hD \hl}(P_i,Z_i) \  \hG^{\tdot (\pf,\qf)}_{s}(P_i,Z_i)
 }{(P_1 \tdot P_1)^{\frac{\D_1}{2}}(P_2 \tdot P_2)^{\frac{\D_2}{2}}} \ .
\ee
The functions  
$\hG^{\pdot}$ and $\hG^{\tdot}$ are eigenfunctions of the parallel and transverse Casimir equations respectively (with appropriate boundary conditions), namely 
\be
\label{Casimir_Spin_Defect}
 - \frac{1}{2} (J_1^{\pdot} )^2  \, \hG^{\pdot (\pf,\qf)}_{\hD \hl}= c^{\pdot}_{\hD \hl}  \, \hG^{\pdot (\pf,\qf)}_{\hD \hl}\ ,
 \qquad \qquad
 -  \frac{1}{2} (J_1^{\tdot} )^2 \, \hG^{\tdot (\pf,\qf)}_{s}=c^{\tdot}_{s} \,  \hG^{\tdot (\pf,\qf)}_{s} \ .
\ee
Here $J_1$ is defined as $J_1^{M N} \equiv P_1^M \partial_{P_1}^N- P_1^N \partial_{P_1}^M+Z_1^M \partial_{Z_1}^N- Z_1^N \partial_{Z_1}^M $ where the suffix $\pdot$ ($\tdot$) means that we consider the indices $M,N$ to be in the parallel (transverse) space. The eigenvalues are
\be
c^{\pdot}_{\hD \hl} =\hD(\hD-p) + \sum_{i=1}^{[\frac{p}{2}]} \hl_i (\hl_i+p-2 i) \ ,
\qquad
\qquad
c^{\tdot}_{s} =\sum_{i=1}^{[\frac{q}{2}]} s_i (s_i+q-2 i) \ .
\ee
Notice that $P_i \tdot P_i$ commutes with both the Casimir operators. This implies that the defect conformal blocks are independent of the dimensions $\D_i$ of the external operators.

%%%%%%%%%%%%%%%%%%%%%%%%%%
\subsubsection{Seed blocks as projectors}
\label{Seed_Projectors}
%%%%%%%%%%%%%%%%%%%%%%%%%%
Our strategy will be to obtain a closed form expression for the so called seed partial waves and then to act on them with differential operators in order to generate the full set of conformal blocks. 

In this subsection we explain how to obtain all the seed conformal blocks  in terms of the mixed symmetry projectors found in \cite{projectors}, schematically
\be
\label{schematic_seed}
\mathtt{
seed\, CB
 =  Projector [SO(p+1,1)] 
 \times
Projector [SO(q)]
 }
\ .
\ee
Let us now explain the ingredients that enter formula \eqref{schematic_seed}. 
%----------------------------
%\paragraph{Seeds  = Projectors\\}
%-----------------------------
From the discussion in subsection \ref{ss:bulk-to-defect}, defect seed conformal blocks appear when the $SO(d)$ representations $l_i=(l_{i \, 1}, \dots, l_{i \, {[\frac{d}{2}]}})$ of the external operators $\Ocal_i$ satisfy the following relation:
\be
\label{SEED_condition}
\mathtt{seed}
\leftrightarrow
l_{1 \, i}=l_{2 \, i}=\hl_i+s_{i+1} \ ,
\ee
where $\hl=(\hl_1, \dots, \hl_{[\frac{p}{2}]})$ and $s=(s_1, \dots, s_{[\frac{q}{2}]})$ are respectively the parallel and transverse spins of the exchanged operator $\hO$.
For the sake of clarity, and in line with the main focus of the paper, from now on we restrict ourselves to the case of external traceless and symmetric primaries with spin $l_1$ and $l_2$,
\be
\label{seed_picture}
\mathtt{
seed \, CB
}
=   
  \raisebox{3.2em}{
$
\xymatrix@=0.3pt{
&&& \ar@{=}[dd]&  \\
&&&& \\
{\Ocal_{\D_1 l}}\ar@{-}[rrr]& &&*+[o][F]{} \ar@{:}[dd] & \\  
&  &&&   \hO_{\hD,\, \hl=l-s_2, \, (s_1s_2)} \\
\Ocal_{\D_2 l} \ar@{-}[rrr]&&&*+[o][F]{} \ar@{=}[dd]& \\
&&& \\
&&&& }
$
}
\  \ \ .
\ee

It is convenient to rephrase the condition \eqref{SEED_condition}  as a property of the parallel and transverse seed partial waves defined in \eqref{Factorization_Spinning_Partial_Wave}. The factorized seeds need to satisfy the following scaling properties:
\be
\label{seeds_scaling}
\hG^{\pdot}_{\hD \hl}(P_i,\a_i Z_i)= (\a_1\a_2)^{\hl} \hG^{\pdot}_{\hD \hl}(P_i,Z_i) \ ,
 \qquad
 \hG^{\tdot}_{s_1 s_2}(P_i,\a_i Z_i)= (\a_1\a_2)^{s_2}  \hG^{\tdot}_{s_1 s_2}(P_i,Z_i) \ .
\ee
This implies in particular that the full seed block $\hG_{\hD \hl (s_1,s_2)}$ satisfies the seed condition \eqref{SEED_condition}.
As we remarked at the end of section \ref{ss:bulk-to-defect}, seed blocks are automatically conserved.

We claim that all the transverse seed blocks can be simply written in terms of the polynomials \eqref{Pnl}. 
For example, if the external operators are symmetric and traceless we can write all the transverse seeds as follows 
\be
\label{q_seed}
 \hG^{\tdot}_{s_1 s_2}(P_i,Z_i) 
 = 
 \frac{
  {\bf P}^{q}_{s_1, s_2}(P_1,Z_1; P_2, Z_2 )
 }{(P_1 \tdot P_1)^{\frac{s_1}{2}}(P_2 \tdot P_2)^{\frac{s_1}{2}}} \ .
\ee
This can be easily seen from the leading defect OPE as we will describe in more detail in section \ref{defect_Spinning_CBs}. From an abstract point of view, one can check that \eqref{q_seed} satisfies all the required properties to be a seed. 
In fact \eqref{q_seed} has the appropriate scaling \eqref{seeds_scaling} and it satisfies the Casimir equation \eqref{Casimir_Spin_Defect}. In addition, \eqref{q_seed} is conserved. All of this immediately follows from the properties of the projectors described in subsection \ref{subsec:proj_son}.

We moreover claim that also the parallel seed blocks can be written in terms of the polynomials \eqref{Pnl}. 
This statement may look less trivial since the parallel seed $\hG^{\pdot}_{\hD \hl}$ is not a polynomial. However, if $\hD$ is a negative integer the parallel seed satisfies the same set of properties as the transverse one, thus, in this special case, we are lead to write
\ba
\label{p_seed}
 \hG^{\pdot}_{\hD \hl}(P_i,Z_i) 
  &=& 
  \frac{
 {\bf P}^{p+2}_{-\hD, \hl}(P_1,Z_1; P_2, Z_2 )
  }{(P_1 \pdot P_1)^{- \frac{\hD}{2}}(P_2 \pdot P_2)^{- \frac{\hD}{2}}} \ .
\ea
On the other hand, we are interested in the case where $\hD$ is a positive real number. We therefore define \eqref{p_seed} as the analytic continuation of \eqref{Pnl} for a Young tableau with a negative real number of boxes in the first row. In practice, this analytic continuation is straightforward: roughly speaking it amounts to replace a Gegenbauer polynomial by  a Hypergeometric function $_2F_1$.\footnote{This kind of analytic continuation applied to different physical problems also appeared in \cite{AdSWeightShifting, Kravchuk:2018htv}. }
As an example, we first revisit the scalar case, where the projector into the traceless and symmetric representation simply reduces to the Gegenbauer polynomial \eqref{Pnl_TST}. The transverse partial wave is in fact $ \hG^{\tdot}_{s} \propto C_s^{\frac{q}{2}-1}(\heta)$. The parallel one can be written as \cite{Billo:2016cpy} 
\begin{align}
\label{Scalar_Parallel_Seed}
 \hG^{\pdot}_{\hD \hl=0}
\propto \chi^\hD
{}_2F_1\left(\frac{1+\hD}{2},\frac{\hD}{2};\hD-\frac{p}{2}+1; \chi^2 \right)
\ ,\quad \chi=-\frac{(P_1 \pdot P_1)^{\frac{1}{2}} (P_2 \pdot P_2)^{\frac{1}{2}}}{P_1 \pdot P_2} \ .
\end{align}
Notice that the parallel block $\hG^{\pdot}$ is equal to the Gengenbauer $C_{-\hD}^{\frac{p}{2}}(1/\chi)$ when $\hD$ is a negative integer, up to an overall normalization. Thus, $\hG^{\pdot}$ is an analytic continuation of the projector \eqref{Pnl_TST} for a number $-\hD$ of boxes in the first row, and for $n=p+2$.
It is easy to check that the function $\hG^{\pdot}$ has the correct asymptotic behaviour to describe a defect conformal block.
We claim that even for more general Young tableaux we can still use the prescription
\be
\label{analytic_continuation}
C_l^{\frac{n}{2}-1}(x) \rightarrow 
\frac{2^l  \left(\frac{n}{2}-1\right)_l }{\Gamma (l+1)}  \; x^l {}_2F_1\left(\frac{1-l}{2},-\frac{l}{2};-l-\frac{n}{2}+2;\frac{1}{x^2}\right)
\ .
\ee
Notice that this replacement is easy to perform since every projector in \cite{projectors, AdSWeightShifting} is written in terms of an explicit differential operator acting on a single Gegenbauer polynomial as shown in \eqref{P_Generic}. 

Equations (\ref{q_seed}-\ref{p_seed}) are powerful formulae. Indeed, just by knowing \eqref{Pnl_TST} and \eqref{Pnl1} we  automatically obtain the seed blocks for the exchange of the operators $\hO_{\hD, \hl=0, s}$ (which appears for scalar external operators, $l_i=0$), $\hO_{\hD, \hl=1, s}$, $\hO_{\hD, \hl=0, (s,1)}$ (which appear for $l_i=1$) and $\hO_{\hD, \hl=1, (s,1)}$  (for $l_i=2$). In \cite{AdSWeightShifting} it is explained how to obtain projectors with an arbitrary number of boxes in the second row by applying differential operators on the traceless and symmetric projector. This implies that from  (\ref{q_seed}-\ref{p_seed}) we can obtain  the defect seed blocks  for any two point function of  traceless and symmetric operators. Moreover from the projectors computed in a closed form in \cite{projectors} one can also extract seed blocks for external operators in mixed symmetric representations of $SO(d)$.

From formulae  (\ref{q_seed}-\ref{p_seed}) it is also possible to argue that, when $p$ is even, the dependence on $\hr$ of all the seed blocks  is of the form $\hr^\hD R(\hr)$, where $R(y)$ is a rational function of $y$. Indeed, from formula \eqref{P_Generic}, we see that all the parallel seed blocks are obtained by acting with a finite number of derivatives on the scalar block.\footnote{To be precise, the derivatives are in the variable $1/\chi$, but the Jacobian of the change of coordinate from $\chi$ to $\hr$ is a rational function of $\hr$. Also, the schematic dependence on $x$ of formula \eqref{P_Generic} is polynomial, as one can see from the example \eqref{Pnl1}.}
Since for even $p$ the radial part of the scalar block is of the form $\hr^\hD R(\hr)$, we conclude that the full result takes the same form.

In the next subsection we define new differential operators of the kind explained in subsection \ref{Spinning_Differential_Operators_Bulk}
 and in \cite{Costa2011}, which generate conformal blocks for external operators with generic spin by acting on a seed block. Knowledge of the seed blocks and of the differential operators allows to compute all the defect conformal blocks for external traceless and symmetric operators.

\subsubsection{Spinning differential operators}
In this subsection we explain how to generate all the spinning defect conformal blocks by acting with differential operators on seed blocks. 
First, we define a set of differential operators $\hat \Dcal$ that create the bulk-defect spinning structures out of the seed ones.
Schematically, we look for an operator $\hat \Dcal_i^{(\pf)}$ such that
\be
\label{Def_Spinning_Diff_Op_Defect}
 \hat \Dcal_i^{(\pf)}
   \raisebox{1.7em}{
$
\xymatrix@=0.3pt{
&&& \ar@{=}[dd] & \\
&&& &\\
{\Ocal_{\D_i l}}\ar@{-}[rrr]& &&*+[o][F]{} \ar@{=}[dd]&  \hO_{\hD,l-s_2,(s_1,s_2)}\\  
&  &&& \\
&&&&\\
}
$
}
=
   \raisebox{1.7em}{
$
\xymatrix@=0.3pt{
&&& \ar@{=}[dd] & \\
&&& &\\
{\Ocal_{\D_i l_i}}\ar@{-}[rrr]& &&*+[o][F]{\mbox{\tiny{$\pf$}}} \ar@{=}[dd]&  \hO_{\hD,l-s_2,(s_1,s_2)}\\  
&  &&& \\
&&&&\\
}
$
}
\ ,
\ee
where $l_i>l$ is a generic spin and the index $\pf$ labels a choice of bulk-defect tensor structure as shown in \eqref{BuildBlocksBulkToDef}.
Following the logic explained in \cite{Costa2011}, the differential operators must be functions of positions $P_i$ and the polarizations $Z_i$ of the external bulk operators $\Ocal_i$. 
 As in the bulk case, we consider $\hat \Dcal_i$ to be a composition of elementary operators, each of them increasing the degree of homogeneity of $Z_i$ of one or two units at a time.  To obtain the form of the operators we first impose that their action preserves the submanifold defined by 
 \begin{equation}
 P_i \cdot P_i = P_i \cdot Z_i = Z_i \cdot Z_i=0 \ .
 \end{equation}
While this condition was sufficient to uniquely fix the bulk differential operator, in the defect case it leaves some freedom. Therefore, we explicitly require that \eqref{Def_Spinning_Diff_Op_Defect} holds, namely that the action of the differential operators on a generic bulk-defect tensor structure \eqref{btodefect2pt} is a linear combination of tensor structures \eqref{btodefect2pt}. This is carefully explained in appendix  \ref{App_Defect_Spinning_Diff_Ops}. The result is that the operator $\hat \Dcal_i$ (for $i=1,2$) is generated by products of three elementary operators. The first two take the differential form 
  \begin{align}\label{diffOP_def1}
  \!\!\!\!
 \widehat{D}^\star_{i} & \equiv (P_i \star Z_i){ }(P_i \star \partial_{ P_i}) -(P_i \star P_i){ }(Z_i \star {\partial_{ P_i}})- (P_i \star Z_i){ }(Z_i \star \partial_{ Z_i})+ (Z_i \star Z_i){ }(P_i \star \partial_{ Z_i}),
 \end{align}
 where $\star=\pdot, \tdot$. The third one is simply the multiplication by $\Hp{ii}$, defined in \eqref{HVdef}, which increases the spin by two at point $P_i$.
 
 In conclusion, we obtain a generic spinning conformal block in the defect channel just by acting with differential operators on a seed block \eqref{p_seed}, namely 
  \be
 \hat\Dcal_1^{(\pf)}
 \hat \Dcal_2^{(\qf)}
 \
 \left[
   \raisebox{2.5em}{
$
\xymatrix@=0.3pt{
&&& \ar@{=}[dd]  \\
&&& \\
{\Ocal_{\D_1 l}}\ar@{-}[rrr]& &&*+[o][F]{} \ar@{:}[dd]  \\  
&  &&  \\
\Ocal_{\D_2 l} \ar@{-}[rrr]&&&*+[o][F]{} \ar@{=}[dd] \\
&&& \\
&&& }
$
}
\mbox{\scriptsize $ \hD,l-s_2,(s_1,s_2)$}
\right]
\
=
\   
  \raisebox{2.5em}{
$
\xymatrix@=0.3pt{
&&& \ar@{=}[dd]  \\
&&& \\
{\Ocal_{\D_1 l_1}}\ar@{-}[rrr]& &&*+[o][F]{\mbox{\tiny $\pf$}} \ar@{:}[dd]  \\  
&  && \\
\Ocal_{\D_2 l_2} \ar@{-}[rrr]&&&*+[o][F]{\mbox{\tiny $\qf$}} \ar@{=}[dd] \\
&&& \\
&&& }
$
}
\mbox{\scriptsize $ \hD,l-s_2,(s_1,s_2)$}
\ ,
\ee
 where each operator $ \hat\Dcal_i^{(\pf)}$ is generated by the composition of the elementary building blocks \eqref{diffOP_def1} and $\Hp{ii}$ 
 \be
 \label{hD_explicit}
 \hat\Dcal_i^{(\bar \pf)} =  (\Hp{ii})^{m_i}(\hat{D}_i^\pdot)^{n_i^\pdot} (\hat{D}^\tdot_{i})^{n_i^\tdot} \Sigma_i^{n_i^\pdot + n_i^\tdot} \ .
  \ee
 The operator $\Sigma_i^{n}$ implements the shift  $\D_i \to \D_i + n$. The index $\bar \pf$ labels the number of ways in which one can  fix  the integers $m_i,\,n_i^\pdot,\,n_i^\tdot$ such that $2 m_i+ n_i^\pdot + n_i^\tdot=l_i-l$. 
We introduced  a barred index $\bar \pf$, since the basis \eqref{hD_explicit} of the differential operators is not the same of \eqref{BuildBlocksBulkToDef} labelled by $\pf$. However one can easily obtain the change of basis by performing the computation sketched in \eqref{Def_Spinning_Diff_Op_Defect} and expressing the right hand side in terms of \eqref{BuildBlocksBulkToDef}. This procedure gives an invertible map between the differential basis \eqref{hD_explicit} and the basis \eqref{BuildBlocksBulkToDef}. For more details we refer to appendix  \ref{App_Defect_Spinning_Diff_Ops}.

The differential operators either act on the parallel space or on the transverse one. Therefore the partial waves preserve the factorized form
\ba
\label{excitedG}
\hG^{\pdot [m,n]}_{\hD \hl}
& \equiv&
 \frac{
 (\hat D^\pdot_{1})^{m} (\hat D^\pdot_{2})^{n}
  {\bf P}^{\, p+2}_{-\hD, \hl}(P_1,Z_1; P_2, Z_2 )
 }{(P_1 \pdot P_1)^{-\frac{\hD}{2}}(P_2 \pdot P_2)^{-\frac{\hD}{2}}}
\ , 
\\
\hG^{\tdot [m,n]}_{s_1 s_2} 
 & \equiv&
  \frac{
 (\hat D^\tdot_{1})^{m} (\hat D^\tdot_{2})^{n}
  {\bf P}^{\, q}_{s_1,s_2}(P_1,Z_1; P_2, Z_2 )
 }{(P_1 \tdot P_1)^{\frac{s_1}{2}}(P_2 \tdot P_2)^{\frac{s_1}{2}}} \ .
\ea
Hence, we can write a very compact and explicit formula for all the conformal partial waves which can appear in the expansion of a two point function of any external traceless and symmetric operator:
\be
\label{All_Blocks}
 \hG^{(\bar \pf,\bar \qf)}_{\hO} \equiv  (\Hp{11})^{m_1} (\Hp{22})^{m_2} \ \frac{
 \
 \hG^{\pdot [n_1^\pdot,n_2^\pdot]}_{\hD \hl} \  \hG^{\tdot [n_1^\tdot,n_2^\tdot]}_{s_1s_2}
 }{(P_1 \tdot P_1)^{\frac{\D_1+n_1^{\pdot}+n_1^{\tdot}}{2}}(P_2 \tdot P_2)^{\frac{\D_2+n_2^{\pdot}+n_2^{\tdot}}{2}}} \ .
\ee
Again, here the labels $(\bar \pf,\bar \qf)$ count the possible ways to choose $n_i^\star$ and $m_i$ subject to the constraints mentioned above. 

In appendix \ref{App_Defect_Spinning_Diff_Ops} we give more details on the construction of these differential operators, and provide a few examples. 

One can in principle generalize this framework in order to find conformal blocks for external operators in any representation of $SO(d)$. It would be also interesting to generalize the formalism of \cite{Karateev:2017jgd}, which obtained differential operators that change the representation of the exchanged operators.

%%%%%%%%%%%%%%%%%%%%%%%%%
\subsubsection{Radial expansion}
\label{defect_Spinning_CBs}
%%%%%%%%%%%%%%%%%%%%%%%%%
In this subsection we show that the defect conformal blocks can be written as a convenient expansion in radial coordinates. The results presented here are a generalization of the radial expansion for scalar blocks obtained in \cite{Lauria:2017wav} and provide a check of formula \eqref{All_Blocks}.

We are interested in computing the functions $\hGcal_{\hO}$ which are obtained by projecting the partial waves \eqref{DEF_CB_Defect_Spin_Generic} onto the defect radial frame of section \ref{ss:real_rad},
\be 
\label{GToGcal_drf_Spin}
\sum_{\pf \qf} b_{\hO}^{(\pf)}b_{\hO}^{(\qf)} \hG^{(\pf,\qf)}_{\hO}(P_i,Z_i) \underset{d.r.f.}{\longrightarrow}  \hr^{-\D_2}\, \hGcal_{\hO}(\hr,\eta, n \circ z_k, n' \circ z_k, z_k \circ z_j, z_k \cdot z_j) \ ,
\ee
as detailed in appendix \eqref{tensor structures to radial frame defect}.
We define the function $ \hGcal$ by inserting a projector in the two-point function written in the defect radial frame,
\be 
\label{Gcalpq_D}
\hGcal_{\hD \hl s}
=
\left\langle
\hat 0 |
\mathcal{O}_1\left(n,z_1\right)
\hr^{H_{cyl}} \mathcal{P}_{\hD \hl s}
\mathcal{O}_2\left(n',z_2\right)
\hat 0 
\right\rangle
 \ .
\ee
$H_{cyl}$ is the Hamiltonian conjugate to the cylinder time $\tau = \log \hr$.
$\mathcal{P}_{\hD \hl s}$ projects onto the conformal family with highest weight labeled by $\hD$, $\hl$ and a Young tableau $s$ which encodes the transverse spin. With traceless and symmetric external operators, the Young tableau  $s$  can have at most two rows, namely  $s=(s_1,s_2)$. 
We then rewrite the projector as a sum over a complete basis of defect states
\begin{align} 
\label{GcalSumStates_spin_D}
&\hGcal_{\hD \hl s} =  \sum_{m=0}^\infty \hr^{\hD+m}
\sum_\hj
%_{\hj=\max[\hl-m,0]}^{\min[\hl+m,l_1,l_2]}
%%%
\left\langle
\hat 0 |
\mathcal{O}_1(n,z_1)
| m, { \scriptsize
\ytableausetup{centertableaux,boxsize=1.2 em}
\hj,
%\begin{ytableau}
%a_1&\, _{\cdots}&a_j \\
%\end{ytableau}
%\, ,
%\begin{ytableau}
%i_1&\, _{\cdots}&i_s \\
%\end{ytableau}
s
}
 ,\dd \rangle
 %%%
   \langle m, { \scriptsize
\ytableausetup{centertableaux,boxsize=1.2 em}
\hj,
%\begin{ytableau}
%a_1&\, _{\cdots}&a_j \\
%\end{ytableau}
%\, ,
%\begin{ytableau}
%i_1&\, _{\cdots}&i_s \\
%\end{ytableau}
s
},\dd 
|
\mathcal{O}_2(n',z_2)
|
\hat 0 
\right\rangle
%%%
\,,
\end{align}
where the sum over the parallel spin $\hj$ runs from $\max[\hl-m,0]$ to $\min[\hl+m,l_1-s_2,l_2-s_2]$.
Requiring Lorentz invariance fixes  the general structure of the bulk-defect overlaps as follows
\ba
 \label{Spin_Radial_Defect_OPE}
\begin{split}
&
\langle
\hat 0 |
\mathcal{O}_1(n,z_1)
| m, { \scriptsize
\ytableausetup{centertableaux,boxsize=1.3 em}
\begin{ytableau}
a_1&\, _{\cdots}&a_{\hj} \\
\end{ytableau}
\, ,
\begin{ytableau}
i_1&\, _{\cdots}&\, _{\cdots}&\, _{\cdots}&\, i_{s_1} \\
i'_1&\, _{\cdots}&\, i'_{s_2}
\end{ytableau}
},
\dd
 \rangle
 =
  z_1^{(a_1} \cdots z_1^{a_{\hj})} \times
 \\
& \quad \times
\sum_p u^{(\pf)}(m,\hj,\dd)  t^{(\pf)}_{\hj}( z_1 \tdot \nabla_n ,z_1  \tdot n , z_1  \tdot z_1) 
\ytableausetup{centertableaux,boxsize=1.2 em}
\
 \pi_{s_1,s_2}^{SO(q)}\left(
 \,
 { \scriptsize
\begin{ytableau}
i_1&\, _{\cdots}&\, _{\cdots}&\, _{\cdots}&\, i_{s_1} \\
i '_1&\, _{\cdots}&\, i '_{s_2}
\end{ytableau}
}
\, , \,
{ \scriptsize
\begin{ytableau}
n&\, _{\cdots}&\, _{\cdots}&\, _{\cdots}&\, n \\
z_1&\, _{\cdots}&\, z_1
\end{ytableau}
}
\,
 \right) 
\ ,
\end{split}
\ea
where $\nabla_n^\m=\partial^\m_n-n^\m \, (n\cdot \partial_n)$, the indices $a_k$ are in parallel directions and the indices $i_k$ in orthogonal directions.
Notice that the structures $t^{(\pf)}$ are generated by three building blocks $z_1 \tdot \nabla_n ,z_1 \circ n , z_1 \circ z_1$, while the extra building block  $ z_1^a$ is already factorized in formula \eqref{Spin_Radial_Defect_OPE}.
The four building blocks plus the projector itself are in fact in correspondence with the embedding space structures defined in formula \eqref{BuildBlocksBulkToDef} (the projector takes into account the contributions of two structures: each column of length $1$ correspond to a $K^2_{1}$ while each column of length $2$ to $S^2_1$) .
As an example, when the operator $\Ocal_1$ has  spin $l_1=1$ and the exchanged operator has $s_2=0$ we have two possible cases: either $\hj=0$ or $\hj=1$.
 When $\hj=0$ there are two possible structures,
\be
\label{JsO_radial}
 t^{(1)}_0=z_1  \tdot n
 \ ,
\qquad
\qquad
t^{(2)}_0=z_1 \tdot \nabla_n  \ .
\ee
On the other hand when $\hj=1$ there is just the trivial structure $t^{(\pf)}_1=1$.

 Putting together the left and right overlaps we obtain the following expression for the conformal blocks:
\ba
\begin{split}
\label{Ansatz_Defect_Radial_Frame}
&\hGcal_{\hD \hl (s_1s_2)}= \sum_{\hj=0}^{\min[l_1,l_2]} {\bf P}^{p}_{\hj}(z_1 , z_2)
\sum_{\pf \qf}  
t^{(\pf)}_{\hj}( z_1 \tdot \nabla_n ,z_1  \tdot n , z_1  \tdot z_1) 
\\
&
\qquad\qquad\qquad\qquad
\times 
t^{(\qf)}_{\hj}( z_2 \tdot \nabla_{n'} ,z_2  \tdot n' , z_2  \tdot z_2)
{\bf P}^{q}_{s_1,s_2}(n,z_1 ; n', z_2)
 \Wcal_{\hj}^{(\pf,\qf)}(\hr)
\ ,
\end{split}
\ea
where the functions $\Wcal^{(\pf,\qf)}_{\hj}$ are defined as
\be
\label{Def_W_Defect}
 \Wcal_{\hj}^{(\pf,\qf)}(\hr) = \sum_{m=0}^\infty w_\hj^{(\pf,\qf)}(m) \hr^{\D+m} \ .
\ee
with $w^{(\pf,\qf)} = \sum_{\dd}u^{(\pf)}\tilde u^{(\qf)}$. Let us now compare the ansatz \eqref{Ansatz_Defect_Radial_Frame} with the counterpart in embedding space \eqref{Factorization_Spinning_Partial_Wave}. 
The transverse piece ${\bf P}^{q}_{s_1,s_2}$ is equal to the transverse seed \eqref{q_seed} after projecting the points to the radial frame \eqref{D_Poin_Conf}. However, the full result \eqref{Ansatz_Defect_Radial_Frame} for fixed $\pf,$ $\qf$ is not factorized in a purely transverse times a purely parallel part. This is expected, since the parallel scalar products in embedding space project onto linear combinations of both parallel and orthogonal products in the defect radial frame of subsection \ref{ss:real_rad}.

At this point, the parallel functions are still unknown since the coefficients $w_\hj^{(\pf,\qf)}$ have not been fixed (in fact so far we only imposed Lorenz symmetry). 
In order to compute the functions $\Wcal_\hj^{(\pf,\qf)}(\hr)$ it is convenient to plug the ansatz 
(\ref{Ansatz_Defect_Radial_Frame}-\ref{Def_W_Defect}) into the Casimir equation. 
This leads to simple recurrence relations for the coefficients $w_\hj^{(\pf,\qf)}$, which we were able to solve and resum in all the cases that we considered. 

In appendix \ref{App_Examples_Defect_CB_Casimir_Expansion} we give two examples of this technique. First we consider the two point function of a vector and a scalar operator, then we study the case of two external vectors. In both cases we obtain closed form expressions for the defect channel conformal blocks which match the results of formula \eqref{All_Blocks}.

\subsubsection{Zamolodchikov recurrence relation}\label{SpinZamDef}

In this subsection we apply Zamolodchikov's recurrence to the defect conformal blocks for spinning external operators.
Again, we focus on external operators in the traceless and symmetric representation, but everything can be easily generalized to other cases.

Spinning defect conformal partial waves have poles at special values of $\hD$ with residues proportional to other conformal partial waves. The expected analytic structure in the poles of a generic spinning defect conformal partial wave is
\be
\label{recrel_defect_CPW}
 \hG^{(\pf,\qf)}_{\hO}(P_i,Z_i)= \sum_{\pf,\qf} \frac{(R_A)_{\pf \pf' \qf \qf'}}{\hD-\hD_A^{\star}}  \hG^{(\pf',\qf')}_{\hO_A}(P_i,Z_i) +O((\hD-\hD_A^{\star})^0) \ .
\ee
Note that $R_A$ is a matrix that mixes the various defect conformal partial waves associated to the exchange of the primary descendant operator $\hO_A$. 

We can then obtain a recurrence relation for the defect conformal blocks defined in \eqref{DEF_CB_Defect_Spin_Generic} by summing over the poles in $\hD$ and on the regular part,
\begin{align}
\begin{split}
\label{D_CB_RecRel_Spinning}
\hat h_{\hD \hl s}^{(\pf,\qf),k}(\hr,\heta)&\equiv (\hr)^{-\D} \hat g_{\hD \hl s}^{(\pf,\qf),k}(\hr,\heta) \underset{\hD \rightarrow \infty}{\longrightarrow} \hat h_{\infty \hl s}^{(\pf,\qf),k}(\hr,\heta) \ ,
\\
\hat h_{\hD \hl s}^{(\pf,\qf),k}(\hr,\heta)&=\hat h_{\infty \hl s}^{(\pf,\qf),k}(\hr,\heta)+\sum_{A}  \sum_{\pf' \qf' } \hr^{n_A}  \frac{(R_A)_{\pf \pf'\qf \qf'}}{\hD-\hD^\star_A} 
\hat h_{\hD_A \hl_A s}^{(\pf',\qf'),k} (\hr,\heta) \ .
\end{split}
\end{align}
The transverse $s$ spin is diagonal in formula \eqref{D_CB_RecRel_Spinning} since $\hO_A$ is a descendant of $\hO$.
The sum over $A$ runs over the types (I,II,III) and the integers $n$.  When the two external operators are traceless and symmetric, the values of $A$ and $n$ can be found in the following table\footnote{When the external operators are not in the traceless and symmetric representation there are more types which match the extra types $\I_k, \II_k$ obtained in \cite{recrel}.}
\be\label{TableDefRes}
\begin{array}{|c | ccc|}
\hline
\phantom{\Big(}		A														&\D^\star_A 					&n_A 										&\hl_A	 \\ 
\hline
\phantom{\Big(} \mbox{Type \I } \quad \ \
n=1,\dots 	n^{\max}_{\I} \quad				& 1-\hl-n 			 				& n 	  	   									 &\hl + n \\ 
%\!\!\!\!\!\!\!
\phantom{\Big(}   \mbox{Type \II } \quad \ 
 %\ \ \ \ \
n=1, \dots n^{\max}_{\II} \quad 
&\quad   \hl+p-1-n  	\quad & \quad n \quad  						&\quad \hl - n \quad  \\
%\!\!\!\!\!\!\!\!\!\!\!\!\!\!
\phantom{\Big(} \mbox{Type \III } \quad 
%\ \ \ 
n=1,\dots \infty \quad \ \
 				& p/2-n  							& 2n       			 						 &\hl       \\ 
\hline
\end{array}
\ee
and $\D_A=\D^\star_A+n_A$. 
The value of $n$ for the type \III runs in general over an infinite range, while the type \I and \II are bounded by the spin of the external states. In fact $n^{\max}_{\I}=\min(l_1,l_2)-s_2$ while $n^{max}_\II=\hl\leq \min(l_1,l_2)-s_2$, where $s=(s_1,s_2)$ is the transverse spin of the exchanged operator. The function $\hat h_{\infty \hl s}^{(\pf,\qf),k}(\hr,\heta)$ can be computed case by case by solving the Casimir equation at leading order in large $\hD$. Finally, $R_A$ is defined as follows:
\be
\label{R=MQM_general_defect}
(R_A)_{\pf \pf' \qf \qf '}= (M^{(L)}_A)_{\pf \pf'} \hat Q_A (M_A^{(R)})_{\qf \qf '} \ ,
\ee
and can be computed following the same logic of \cite{recrel}. In particular the $\hat Q_A$'s are obtained by comparing the normalization of the two-point function of primary descendant operators with the one of canonically normalized primaries. The expression of $\hat Q_A$ is the same as the one computed in \cite{recrel}, once we replace $h\rightarrow p/2$ and $l\rightarrow \hl$ and the result is reported for completeness in appendix \ref{SpinningZam}. The coefficients $M_A$ are obtained by computing the normalization of a bulk-defect two-point function where the defect operator is a primary descendant,
\begin{equation}
\label{Mdef}
\langle \Ocal_1 \hOcal_A \rangle^{(\pf)} = \sum_{\pf'} (M^{(L)}_A)_{\pf \pf'} \langle \Ocal_1 \hOcal' \rangle^{(\pf')} \ ,
\end{equation}
where $\hOcal'$ is a canonically normalized primary with the same quantum numbers of $\hO_A$. In appendix \ref{SpinningZam} we present all the ingredients to obtain the recurrence relation for the scalar-vector and the vector-vector two-point functions.

 In appendix \ref{SpinningZam} - see equation \eqref{type3onmax} - we also show that when $p$ is even the poles of type \III of any seed block have zero residue for $n \geq p/2-1$. Thus, in this case there is a finite number of poles for any conformal block. We recover the expectation that all the defect conformal blocks $\hat h_{\hD \hl s}$ drastically simplify and become rational functions of $\hr$. 

Let us make a general comment on the matrix $R_A$ in equation \eqref{R=MQM_general_defect}. 
From \eqref{Mdef} it is easy to see that $(M^{(L)}_A)_{\pf \pf'}=0$ if the two point functions $\langle \Ocal_1 \hOcal \rangle^{(\pf)}$  and $\langle \Ocal_1 \hOcal' \rangle^{(\pf')}$ do not share the same transverse tensor structures (\emph{i.e.} they need to share the same $s_2,n_2,m_{12}$  in the notation of \eqref{BuildBlocksBulkToDef}). In fact the descendant $\hOcal_A$ is obtained by acting with  parallel derivatives on $\hOcal$, which commute with all transverse products $x \tdot y$.  Thus, the block $\hG^{(\pf,\qf)}_{\hO}$ can only be proportional to $\hG^{(\pf',\qf')}_{\hO_A}$ for special values of $\pf,\pf'$  and $\qf,\qf'$ when their transverse part exactly matches. 
This means that just few of the  $\hat h^{(\pf,\qf)}$ (with different $\pf,\qf$) will actually be coupled in \eqref{D_CB_RecRel_Spinning}, or in other words that $R_A$ is going to be very sparse. We will explicitly see this phenomenon happening in the examples computed in appendix \ref{SpinningZam}.

In order to avoid the redundancy explained above it is possible to directly write  a recurrence relation for any linear combination (in $m$ and $n$) of $\hG^{\pdot [m,n]}_{\hD \hl}$ defined in \eqref{excitedG}, which by definition involves only the parallel part of the partial waves. We chose however to present the recurrence relation without introducing this optimization step for the sake of clarity.
On the other hand, from this point of view we can give a new interpretation to the Zamolodchikov recurrence relation for defect blocks. Roughly speaking it can be understood  as a recurrence relation for the (analytically continued) projectors themselves.
For example the recurrence relation for the scalar CB (obtained in \cite{Lauria:2017wav}) can be exactly interpreted as a property of ${\bf P}^{n}_{-\hD}$ when we continue into the complex plane the number  $-\hD$ of boxes in the first row of the projector. It would be interesting to expand on this point of view and see whether it may help in the construction of more generic projectors.

%%%%%%%%%%%%%%%%%%%%%
\section{Example: the scalar Wilson line}
\label{sec:example}

As a further check of our formulae, we would like to apply the formalism to a specific example. 
We consider a theory of a complex free boson in 4d coupled to a line defect of the following form:
\beq
\mathcal{D}=\exp \left( \frac{1}{\pi} \int_\Sigma \bar{\l} \varphi+\l \bar{\varphi}\right)
\label{scalarWL_2}
\eeq
where $\Sigma$ is a straight line. $\mathcal{D}$ is inserted in the path-integral of the free theory. The theory has a conserved current for the $U(1)$ symmetry of the bulk:
\begin{equation}\label{U1current}
J_\mu=\frac{\ii}{2}(\varphi\partial_\mu \bar{\varphi}-\bar{\varphi}\partial_\mu\varphi)~,
\end{equation} while the next vector primary $O_\mu$ has dimension $\Delta_O=3(d-2)/2+1=4$ in $d=4$:
\beq
O_\m=\frac{1}{2}\bar{\varphi}^2 \partial_\m \varphi-\frac{1}{2}\varphi\bar{\varphi}\partial_\m \bar{\varphi}.
\label{D4J1}
\eeq
In the rest of the section, we present the conformal block decomposition of correlators involving \eqref{U1current} and \eqref{D4J1}. We shall take $\varphi$ to be canonically normalized: $\braket{\varphi \bar{\varphi}}\sim 1$.

\subsection{Decomposition of $\langle O O \rangle$}

We begin with the non-conserved vector primary \eqref{D4J1}. We choose to exhibit the two-point function $\braket{O_\mu O_\nu}$ rather than $\braket{O_\mu \bar{O}_\nu}$. Notice that the former does not vanish since the defect transforms under $U(1)$. After some simple Wick contractions, one finds:
\begin{multline}
\label{2ptNOnCons}
\braket{O(P_1,Z_1)O(P_2,Z_2)}= 
\frac{ 
1}{(P_1\wbullet P_1)^2 (P_2\wbullet P_2)^2} \ \frac{\bar{\l} ^2}{128 \, \xi ^3} \\
\times \Big\{ \left(- c \left(16 \lambda  \bar{\lambda} \xi +3\right)-16 \lambda  \bar{\lambda} \xi ^2+2 \xi \right) Q_1 - c \left(16  \lambda  \bar{\lambda} \xi +3\right)(Q_2+Q_3) \\
  -\frac{c^2 \left(16  \lambda  \bar{\lambda} \xi +3\right)+8\xi c \left(2  \lambda  \bar{\lambda} \xi +1\right)}{ ( 2\xi+c )}Q_4 - 4\left(2 \lambda  \bar{\lambda} \xi +1\right)(Q_5+Q_6) 
\Big\}~.
\end{multline}
Where the two cross ratios $\xi$ and $c\equiv \cos\phi$ are defined as \cite{Lauria:2017wav}
\begin{align}
\xi= -\frac{P_1 \cdot P_2}{2(P_1 \tdot P_1 )^{\frac{1}{2}}(P_2 \tdot P_2 )^{\frac{1}{2}}},\quad \cos\phi =\frac{P_1 \tdot P_2}{(P_1 \tdot P_1 )^{\frac{1}{2}}(P_2 \tdot P_2 )^{\frac{1}{2}}}.
\label{xicosphi}
\end{align}
We recall that the structures $Q_k$ are defined as follows
\begin{align}\label{JJbulkStructures}
Q_1=\Vp{1} \Vp{2}, 
\; 
Q_2=\Vp{1} \Vt{2},
\;
Q_3=\Vt{1} \Vp{2},
 \;
Q_4=\Vt{1} \Vt{2},
 \;
Q_5=\Hp{12},
 \;
Q_6= \Ht{12}~,
\end{align}
in accordance with the discussion in section \ref{SO(d+1,1)SO(q)}.

\subsubsection*{Bulk channel}
The bulk OPE is constrained by charge conservation, and since $O_\mu$ has unit charge, the only contributing primaries have charge 2. Wick theorem restricts the choice to primaries built out of $\varphi^m \bar{\varphi}^{m+2}$, with $m=0,1,2$. Their one-point functions are proportional to $\l^m\bar{\l}^{m+2}$. We learn from eq. \eqref{2ptNOnCons} that operators with $m=2$ are not exchanged.  As we shall see, this is enforced by the structure of the defect channel. The exchanged operators are either  symmetric traceless tensors with even spin or mixed symmetric tensors with labels $(l,2)$: this is presented in table \eqref{TAB_2pt_Spin_Example_bulk}, and it is easily verified using eqs. (\ref{three_point_BBB}).
Let us first consider the symmetric and traceless tensors $\Ocal$ of even spin $l=2m$ and conformal dimensions 
$\D=l+ 2n+2$ ($n\in \mathbb{N}$). It is immediate to derive the following table for the one-point functions of $\Ocal$:
\beq
\begin{array}{ll}
\D-l=2,\ \ &a_\Ocal \propto \bar{\lambda}^2 \\
\D-l\in 2\mathbb{N}+4,\ \ & a_\Ocal \propto \lambda\bar{\lambda}^3~.
 \label{ex_bulkops}
\end{array} 
\eeq
As it turns out, only the subset of operators with $\D-l\in 4\mathbb{N}+4$ is exchanged in the correlator \eqref{2ptNOnCons}.
Moreover, no $SO(d)$ representation with labels $(l,2)$ is exchanged. 
It can be checked that primaries in this representation exist in the spectrum of theory, with the right quantum numbers to be exchanged in the OPE of $O_\mu$ with itself.\footnote{One can use the conformal characters \cite{Dolan:2005wy}, see \emph{e.g.} \cite{deMelloKoch:2017dgi} for the explicit computation relevant here.}
 A direct computation for a low lying example shows that they also acquire a one-point function. The resolution is that their three-point function with the external operators vanishes. Therefore, we only compute the partial waves for symmetric traceless exchanges in appendix \ref{SpinCB_bulk_examples}, and in particular we use the basis $G_{\Delta,l}^{OO,(\pf)}$ for equal external operators defined in \eqref{GOO}. 
 
The decomposition reads
\begin{align}
\braket{O(P_1,Z_1)O(P_2,Z_2)}= &  \frac{\bar{\lambda}^2}{4^5}\sum_{l=2\mathbb{N}}
\a_{l} \left(G^{OO,(4)}_{\D=2+l,l}+\frac{1}{3}G^{OO,(3)}_{\D=2+l,l}\right)+\nonumber\\
&+  \frac{\bar{\lambda}^3 \lambda}{4^5} \sum_{l=2\mathbb{N}}  \sum_{\D=4+4\mathbb{N}+l}
\b_{\D,l} \left(G^{OO,(4)}_{\D,l}-\frac{1}{6}G^{OO,(3)}_{\D,l}\right) \ ,
\end{align}
where we factored some powers of $4$ for convenience.
It is interesting to notice that only the structures $\pf=3,4$ appear, out of the four possible.

All the exchanged operators of twist $\Delta-l=2$ appear in the CBs decomposition weighted by coefficients  $ \a_{l} $ which take the following form
 \begin{equation}
\a_{l} =3\ \frac{ \left(\frac{1}{2}\right)_{l/2}^2}{  (\frac{l}{2})! \left(\frac{l+1}{2}\right)_{\frac{l}{2}}} .
 \end{equation} 
We could not guess a general form for the coefficients $\b_{\D,l}$, associated to the exchange of operators with twist $\Delta-l=4+4 n$. We propose a closed form result for $n=0,1$ which is compatible with the conformal block expansion up to values of $\D \leq 22$, 
 \be
   \b_{\D=4+l,l}= 3 \frac{ (l+1)!}{2^{l}\left(\frac{3}{2}\right)_{l}}, \qquad 
  \b_{\D=8+l,l}=   \frac{  (l+2) (l+5) (l+2)!}{ 2^{l+7}(l+4)  \left(\frac{3}{2}\right)_{l+2}} \ .
 \ee
 For completeness we also report the firsts few coefficients for higher twist operators $\Delta-l=4+4 n$ (with $n\geq3$),
 \be
 \begin{array}{|c | cccccc c |}
\hline
\phantom{\Big|}l&\quad 0 \qquad&\qquad 2 \qquad&\qquad 4 \qquad&\qquad 6\qquad&\qquad8 \qquad&\qquad10 \qquad& \ \ \geq 12  \ \ \\
\hline
\phantom{\Big|} \b_{\D=12+l,l} \phantom{\Big|}& \frac{3}{89600} & \frac{243}{20384000} & \frac{3}{833000} & \frac{5}{4874716} & \frac{243}{857376520} & \frac{1}{13006500} & \dots \\
\phantom{\Big|}\b_{\D=16+l,l} \phantom{\Big|}& \frac{1}{8830976} & \frac{1}{24630144} & \frac{625}{50871389184} & \frac{25}{7147812672} & \dots & \text{}& \\
\phantom{\Big|}\b_{\D=20+l,l} \phantom{\Big|}& \frac{1}{2530344960} & \frac{25}{175352905728} & \dots & \text{} & \text{} & \text{} & \\
\hline
\end{array}
\nonumber
\ .
 \ee

\subsubsection*{Defect channel}

As summarized in table \eqref{TAB_2pt_Spin_Example_def}, this correlator can exchange defect operators in representations labeled by $(\hD,\hl=0,s)$, $(\hD,\hl=0,(s,1))$ or $(\hD,\hl=1,s)$. For a line defect, $\hl$ is in fact only defined mod2, and measures parity under the reflection of the coordinate parallel to the defect. Again, the $U(1)$ charge imposes selection rules. Let us promote $\lambda$ and $\bar{\lambda}$ to background fields with the appropriate charge, so that $U(1)$ is conserved. $O_\mu$ couples to defect operators with unit charge. Since $O_\mu\sim  \varphi\bar{\varphi}^2$, Wick theorem leaves four possibilities: $\lambda\bar{\lambda}^2$, $\bar{\lambda}^2\varphi$, $\lambda \bar{\lambda} \bar{\varphi}$, $\bar{\lambda}  \varphi\bar{\varphi}$, which can be decorated by derivatives. We already disregarded operators that appear in the defect OPE of $O_\mu$, but cannot be exchanged in the two-point function \eqref{2ptNOnCons}, \emph{e.g.} operators of the form $\sim \varphi\bar{\varphi}^2$. The operator $\lambda\bar{\lambda}^2$ is the identity, which is not present in the defect OPE of a vector. The operators with one power of $\varphi$ only contribute to the $\hl=0$ spectrum, since $\partial_\parallel\varphi$ is a descendant. Furthermore, all the operators of the kind $(\partial_i\partial^i)^n\varphi$, where $i,j$ are indices orthogonal to the defect, are descendants up to the equations of motion. Finally, in order to anti-symmetrize the derivatives in transverse directions we need to apply them to $\varphi\bar{\varphi}$, \emph{e.g.} $\partial_{[i}\varphi\partial_{j]}\bar{\varphi}$. All in all, we find the exchanged spectrum, and the powers of the couplings in the CFT data:
\beq
\begin{array}{lll}
(\hD,\hl=0,s)\,:\qquad 
                       &\hD-s = 1,\ \ &b_\hO \bar{b}_\hO \propto \bar{\lambda}^3\l \\
                       &\hD-s\in 2\mathbb{N}+2,\ \ &b_\hO \bar{b}_\hO \propto \bar{\lambda}^2  \\
 (\hD,\hl=1,s)\,:\qquad &\hD-s\in 2\mathbb{N}+3,\ \ &b_\hO \bar{b}_\hO \propto \bar{\lambda}^2  \\
 (\hD,\hl=0,(s,1))\,:\qquad &\hD-(s+1)\in 2\mathbb{N}+2,\ \ &b_\hO \bar{b}_\hO \propto \bar{\lambda}^2~,
 \label{ex_defectops}
\end{array} 
\eeq
with $s=0,1,\dots$ except for the last line where $s=0$ is excluded. Let us check that this is indeed what happens.

In accordance with the discussion above, three families of defect partial waves contribute to \eqref{2ptNOnCons}, for a total of six partial waves, as presented in the Table \eqref{TAB_2pt_Spin_Example_def}. Their explicit form can be found in Appendix \ref{FinalRes}. There are two defect OPE structures associated to the exchange of a spin $s$ primary with $\hl=0$ and correspondingly four defect partial waves, $\hG_{\hD 0 s}^{(\pf,\qf)}$ with $\pf,\qf=1,2$. There is a unique defect OPE structure associated to the exchange of a mixed symmetric representation $(s,1)$ and therefore a unique partial wave associated to it: the seed block $\hG_{\hD 0 (s,1)}$. Similarly, when $\hl=1$ there is a unique defect OPE structure and its associated partial wave is the seed $\hG_{\hD 1 s}$.

The conformal block decomposition precisely obeys table \eqref{ex_defectops}:
\begin{align}
\begin{split}
\braket{O(P_1,Z_1)O(P_2,Z_2)}=& \sum_{\pf,\qf=1}^2 \sum_{s=0}^{\infty} \a_{s}^{(\pf,\qf)}\hG_{\hD=s+1,\hl=0,s}^{(\pf,\qf)}+
 \sum_{\pf,\qf=1}^2\sum_{s=0}^{\infty}\sum_{ \hD= 2\mathbb{N}+2+s} \b_{\hD,s}^{(\pf,\qf)}\hG_{\hD,\hl=0,s}^{(\pf,\qf)} \\
&+\sum_{s=0}^{\infty}\sum_{\hD= 2\mathbb{N}+3+s} \g_{\hD,s}\hG_{\hD,\hl=1,s}+ \sum_{s=1}^{\infty}\sum_{\hD= 2\mathbb{N}+3+s} \d_{\hD,s}\hG_{\hD,\hl=0,(s,1)} ~.
\end{split}
\end{align}
 The equality of the external operators implies at the level of defect OPE structures that $\a_{s}^{(1,2)}=\a_{s}^{(2,1)}$ and similarly  $\b_{\hD,s}^{(1,2)}=\b_{\hD,s}^{(2,1)}$.
In the \emph{transverse twist} $\hD-s=1$ sector, the coefficients take the following form 
\begin{align}
\a_{s}^{(1,1)} &=- \bar{\lambda}^3 \lambda\, 2^{s-1}\, (s+1)^2,
\quad % \notag\\
 \a_{s}^{(1,2)}&=\bar{\lambda}^3\lambda \,2^{s-1} s (s+1), 
 \quad
 %\notag\\
\a_{s}^{(2,2)} &=- \bar{\lambda}^3 \lambda \,2^{s-1} s^2.
\end{align}
In the even transverse twist sector ($\hD-s=2\mathbb{N}+2$), we find
\be
\begin{array}{ll}
\b_{\hD,s}^{(1,1)}&=\frac{ \bar{\lambda}^2}{ 2^{4-s}}\hD \left( (\hD-1) \hD-3 s (s+1)\right)  \frac{(s+1)_{\hD-s -2}}{\left(s+\frac{3}{2}\right)_{\hD-s -2}}
,\\
\b_{\hD,s}^{(1,2)}&=\frac{ \bar{\lambda}^2}{ 2^{3-s}} \, 
\frac{(s)_{\hD-s +1}}{\left(s+\frac{3}{2}\right)_{\hD -s-2}}
,\\
\b_{\hD,s}^{(2,2)}&=\frac{ \bar{\lambda}^2}{ 2^{4-s}}\,  s \left(s(s+1)-3 (\hD-1) \hD\right) \frac{(s+2)_{\hD-s -2}}{\left(s+\frac{3}{2}\right)_{\hD-s -2}} .
\end{array}
\ee
Finally, as expected the seed blocks only contribute to the transverse twist $\hD-s \in 2\mathbb{N}+3$ sector, as follows:
\begin{align}
\g_{\hD,s}= \frac{3\bar{\lambda}^2}{2^{4-s}}\, \hD(-\hD+s+1) (\hD+s) 
\frac{(s+1)_{\hD-s -2}}{\left(s+\frac{3}{2}\right)_{\hD-s -2}}\ ,
\qquad 
\d_{\hD,s}=\frac{\hD -1}{\hD } \g_{\hD,s}.
\end{align}

%%%%%%%%%%%%%%%%%%%%%%%%%%%%%%%%%%%%%%%%%%%%%%%%%%%%%%%%%%%%%%%%%%%%
\subsection{Decomposition of $\langle JJ \rangle$}

Let us turn to the 2-point function of the current \eqref{U1current}:
\begin{multline}\label{conservedJJ}
\braket{J(P_1,Z_1)J(P_2,Z_2)}= \frac{1}{(P_1\wbullet P_1)^{3/2} (P_2\wbullet P_2)^{3/2}} \ \frac{1}{64\, \xi ^3}
\\
\times
\Big\{(8  c \lambda  \bar{\lambda} \xi + c+8 \lambda  \bar{\lambda} \xi ^2)  Q_1
+(8 c \lambda  \bar{\lambda} \xi +c)(Q_2+Q_3) \\
+\frac{c \left(8  c \lambda  \bar{\lambda} \xi + c+8 \lambda  \bar{\lambda} \xi ^2+2\xi \right)}{ ( 2\xi+c )}Q_4 
+(4  \lambda  \bar{\lambda} \xi +1)(Q_5+Q_6)
\Big\},
\end{multline}
where we the structures $Q_k$ are defined in \eqref{JJbulkStructures} and the cross ratios $\xi,\, c\equiv\cos\phi$ in \eqref{xicosphi}. As a first simple check, it is easy to verify that setting to zero the couplings $\l$, $\bar{\l}$, and adjusting the normalization, one obtains the correct central charge $C_J$, see \emph{e.g.} \cite{Osborn:1993cr}.  

\subsubsection*{Bulk channel}
The fusion rule of the $U(1)$ current involves higher twist operators. However, the bulk channel decomposition only includes, besides the identity, primaries with $l=2m$ and $\Delta=2+2m$ ($m\in \mathbb{N}$), \emph{i.e.} twist two. This is enforced by crossing. As we shall point out, it is clear from the defect channel that the coupling to the defect is proportional to $\lambda\bar{\lambda}$, and it is easy to see that only the one-point functions of operators of with twist two are compatible with the requirement.
We write this decomposition in terms of the conserved blocks, $G_{\Delta,l}^{JJ,a}$ presented in appendix \ref{bulk_cons}. We find
\beq
\braket{J(P_1,Z_1)J(P_2,Z_2)}=\a_{\text{Id}}G^{JJ,1}_{0,0}+ \lambda  \overline{\lambda}\,\sum_{l=2\mathbb{N}}
\a_l \left(G^{JJ,1}_{\D=2+l,l}+G^{JJ,2}_{\D=2+l,l}\right),
\eeq
with coefficients
\begin{align}
\a_l = \frac{(l-1) \left(\frac{1}{2}\right)_{\frac{l}{2}-1}^2}{ 2^{13} (\frac{l}{2})! \left(\frac{l-1}{2}\right)_{\frac{l}{2}+1}} \, , 
\qquad 
\a_{\text{Id}}=\frac{1}{3 \times 2^{11} }\, .
\end{align}

%%%%%%%%%%%%%%%%%%%%%%
\subsubsection*{Defect channel}

The spectrum exchanged in the defect channel can be guessed in a way analogous to the previous subsection. Part of the spectrum just coincides with the one of a trivial defect: it consists of the Taylor expansion of $J_\mu$ evaluated on the defect, and its purpose in life is to decompose the bulk channel identity \cite{Lemos:2017vnx}. The coupling to the defect happens through the operators $\lambda\bar{\varphi}$ and $\bar{\lambda}\varphi$, which form a sector with $\hD-s=1$ and OPE coefficients proportional to $\lambda\bar{\lambda}$. 

Conservation reduces the number of independent OPE coefficients and, as a consequence, not all the blocks involved in the decomposition of \eqref{2ptNOnCons} are separately consistent with the conservation in the bulk. As argued in section \ref{Seed_Projectors} the seed blocks $\hG_{\hD,\hl=1,s}$ and $\hG_{\hD,\hl=0,(s,1)}$ are always separately conserved and are therefore generically exchanged in the defect decomposition of \eqref{conservedJJ}. On the other hand, only a precise combination of blocks associated to $\hl=0$ defect primaries of transverse spin $s$ is consistent with conservation. This is denoted as $\hG^{JJ}_{\hD,\hl=0,s}$ and its form can be found in \ref{ExamplesDiffDefBlocks}.
The decomposition in terms of the defect channel conserved blocks presented in \ref{FinalRes} reads
\begin{align}
\braket{J(P_1,Z_1)J(P_2,Z_2)}=&\sum_{s=0}^{\infty} \a_{s}\hG^{JJ}_{\hD=s+1,\hl=0,s}+\sum_{s=0}^{\infty} \sum_{\hD= 2\mathbb{N}+2+s} \b_{\hD,s}\hG^{JJ}_{\hD,\hl=0,s}\\
&+\sum_{s=0}^{\infty} \sum_{\hD= 2\mathbb{N}+3+s} \g_{\hD,s}\hG_{\hD,\hl=1,s}+ \sum_{s=1}^{\infty} \sum_{\hD= 2\mathbb{N}+3+s} \d_{\hD,s}\hG_{\hD,\hl=0,(s,1)},\nonumber\\
\end{align}
As anticipated, the only coefficients that depend on the couplings are those with $\hD-s=1$:
\begin{equation}
\a_{s}= \lambda  \overline{\lambda} \  2^{s-1} (s+1)^2 \, .
\end{equation}
Notice in particular the presence of a primary with $(\hD,s)=(1,0)$: this operator is in fact protected, and required when the defect breaks a global symmetry, see \emph{e.g.} \cite{Bianchi:2018scb,Bianchi:2018zpb}.
The coefficients $\b_{\hD,s}$ are  in correspondence of primaries with $\hD-s=2,4,6,\dots$. We present them in closed form:
\begin{align}
\b_{\hD,s}=&\frac{2^{s-2} (s+1)  }{3 (\hD-1)} \frac{(s)_{\Delta -s+1}}{\left(s+\frac{3}{2}\right)_{\Delta-s -2}} \ .
\end{align}
Finally, $\g_{\hD,s}$ and $\d_{\hD,s}$ receive contribution from the odd transverse twist sector $\hD-s~=~3,5,7,\dots$,
\begin{align}
\g_{\hD,s}=\frac{2^{s-2}  (\hD-s-1) (\hD+s) }{3 (\hD-1) } \frac{(s+1)_{\Delta -s}}{\left(s+\frac{3}{2}\right)_{\Delta-s -2}},
\qquad \ \ 
\d_{\hD,s}=\frac{(\hD-1)}{\hD}\g_{\hD,s} \, .
\end{align}
%%%%%%%%%%%%%%%%%%%%%%%%%%%%%%%%%

\section{Conclusions}

In this work, we studied correlation functions involving local operators which transform in mixed symmetry representations of $SO(d)$. Firstly, we described a constructive procedure to obtain the tensor structures that appear in any correlation function of bosonic operators in a CFT. The procedure associates to each local operator the tensors \eqref{Cbuilding}. In the defect CFT scenario, the additional building blocks are the tensors \eqref{CWbuilding}, which take care of the quantum numbers of defect operators associated to the transverse rotations, and the projectors (\ref{Pdot} - \ref{Tdot}) onto the spaces parallel and transverse to the defect. 

These ingredients allowed us to define explicitly all the tensor structures relevant to the conformal block decomposition of a two-point function of symmetric traceless primaries in the presence of a defect. In the second part of the paper we precisely focused on this decomposition. We explained how to generate the bulk channel conformal blocks in a radial expansion, adapting various techniques present in the literature, with special emphasis on the Zamolodchikov recurrence relation and on the differential operators first defined in \cite{SpinningCC}. For the defect channel, on the other hand, we proposed a complete solution: the seed blocks can be found in closed form performing an analytic continuation of the projectors onto representations of the orthogonal group. The generic block is then obtained by application of a set of differential operators. The structure of the radial expansion and of the Zamolodchikov recurrence relation are also explained in the defect channel, and used to check the closed form of the bocks.

With the toolbox provided here, the kinematics of a two-point function of bosonic operators with a conformal defect is tamed. Hopefully, this will be useful in all the situations in which such a correlator must be computed, whatever the technique employed.

%%%%%%%%%%%%%%%%%%%%%%%%%%%%%%%%
\section*{Acknowledgments}
%%%%%%%%%%%%%%%%%%%%%%%%%%%%%%%%

We thank Tobias Hansen, Slava Rychkov,  Petr Kravchuk, Denis Karateev and Jo\~ao Penedones for numerous useful discussions. We particularly thank Vasco Gonçalves for collaboration in the early stages of the project. We also thank the Simons Collaboration on the Non-perturbative Bootstrap for organizing many stimulating conferences and workshops where this work was carried out.

EL is grateful to Nikolay Bobev and Lorenzo Di Pietro for many illuminating discussions and suggestions. In addition, he thanks the Perimeter Institute for Theoretical Physics, where part of the project was carried out. Research at Perimeter Institute is supported by the Government of Canada through Industry Canada and by the Province of Ontario through the Ministry of Research \& Innovation.
EL is supported by the Belgian Federal Science Policy Office through the Inter-University Attraction Pole P7/37, by the COST Action MP1210,  European Research Council grant no. ERC-2013-CoG 616732 HoloQosmos, as well as the FWO  Odysseus  grants G.001.12 and G.0.E52.14N\@.

MM is supported by the Simons Foundation grant 488649 (Simons collaboration on the non-perturbative bootstrap) and by the National Centre of Competence in Research SwissMAP funded by the Swiss National Science Foundation.

ET is supported by the Simons Foundation grant 488655 (Simons Collaboration on the Nonperturbative Bootstrap).

%%%%%%%%%%%%%%%%%%%%%%%%%%%
\appendix
%%%%%%%%%%%%%%%%%%%%%%%%%%%

%%%%%%%%%%%%%%%%%%%
\section{Polynomials from projectors}
\label{poly_Proj_j2}
%%%%%%%%%%%%%%%%%%%
This appendix is dedicated to an example of the polynomials ${\bf P}^{\pdot d}_{l_1,\dots,l_k}$ which appear in the one-point function of primaries -- equation \eqref{1ptproj} -- and in the bulk channel blocks -- equation \eqref{Wk_Bulk_Spin}. The polynomials may of course be constructed using the definition \eqref{Ptnl} in terms of explicit projectors, but the alternative method presented here directly yields expressions analytic in the label $l_1$.

In order to obtain the polynomial ${\bf P}^{\pdot d}_{j,2}$ in closed form, we define the following ansatz 
$$
{\bf P}^{\pdot d}_{j,2}(X_1, X_2)=(X_1 \cdot X_1)^{j/2} \sum_{i=1}^5 a_i \; F_i\left(\sqrt{\frac{X_1\pdot X_1}{X_1 \cdot X_1}}\right) \mathcal{T}_i~,
$$
where
\be
\nonumber
\begin{array}{l}
\mathcal{T}_1\equiv X_2\pdot X_2 \, ,\quad 
\mathcal{T}_2\equiv X_2\tdot X_2,\, ,\quad 
\mathcal{T}_3\equiv \frac{\left(X_1\pdot X_2\right){}^2}{X_1\cdot X_1} \, ,\quad 
\mathcal{T}_4\equiv \frac{X_1\tdot X_2 X_1\pdot X_2}{X_1\cdot X_1}\, ,\quad 
\mathcal{T}_5\equiv \frac{\left(X_1\tdot X_2\right){}^2}{X_1\cdot X_1} \ ,
\end{array}
\ee
and $F_i$ are functions of the ratio $\frac{X_1\pdot X_1}{X_1 \cdot X_1}$. We introduced coefficients $a_i$ (which could be absorbed in the definition of $F_i$) for convenience.
We then require this ansatz to be compatible with the definition \eqref{Ptnl}, \emph{i.e.} to satisfy the properties of scaling, mixed symmetry and traceleness \eqref{scalingSOn}, \eqref{antisym}, \eqref{tracelessness}.
These properties imply
\be
\begin{array}{l}
a_1\equiv -\frac{1}{2 p (p+1) (p+3)} a_4\, ,\quad 
a_2\equiv\frac{ (p+1)}{2 (q-2) (q-1)} a_4 \, ,\quad 
a_3\equiv-\frac{ (j+p+1) (j+q-3)}{2 p (p+3)} a_4 \, ,\quad 
a_5\equiv  (1-q) a_2\, ,
\\
a_4\equiv-\frac{2 (-1)^{j/2} j p (q-2) }{(j-1) (p+q-2) (j+p+q-3) } \frac{\left(\frac{p+3}{2}\right)_{\frac{j-2}{2}}}{\left(\frac{j+p+q}{2}\right)_{\frac{j-2}{2}}}\ ,
\\
\\
F_1(\eta)\equiv p (p+1) (p+3) f_{j+2,-4,1}(\eta )-\eta ^2 (j+p+1) (j+q-3) f_{j,0,5}(\eta )\ , \\
F_2(\eta)\equiv (q-2) f_{j+2,-4,1}(\eta )-\left(\eta ^2-1\right) f_{j,0,1}(\eta ) \ ,
\\
F_3(\eta)\equiv f_{j,0,5}(\eta ) \ ,
\quad 
F_4(\eta)\equiv f_{j,0,3}(\eta ) \ ,
\quad F_5(\eta)\equiv f_{j,0,1}(\eta ) \ ,
\\
\end{array}
\ee
where we defined
\be
f_{j,m,n}(\eta)\equiv\, _2F_1\left(-\frac{j-2}{2},\frac{d +j+m}{2};\frac{n+p}{2};\eta ^2\right) \ .
\ee

Note that the explicit form of this projector can be also obtained studying the analytic structure in $\Delta$ of the bulk blocks for two external vectors \eqref{GVV_pTOpbar}. Indeed, as we explain in \ref{BulkZamolodchikovSpinning}, the latter exhibit a I$_{2}$-type pole at $n=2$ (which corresponds to $\Delta=\D^\star_A=2$) with residue proportional to ${\bf P}^{\pdot d}_{l,2}$. This is a consistency check of our results. Notice that, since \eqref{GVV_pTOpbar} are computed as differential operators acting on the scalar bulk conformal blocks, this procedure gives the projector ${\bf P}^{\pdot d}_{l,2}$ as a combination of derivative acting on the symmetric and traceless projector ${\bf P}^{\pdot d}_{l}$.

%%%%%%%%%%%%%%%%%%%
\section{Spinning differential operators - bulk channel}
\label{SpinCB_bulk_examples}
%%%%%%%%%%%%%%%%%%%
%%%%%%%%%%%%%%%%%%%

\subsection{Recurrence relation for the differential basis}
\label{recrelBULK}
Following the same logic as \cite{Costa2011}, we  now derive a set of recurrence relations which allow to build generic three-point functions in terms of the action of differential operators on seed three-point functions.

The list of the relevant tensor structures was presented in equations \eqref{three_point_BBB} and \eqref{3ptallstructures}. For convenience, in this appendix we denote them as follows:
\be
\label{BB_BuildingBlocks}
\left[
\begin{array}{ccc}
\D_1 &\D_2&\D_3 \\
n_1&n_2&n_3\\
n_{12}&n_{13}&n_{23}\\
k_1&k_2&k
\end{array}
\right]
\equiv \frac{\prod_{i=1}^3(V_i)^{n_i}  \prod_{i<j}(H_{ij})^{n_{ij}}  (T^{3,2 1}_{3, 1 2})^{k_{1}} (T^{3,2 1}_{3, 2 1})^{k_{2}} (T^{4, 2 2}_{3,1 2})^{k}}{P_{12}^{{\D_{123}}{}}P_{13}^{{\D_{132}}{}}P_{23}^{{\D_{231}}{}}[- 2 (P_1\cdot P_2)( P_2\cdot P_3)( P_1\cdot P_3)]^{\frac{k_1+k_2+k}{2}}} \, ,
\ee
where $\D_{ijk}=\frac{1}{2}(\D_i+\D_j-\D_k)$ and the integers satisfy the condition \eqref{3ptconditions}. 
From  \eqref{3ptconditions} we obtain that the seed three-point functions take the form \eqref{3ptseed},
\be
\label{seed_BB_BuildingBlocks}
\left[
\begin{array}{ccc}
\D_1 &\D_2&\D_3 \\
0&0&l_3^{(1)}-l_3^{(2)}\\
0&0&0\\
l_1-l_3^{(3)}&l_2-l_3^{(3)}&l_3^{(3)}
\end{array}
\right] \ .
\ee
Notice that for a given operator $\Ocal_3$ (which we want to think of as the exchanged operator in the OPE), there are many possible seed three-point functions. 
In fact, from the conditions \eqref{existence3pt} and \eqref{3ptseed}, they are labelled by the choices of $l_1\geq l_3^{(3)}$ and $l_2\geq l_3^{(3)}$ such that $l_1+l_2=l_3^{(2)}+l_3^{(3)}$.
Their total number is therefore $l_3^{(2)}-l_3^{(3)}+1$.
For example, if $\Ocal_3$ has spin $l_3^{(2)}=4,\,l_3^{(3)}=1$ there are four possible seeds which correspond to the following choices of spin for the external operators $(l_1,l_2)=(4,1),(3,2),(2,3),(1,4)$.

There exist two \emph{spin transfer} differential operators $D^{(T)}_{12},D^{(T)}_{21}$, first defined in equation (D.7) of \cite{projectors} and reviewed  in the following subsection, which map seed three-point functions in seed three-point functions as follows, 
\begin{align}
D^{(T)}_{12} \left[
\begin{array}{ccc}
\D_1 &\D_2&\D_3 \\
0&0&n_3\\
0&0&0\\
k_1&k_2&k
\end{array}
\right]={k_2  (\tfrac{d}{2}+k_2-3+k) (\Delta_2-d-k_2+4-k)}{}\left[
\begin{array}{ccc}
\D_1 &\D_2&\D_3 \\
0&0&n_3\\
0&0&0\\
k_1+1&k_2-1&k
\end{array}
\right],
\end{align}
\begin{align}
D^{(T)}_{21}\left[
\begin{array}{ccc}
\D_1 &\D_2&\D_3 \\
0&0&n_3\\
0&0&0\\
k_1&k_2&k
\end{array}
\right]={k_1  (\tfrac{d}{2}+k_1-3+k) (\Delta_1-d-k_1+4-k)}{}\left[
\begin{array}{ccc}
\D_1 &\D_2&\D_3 \\
0&0&n_3\\
0&0&0\\
k_1-1&k_2+1&k
\end{array}
\right].
\end{align}
%%%%%%%%%%%%%%%%
The existence of such operators allows us to obtain all the $l_3^{(2)}-l_3^{(3)}+1$ seed three-point functions associated to a fixed operator $\Ocal_3$, just by knowing one representative.  This can be in turn used to compute all the seed conformal blocks for the exchanged operator just by knowing one of them. A natural choice for the representative is the three-point function $l_1= l_3^{(2)}$, $l_2=l_3^{(3)}$ and generic $l_3^{(1)}$.

In the following, we generalize the computation of  \cite{Costa2011} by writing the action of the differential operators \eqref{bulkdiff} on the basis \eqref{BB_BuildingBlocks},
%%%%%%%%%%%%%%%%
\be
\resizebox{.9 \textwidth}{!}{
$
\begin{array}{ll}
D_{11}\left[
\begin{array}{ccc}
\D_1 &\D_2&\D_3 \\
n_1&n_2&n_3\\
n_{12}&n_{13}&n_{23}\\
k_1&k_2&k
\end{array}
\right]
=&
\frac{n_2}{2}  \left[
\begin{array}{ccc}
\D_1-1 &\D_2&\D_3 \\
n_1&n_2-1&n_3\\
n_{12}+1&n_{13}&n_{23}\\
k_1&k_2&k
\end{array}
\right]
+\frac{k_2+n_3}{2} \left[
\begin{array}{ccc}
\D_1-1 &\D_2&\D_3 \\
n_1&n_2&n_3-1\\
n_{12}&n_{13}+1&n_{23}\\
k_1&k_2&k
\end{array}
\right]
+
\frac{\Delta -\Delta_{12}+l_3^{(2)}+\sum_{i=1}^3 n_i -1}{2}  \left[
\begin{array}{ccc}
\D_1-1 &\D_2&\D_3 \\
n_1+1&n_2&n_3\\
n_{12}&n_{13}&n_{23}\\
k_1&k_2&k
\end{array}
\right]
\\
\\
%%%%%%%%%%%%%%
&
-n_{23} \left[
\begin{array}{ccc}
\D_1-1 &\D_2&\D_3 \\
n_1&n_2&n_3+1\\
n_{12}+1&n_{13}&n_{23}-1\\
k_1&k_2&k
\end{array}
\right]
-n_{23} \left[
\begin{array}{ccc}
\D_1-1 &\D_2&\D_3 \\
n_1&n_2+1&n_3\\
n_{12}&n_{13}+1&n_{23}-1\\
k_1&k_2&k
\end{array}
\right]
-2 n_{23} \left[
\begin{array}{ccc}
\D_1-1 &\D_2&\D_3 \\
n_1+1&n_2+1&n_3+1\\
n_{12}&n_{13}&n_{23}-1\\
k_1&k_2&k
\end{array}
\right]
\\
\\
&
+\frac{k_2}{2}  \left[
\begin{array}{ccc}
\D_1-1 &\D_2&\D_3 \\
n_1&n_2&n_3-1\\
n_{12}&n_{13}&n_{23}+1\\
k_1+1&k_2-1&k
\end{array}
\right]
+k_2 \left[
\begin{array}{ccc}
\D_1-1 &\D_2&\D_3 \\
n_1&n_2+1&n_3\\
n_{12}&n_{13}&n_{23}\\
k_1+1&k_2-1&k
\end{array}
\right]
,
\end{array}
$
}
\ee
%%%%%%%%%%%%%%%%%%%%%%%%%%%%%%%%%%%%%%%%%%%%%%%%%%%%%%%%%%%%%%%%%%%%%%
\be
\resizebox{.9 \textwidth}{!}{
$
\begin{array}{ll}
D_{21}\left[
\begin{array}{ccc}
\D_1 &\D_2&\D_3 \\
n_1&n_2&n_3\\
n_{12}&n_{13}&n_{23}\\
k_1&k_2&k
\end{array}
\right]=&
-\frac{n_1}{2}  \left[
\begin{array}{ccc}
\D_1-1 &\D_2&\D_3 \\
n_1-1&n_2&n_3\\
n_{12}+1&n_{13}&n_{23}\\
k_1&k_2&k
\end{array}
\right]
+\frac{k_1+n_3}{2} \left[
\begin{array}{ccc}
\D_1-1 &\D_2&\D_3 \\
n_1&n_2&n_3-1\\
n_{12}&n_{13}&n_{23}+1\\
k_1&k_2&k
\end{array}
\right]
+\frac{k_1}{2}  \left[
\begin{array}{ccc}
\D_1-1 &\D_2&\D_3 \\
n_1&n_2&n_3-1\\
n_{12}&n_{13}+1&n_{23}\\
k_1-1&k_2+1&k
\end{array}
\right]
\\ \\
%%%%%%%%%%%%%%%%%%%%
&+\frac{-\Delta +\D_{12}+k_1-k_2-k-n_1-n_2+n_3-2 n_{23}+1}{2}  \left[
\begin{array}{ccc}
\D_1-1 &\D_2&\D_3 \\
n_1&n_2+1&n_3\\
n_{12}&n_{13}&n_{23}\\
k_1&k_2&k
\end{array}
\right] \ ,\\
%%%%%%%%%%%%%%%%%
\end{array}
$
}
\ee
where we used $k_1+k_2+k=l_3^{(2)} $ from  \eqref{3ptconditions}. Similarly we can apply the operators $D_{21}$ and $D_{22}$ which give analogous results (their action is the same as the one of $D_{12}$ and $D_{11}$ once we replace $1$ and $2$). Finally, of course, the multiplication by $H_{12}$  increases the label $n_{12}$ of \eqref{BB_BuildingBlocks} by one.
As expected all the differential operators \eqref{bulkdiff} act internally to the basis \eqref{BB_BuildingBlocks}. 

From the previous formulae one can easily see that acting with \eqref{bulkdiff} on the set  of seed three-point functions \eqref{seed_BB_BuildingBlocks}  it is possible to generate all the possible three-point functions. Moreover, by using the operator $D^{(T)}$ we can obtain all the three-point functions just by acting on the seed representative defined by $l_1= l_3^{(2)}$ and $l_2=l_3^{(3)}$,
\be
\label{DifferentialBasisBulk}
\left\{
\begin{array}{ccc}
\D_1 &\D_2&\D_3 \\
n_1&n_2&n_3\\
n_{12}&n_{13}&n_{23}\\
k_1&k_2&k
\end{array}
\right\}
\equiv
H_{12}^{n_{12}}D_{12}^{n_{13}}D_{21}^{n_{23}}D_{11}^{n_{1}}D_{22}^{n_{2}} D^{(T)\, k_2}_{12} \ \Sigma^{n_1+n_{23},n_2+n_{13}}
\left[
\begin{array}{ccc}
\D_1 &\D_2&\D_3 \\
0&0&l_3^{(1)}-l_3^{(2)}\\
0&0&0\\
 l_3^{(2)}-l_3^{(3)}&0&l_3^{(3)}
\end{array}
\right]
\ ,
\ee
where the integers $n_i,n_{ij}$  satisfy \eqref{3ptconditions} and $\Sigma^{a_1,a_2}$  shifts the external dimensions as $\Delta_i \rightarrow \Delta_i+a_i$. The curly bracket basis on left hand side of \eqref{DifferentialBasisBulk}  is the  differential basis defined in \eqref{diffBulk}.  The basis \eqref{BB_BuildingBlocks} and \eqref{DifferentialBasisBulk}   are related by an invertible linear map. In appendix \ref{examples_spinning_CB_bulk} we explicitly compute this change of  basis in few examples.

\subsection{Spin transfer operators}
Here we define the \emph{spin transfer operators}  $D^{(T)}_{ij}$ which act by transferring one unit of spin from the operator $j$ to the operator $i$ . By this we mean that $D^{(T)}_{ij}$ has homogeneity $-1$ in $Z_i$ and $+1$ in $Z_j$. 
The spin transfer operators are defined as  \cite{projectors}
\be
\label{def:SpinTransfer}
D^{(T)}_{ij}= \frac{(P_i\cdot Z_j) P_{j \, M}- (P_i\cdot P_j) Z_{j \, M}}{P_i \cdot P_j}  D^{M}_{Z_i ,P_i} \ ,
\ee
where
\be
 D^{M}_{X,Y}\equiv d_{00}(X,Y) \partial^{M}_{X}+ d_{-1 1}(X,Y) \partial^{M}_{Y}+X^{M} d_{-2 0}(X,Y) +Y^{M} d_{-1 -1}(X,Y)\,,
 \ee
and $d_{m n}(X,Y)$ are differential operators with weight $m$ in the variable $X$ and $n$ in the variable $Y$. They are defined as follows
\begin{align}
 d_{00}(X,Y)& \equiv -\tfrac{d}{2}  \big[ (d-1)+3  (X\cdot \partial_{X})+  (Y\cdot \partial_{Y})\big]-(X\cdot \partial_{X})(Y\cdot \partial_{Y})-X^{M}(X\cdot \partial_{X}) \partial_{X\, M} \,,
 \nonumber\\
 d_{-2 0}(X,Y)& \equiv   \tfrac{1}{2} \big[ d+(Y\cdot \partial_{Y}) +(X\cdot \partial_{X})\big](\partial_{X}\cdot \partial_{X}) \,,
 \nonumber\\
d_{-1 1} (X,Y)&\equiv- d (Y\cdot \partial_{X})-(Y\cdot \partial_{X})(Y\cdot \partial_{Y})-(X\cdot \partial_{X})(Y\cdot \partial_{X})\,,
\\
d_{-1 -1}(X,Y)& \equiv \big[ \tfrac{d}{2}+(X\cdot \partial_{X}) \big] (\partial_{X}\cdot \partial_{Y})+
\tfrac{1}{2}\big[ (Y\cdot \partial_{X})(\partial_{Y}\cdot \partial_{Y})- (X\cdot \partial_{Y}) (\partial_{X}\cdot \partial_{X})\big]\,.
\nonumber
\end{align}
The main property of $ D^{M}_{X,Y}$ is that it acts as a derivative in $Y$ while preserving the conditions $X^2=Y^2=X\cdot Y=0$. These differential operators can in fact be used to write projectors into $SO(n)$ representations labelled by Young tableaux with two rows \cite{projectors}.

%%%%%%%%%%%%%%%%%%%

%%%%%%%%%%%%%%%%%%%%
\subsection{Examples}
\label{examples_spinning_CB_bulk}
%%%%%%%%%%%%%%%%%%
In this section we exemplify how to use the spinning differential operators \eqref{DifferentialBasisBulk}. We will consider simple cases in which we act on the scalar partial wave $G_{\Ocal}$. This can be computed as an expansion in radial coordinates, as explained in  \cite{Lauria:2017wav}.

%%%%%%%%%%%%%%%%%%
\paragraph{Vector-scalar\\}
%%%%%%%%%%%%%%%%%%
We consider the case of the bulk two-point function of one vector operator $\Ocal_1$ and one scalar operator $\Ocal_2$.
In this case there are two independent conformal partial waves $G_{\Ocal}^{(\pf)}$ with $\pf=1,2$ associated to the exchange of the symmetric traceless operator ${\Ocal}$ with dimension $\D$ and spin $l$.
The label $\pf$ is associated to the OPE tensor structures $Q^{(\pf)}$ defined in \eqref{3ptallstructures}, which we choose as follows
\be
 Q^{(1)}= H_{13} V_3^{l-1}\ ,
\qquad
Q^{(2)}= V_1 V_3^l   \ .
\ee
The conformal partial waves are conveniently computed using the differential operators of \eqref{diffBulk}
\begin{align}
G_{\Ocal}^{(\bar 1)}= \Dcal^{(\bar 1)} G_{\Ocal} \equiv D_{11} \Sigma^{1,0} G_{\Ocal}\ , 
\qquad 
G_{\Ocal}^{(\bar 2)}=\Dcal^{(\bar 2)} G_{\Ocal} \equiv D_{12} \Sigma^{0,1} G_{\Ocal}\ ,
\end{align}
where $G_{\Ocal}$ is the scalar partial wave \cite{Lauria:2017wav}.
As we already mentioned, the partial waves $G_{\Ocal}^{(\bar \pf)}$ are in a different basis with respect to $G_{\Ocal}^{(\pf)}$. The two bases are related by an invertible linear map
\be
G_{\Ocal}^{(\pf)}=\sum_{\bar \pf=1}^{2} (a_{\pf  \bar \pf})^{-1} G_{\Ocal}^{(\bar \pf)} \ .
\ee
The matrix $a_{\pf  \bar \pf}$ is obtained performing the following computation
\begin{align}\label{ddddd}
\Dcal^{(\bar \pf)} \frac{V_3^l}{(-2P_1 \cdot P_2)^{-\frac{1}{2} (\Delta -\D_1-\D_2)} (-2P_1 \cdot P_3)^{\frac{1}{2} (\Delta +\Delta_{12})} (-2P_2 \cdot P_3)^{\frac{1}{2} (\Delta -\Delta_{12})}}=\nonumber\\
= \sum_{\pf} a_{\bar \pf \pf } \frac{Q^{(\pf)}}{(-2P_1 \cdot P_2)^{-\frac{1}{2} (\Delta -\D_1-\D_2)} (-2P_1 \cdot P_3)^{\frac{1}{2} (\Delta +\Delta_{12})} (-2P_2 \cdot P_3)^{\frac{1}{2} (\Delta -\Delta_{12})}} \ ,
\end{align}
which gives
\begin{equation}
 (a_{\pf  \bar \pf})^{-1}=\frac{1}{l (\Delta-1)}\left(
 \begin{array}{cc}
  l & l \\
  -l+\Delta +\Delta_{12}-1 &\quad  -l-\Delta +\Delta_{12}+1 \\
 \end{array}
 \right)\ .
\end{equation}

%%%%%%%%%%%%%%%%%%
\paragraph{Vector-vector\\}
%%%%%%%%%%%%%%%%%%
Let us now consider the case of the two-point function of vector operators $\Ocal_i$ with dimensions $\D_i$. The two vectors exchange symmetric traceless primaries and an additional seed block -- see table \ref{TAB_2pt_Spin_Example_bulk}. Here we take care of the first class of operators, which have dimension $\D$ and spin $l$. There are five possible partial waves  $G_{\Ocal}^{(\pf)}$  labelled by the OPE index $\pf=1,\dots,5$. The index $\pf$ is related to the following choice of OPE tensor structures:
 \be
 \label{QpJJO}
Q^{(\pf)} = \{H_{12} V_3^l, V_1V_2 V_3^l, H_{23} V_1 V_3^{l-1}, H_{13} V_2 V_3^{l-1} , H_{13} H_{23} V_3^{l-2}  \}  \ .
 \ee
We can compute the  partial waves by acting with differential operators on the scalar partial wave as shown in  \eqref{diffBulk}
\be
\label{twoEXTvectBULK}
\begin{array}{ll}
G_{\Ocal}^{(\bar 1)}=D_{11}D_{22} \Sigma^{1,1} G_{\Ocal}, \qquad
&G_{\Ocal}^{(\bar 2)}=D_{21}D_{11} \Sigma^{2,0} G_{\Ocal}, \\
G_{\Ocal}^{(\bar 3)}=D_{12}D_{22} \Sigma^{0,2} G_{\Ocal}, 
&G_{\Ocal}^{(\bar 4)}=D_{12}D_{21} \Sigma^{1,1} G_{\Ocal}, \\
G_{\Ocal}^{(\bar 5)}=H_{12} G_{\Ocal}.
\end{array}
\ee
The partial waves $G_{\Ocal}^{(\pf)}$ are obtained after the following change of basis:
\be
\label{GVV_pTOpbar}
G_{\Ocal}^{(\pf)}= \sum_{\bar \pf=1}^5 (a_{\bar \pf \pf})^{-1} G_{\Ocal}^{(\bar \pf)} \ ,
\ee
where the matrix $(a_{\bar \pf \pf})^{-1} $, obtained similarly to the previous case, is 
\begin{equation}\label{matrixVV}
(a_{\bar \pf \pf})^{-1}=-\frac{1}{\Delta(\Delta-1)}\left(
\begin{array}{ccccc}
 0 & 0 & 0 & 0 & 2 (\Delta -1) \Delta  \\
 1 & 1 & 1 & 1 & 2-2 \Delta  \\
 \frac{\Delta }{l}-1 & -\frac{l+\Delta }{l} & \frac{\Delta }{l}-1 & -\frac{l+\Delta }{l} & 2 (\Delta -1) \\
 \frac{\Delta }{l}-1 & \frac{\Delta }{l}-1 & -\frac{l+\Delta }{l} & -\frac{l+\Delta }{l} & 2 (\Delta -1) \\
 \frac{(l-\Delta )^2}{(l-1) l} & \frac{(l-\Delta ) (l+\Delta )}{(l-1) l} & \frac{(l-\Delta ) (l+\Delta )}{(l-1) l} & \frac{(l+\Delta )^2-4 \Delta }{(l-1) l} & \frac{2 (\Delta -1) (\Delta -l)}{l-1} \\
\end{array}
\right)
\end{equation}

%%%%%%%%%%%%%%%%%%
\paragraph{Identical vectors\\}
%%%%%%%%%%%%%%%%%%
When the external vector operators $\Ocal_i$ are equal, we must impose the symmetry under the exchange of $\Ocal_1 \leftrightarrow \Ocal_2$ in the set of structures \eqref{QpJJO}, or equivalently in the differential basis \eqref{twoEXTvectBULK}.
For the exchange of an operator with spin $l$, only four of the five possible conformal blocks survive (for more details see for example \cite{Dymarsky:2017xzb}),
\be
\label{GOO}
G^{OO,(1)}_{\Delta l}\equiv G_{\Delta l}^{(\bar 1)} \, , \qquad 
G^{OO,(2)}_{\Delta l}\equiv G_{\Delta l}^{(\bar 2)}+G_{\Delta l}^{(\bar 3)} \, , \qquad
G^{OO,(3)}_{\Delta l}\equiv  G_{\Delta l}^{(\bar 4)} \,  , \qquad
G^{OO,(4)}_{\Delta l}\equiv  G_{\Delta l}^{(\bar 5)} \, .
\ee

%%%%%%%%%%%%%%%%%%%%%%%%%%%%%
\paragraph{Identical conserved currents\\}\label{bulk_cons}
%%%%%%%%%%%%%%%%%%%%%%%%%%%%
We shall now focus on the case when the external operators $\Ocal_i$ are conserved currents.
It is easy to see that only two combinations of the five tensor structures \eqref{QpJJO} are conserved. As a consequence, there are only two conserved bulk blocks (see for example \cite{Dymarsky:2017xzb}),
\begin{align}\label{conservedBulkCB}
G^{JJ,(1)}_{\Delta l} \equiv \sum_{\pf=1}^5\alpha_{\pf} \;  G_{\Delta l}^{(\pf)} \, ,
 \qquad 
 G^{JJ,(2)}_{\Delta l} \equiv \sum_{\pf=1}^5\beta_{\pf} \; G_{\Delta l}^{(\pf)} \, ,
\end{align}
where $l$ must be even.
Here the $G_{\Delta l}^{(\pf)}$ are associated to the basis \eqref{QpJJO} (of course one can rewrite this relation in the differential basis using \eqref{GVV_pTOpbar}). 
The coefficients are defined as follows,
\be
\begin{array}{ll}
\alpha _1=-2 (l-1) (d-\Delta -1) (2 d-\Delta +l-4)
\qquad
\qquad
&
\beta _1=0 
\\
\alpha _2=-\frac{(\Delta +l)}{d-\Delta -1}\alpha _1 
&
\beta _2=\frac{(2 d-\Delta +l-2)}{d-\Delta -1}\beta _3 
\\
\alpha _3=\alpha _4=0
&
\beta _3=\beta _4=l \alpha _1 
\\
\alpha _5=\frac{ l}{2 d-\Delta +l-4}\alpha _1
&
\beta _5=  (d-\Delta -2)\alpha _5
\\
\end{array}
\ .
\ee
As a last remark, we mention that for $l=0$ only the block $G^{JJ,(1)}_{\Delta l}$ survives:  $G^{JJ,(2)}_{\Delta l}$ identically vanishes.

%%%%%%%%%%%%%%%%%%%

%%%%%%%%%%%%%%%%%%%%%%%%%%%%%%%%%%%%%%%%%%
\section{Conformal blocks in the radial frame}
%%%%%%%%%%%%%%%%%%%%%%%%%%%%%%%%%%%%%%%%%%
\label{RadialCoordForCB}
%%%%%%%%%%%%%%%%%%%%%%%%%%%%%%%%%%%%%%%%%%
\subsection{Two-point function in the radial frame}
%%%%%%%%%%%%%%%%%%%%%%%%%%%%%%%%%%%%%%%%%%
In the main text we explained that a two-point function  of bulk operators can be decomposed in the tensor structures \eqref{BB_tensor_structures}.
In this appendix, we explain how to decompose the same two-point functions in terms of tensor structures directly in the bulk or defect radial frames.
Two sets of tensor structures will naturally appear, depending on the choice of frame. We shall see that two linear maps exist, which relate the embedding space structures \eqref{BB_tensor_structures} to both the bulk and defect radial frames structures.

%%%%%%%%%%%%%%%%%%%%%%%%%%%%%%%%%%%%%%%%%%
\paragraph{Bulk radial frame\\}
\label{tensor structures to radial frame}
%%%%%%%%%%%%%%%%%%%%%%%%%%%%%%%%%%%%%%%%%%
The partial waves in the bulk radial frame,  as defined in \eqref{GToGcal_brf_Spin}, are obtained by evaluating $P_i$ and $Z_i$ as in \eqref{B_Poin_Conf} and \eqref{B_Poin_Z}. 
In the bulk radial frame it is convenient to expand the two-point function in terms of a set of tensor structures $\Qcal_{k}$,
\be \label{GcalToFGeneric}
\Gcal_{\D l}(r, \eta , n\cdot z_i, n  \pdot z_i ,z_i\cdot z_j, z_i  \pdot z_j)=\sum_{k=1}^{k_{max}} F_k(r,\eta) \Qcal_{k}(n\cdot z_i, n  \pdot z_i ,z_i\cdot z_j, z_i  \pdot z_j)  \  .
\ee
The coefficient of each tensor structure is a function $F_k$ of the radial coordinates. The index $k$ runs over a finite range which depends on the choice of the two external operators. 
The tensor structures $\Qcal_{k}$ are  polynomials in the eight variables
\be
\label{BB_Structures_BRF}
n\cdot z_1, \ \
n\cdot z_2, \ \
n\pdot z_1, \ \
n\pdot z_2, \ \
z_1\cdot z_2, \ \
z_1\pdot z_2, \ \
z_1\pdot z_1, \ \
z_2\pdot z_2~.
\ee
The $\Qcal_{k}$ have degree $l_i$ in each polarization $z_i$.
For instance, for one external vector the decomposition \eqref{GcalToFGeneric} takes the form
\be
\label{GacalToF1Vector}
\Gcal_{\D l}=   (n\cdot z_1) F_1(r,\eta)+(n \pdot z_1) F_2(r,\eta)  	\ .
\ee
Notice that the first four structures in \eqref{BB_Structures_BRF} have weight 1 either in $z_1$ or in $z_2$ and the last three have weight $2$ (two of them in the same $z_i$ and two of them in both $z_1$ and $z_2$). The number of building blocks and their weights match the embedding space definitions \eqref{2pt_Structures}.
In fact, there is a map between the building blocks \eqref{2pt_Structures} and \eqref{BB_Structures_BRF}, which is obtained by projecting \eqref{2pt_Structures} onto the bulk radial frame. For instance,
\begin{align}
\Vp{1} \underset{b.r.f.}{\longrightarrow}  &-\frac{\sqrt{2}\left(r^2+1\right)}{\sqrt{\left(r^2+1\right)^4-16 \eta ^4 r^4}} \left[r^2\left(2 \eta ^2-1\right)n \pdot z_1-n \pdot z_1 +2 \eta ^2 r^2 n \circ z_1\right].
\end{align}
 It is important to notice that with the definitions (\ref{2pt_Structures} - \ref{NormalizationHV}), the small $r$ limit of the tensor structures is regular and non-degenerate,
 \beq
\begin{array}{lll}
&\Vs{1}  \underset{b.r.f.}{\longrightarrow}  \sqrt{2}\, n\star z_1 +O(r^2) \ , &\qquad \Hs{12}  \underset{b.r.f.}{\longrightarrow}  z_1\star z_2 +O(r^2) \ ,\\
& \Vs{2}  \underset{b.r.f.}{\longrightarrow}  -\sqrt{2}\, n\star z_2 +O(r^2) \ , &\qquad \Hp{ii}  \underset{b.r.f.}{\longrightarrow}  z_i\pdot z_i +O(r^2) \ .
\end{array}
\eeq
Therefore there exists a linear map between the structures $\Qcal_k$ and   $Q_k$ of \eqref{BB_tensor_structures}, which is invertible also in the leading bulk OPE limit (when $r=0$).

%%%%%%%%%%%%%%%%%%%%%%%%%%%%%%%%%%%%%%%%%%
\paragraph{Defect radial frame\\}
\label{tensor structures to radial frame defect}
%%%%%%%%%%%%%%%%%%%%%%%%%%%%%%%%%%%%%%%%%%
Similarly, one can define the two-point function in the defect radial frame \eqref{GToGcal_drf_Spin} via equations \eqref{D_Poin_Conf} and \eqref{D_Poin_Z}. Therefore, we can expand $\hGcal_{\hO}$ in a basis of tensor structures 
\be
\label{GcalToF}
\hGcal_{\hO}=\sum_{k=1}^{k_{max}}  \hat F_k(\hr,\eta) \ \hat \Qcal_k( n \circ z_i, n' \circ z_i, z_i \circ z_j, z_i \cdot z_j)  \ .
\ee
As in the bulk case, the tensor structures $\hat \Qcal_k$ are generated by $8$ building blocks:
\be
n \circ z_1 \, , \ 
n \circ z_2 \, , \ 
n' \circ z_1 \, , \ 
n' \circ z_2 \, , \ 
z_1 \circ z_1 \, , \ 
z_1 \circ z_2 \, , \ 
z_2 \circ z_2 \, , \ 
z_1 \cdot z_2 \, .  
\label{2pt_Structures_defect}
\ee
For instance, for one external vector we have
\be
\label{GacalToF1VectorDefect}
\hGcal_{\hO}=   (n\circ z_1) \hat F_1(\hr,\eta)+(n' \circ z_1) \hat F_2(r,\eta)  	\ .
\ee
Again, we can map the building blocks \eqref{2pt_Structures_defect} to the ones defined in \eqref{2pt_Structures} by projecting the latter onto the defect radial frame. For instance,
\begin{align}
\Vp{1} \underset{d.r.f.}{\longrightarrow}\frac{\left(1-\hr^2\right)n\circ z_1}{\left(\hr^2+1\right) \sqrt{1-\frac{2 \heta  \hr}{\hr^2+1}}} \, ,
\ \  \qquad 
\Vt{1} \underset{d.r.f.}{\longrightarrow}\frac{ n\circ z_1-\heta^{-1}n'\circ z_1}{\sqrt{1-\frac{2 \heta  \hr}{\hr^2+1}}} \, .
\end{align}

The normalization of the structures \eqref{2pt_Structures} was chosen such that  this map is also invertible  at $\hr=0$,  
\beq
\begin{array}{lll}
&\Vp{1}  \underset{d.r.f.}{\longrightarrow} n\tdot z_1 +O(\hr) \ , \qquad
&\Hp{12}  \underset{d.r.f.}{\longrightarrow}  z_1\pdot z_2 +O(\hr) \ ,\\
&\Vp{2}  \underset{d.r.f.}{\longrightarrow} -n'\tdot z_2 +O(\hr) \ , \qquad
&\Ht{12}  \underset{d.r.f.}{\longrightarrow}  z_1\tdot z_2  -\heta^{-1}\, n\tdot z_2 \,n'\tdot z_1+O(\hr) \ ,\\
&\Vt{1}  \underset{d.r.f.}{\longrightarrow} n\tdot z_1 -\heta^{-1}\, n'\tdot z_1+O(\hr) \ , \qquad
&\Hp{11}  \underset{d.r.f.}{\longrightarrow} z_1\pdot z_1+(n\tdot z_1)^2 +O(\hr) \ , \\
&\Vt{2}   \underset{d.r.f.}{\longrightarrow} n'\tdot z_2 -\heta^{-1}\, n\tdot z_2+O(\hr) \ , \qquad
&\Hp{22}  \underset{d.r.f.}{\longrightarrow} z_2\pdot z_2+(n'\tdot z_2)^2 +O(\hr) \ .
\end{array}
\eeq
In turns, these relations imply the existence of an invertible and non-degenerate linear map between the structures $\hat \Qcal_k$ and the $Q_k$ of \eqref{BB_tensor_structures}.

%%%%%%%%%%%%%
\subsection{Examples - defect channel}
%%%%%%%%%%%%%
\label{App_Examples_Defect_CB_Casimir_Expansion}
Here we present two explicit examples of application of the Casimir recurrence relation, mentioned in subsection \ref{defect_Spinning_CBs}, to the computation of defect channel conformal blocks.

%%%%%%%%%%%%%%%%%%%%%%%%%%%
\paragraph{Vector-scalar \\}
%%%%%%%%%%%%%%%%%%%%%%%%%%%
Consider the two-point function of a vector $\Ocal_1$ and a scalar $\Ocal_2$. The tensor structures are written in \eqref{JsO_radial}.  We therefore get that formula (\ref{Ansatz_Defect_Radial_Frame}) reduces to
\be
\label{Gcal_10_Defect}
\Gcal_{\hD 0 s}=  
(z_1 \tdot n) \Ccal_s(\heta)  \Wcal^{(1)}(\hr) 
+
(z_1 \tdot \nabla_n) \Ccal_s(\heta)  \Wcal^{(2)}(\hr) 
\ ,
\ee
where we dropped the index $\hj$, since it can only take the value $\hj=0$. The function $\Ccal_s$ is written in terms of a Gegenbauer polynomial as follows 
\be
\label{Ccals}
\Ccal_s(\heta )\equiv {\bf P}_ s^{q}(n;n')=\frac{s!}{2^s \left(\frac{q}{2}-1\right)_s}C_s^{\frac{q}{2}-1}(\heta).
\ee
Using the identity
\begin{equation}
\nabla_n^\m \, f(\heta)=(n'^\m -\heta \, n^\m)\partial_\heta f(\heta),
\end{equation}
where $f$ is a generic function, we can rewrite \eqref{Gcal_10_Defect} in the basis of the two structures $ (z_1\circ n)$ and $ (z_1\circ n')$ to match \eqref{GacalToF1VectorDefect},
\begin{align} 
\label{D_Ansatz_BF_1Vec}
\Gcal_{\hD 0 s} &=  \left[(z_1\circ n) \left( \Wcal^{(1)}(\hr) -   \Wcal^{(2)}(\hr)\, \heta \partial_\heta \right)+ (z_1\circ n')    \Wcal^{(2)}(\hr)\partial_\heta \right]\Ccal_s(\heta)   \ .
\end{align}
The basis of $ (z_1\circ n)$ and $ (z_1\circ n')$ can be mapped to the usual $H$ and $V$ basis by a linear transformation. It is therefore trivial to write the Casimir equation as a differential equation for the functions $ \Wcal^{(\pf)}(\hr) $. 
The transverse part of the Casimir is solved by the ansatz \eqref{D_Ansatz_BF_1Vec}.
On the other hand, the parallel part of the Casimir acts on the functions $\Wcal^{(\pf)}(\hr)\equiv \sum_{m=0}^\infty w^{(\pf)}(m) \hr^{\hD+m} $ and implies recurrence relations for the coefficients $w^{(\pf)}$. Notice that using this basis of functions the recurrence relations decouple. For $w^{(1)}$ we get
\ba
\begin{split}
0=&-2  \left(m^2+\hD  (2 m+p-4)-4 m+2 p+4\right) w^{(1)}(m-2) \\
&+(2 \hD +m-4) (m+p-4) w^{(1)}(m-4)+m  (2 \hD +m-p) w^{(1)}(m) \ , 
\end{split}
\ea
while for $w^{(2)}$ we get the same recurrence relation of the scalar conformal block \cite{Lauria:2017wav} and therefore the solution is 
\begin{align}
w^{(2)}(2m)=\frac{\left(\frac{p}{2}\right)_m (\hD)_m}{m! \left (-\frac{p}{2}+\hD+1\right)_m}, \quad w^{(2)}(2m+1)=0.
\end{align}
 The recurrence relation for $w^{(1)}$ in slightly more involved and it can be solved to find
\be
w^{(1)}(2m)=
w^{(1)}(0)
\frac{ (\hD+2 m) \left(\frac{p}{2}\right)_m (\hD+1)_{m-1}}{m! \left(-\frac{p}{2}+\hD+1\right)_m}
\ ,\qquad w^{(1)}(2m+1)=0 
\ ,
\ee
in terms of the initial condition $w^{(1)}(0)$. It is then straightforward  to resum the series in $\hr$ obtaining
\begin{align}
\begin{split}
\Wcal^{(1)}(\hr)&=
w^{(1)}(0) \left[
\frac{2 p \hr^2 \, _2F_1\left(\frac{p}{2}+1,\hD+1;-\frac{p}{2}+\hD+2;\hr^2\right)}{2 \hD-p+2}+\, _2F_1\left(\frac{p}{2},\hD;-\frac{p}{2}+\hD+1;\hr^2\right)
\right]
\ ,
\\
\Wcal^{(2)}(\hr)&=
w^{(2)}(0)
\, _2F_1\left(\frac{p}{2},\hD;\hD-\frac{p}{2}+1;\hr^2\right)\ .
\end{split}
\label{D_block_1vec}
\end{align}
The  coefficients  $w^{(\pf)}(0)$ set the normalization of the conformal blocks. They can be fixed to reproduce the defect OPE limit.

%%%%%%%%%%%%%%%%%%
\paragraph{Vector-vector \\}
%%%%%%%%%%%%%%%%%%
We repeat the previous exercise for two external vectors. In this case $\hj$ can be either $0$ or $1$. When $\hj=0$ we have the structures 
\eqref{JsO_radial} for both the left and the right overlaps (of course for the right overlap we need to replace $z_1 \rightarrow z_2$ and $n \to n'$). When $\hj=1$ there is a unique structure given by $(z_1 \pdot z_2)$.
If the exchanged operator is in a traceless and symmetric representation of $SO(q)$, the decomposition \eqref{GcalToFGeneric} becomes
\ba
\label{Gcal_11_Defect}
\Gcal_{\hD \hl s}
&=&  
\Big[(z_1 \tdot n)(z_2 \tdot n') \Wcal^{(1,1)}_0(\hr) 
+
(z_1 \tdot n)(z_2 \tdot \nabla_n')\Wcal^{(1,2)}_0(\hr) 
\nonumber
\\
&&
+
(z_1 \tdot \nabla_n)(z_2 \tdot n') \Wcal^{(2,1)}_0(\hr) 
+
(z_1 \tdot \nabla_n)(z_2 \tdot \nabla_{n'})\Wcal^{(2,2)}_0(\hr)
\nonumber
\\
&&+(z_1 \pdot z_2)\Wcal_1(\hr)\Big]  \Ccal_s(\heta)
\ ,
\ea
where we dropped the indices $(\pf,\qf)$ of $\Wcal_1(\hr)$ since there is a single tensor structure in this case. We remind that the transverse part of these functions is already solved by the ansatz  \eqref{Gcal_11_Defect}, we only need to find $\Wcal^{(\pf,\qf)}_0(\hr)$ and $\Wcal_1(\hr)$. To do so we make use of the parallel Casimir equation.\\

In this case, according to formula \eqref{BB_tensor_structures}, the tensor structures in the bulk two-point function are six, matching the six radial frame structures:
\begin{align}
(z_1\circ n)  (z_2\circ n'), \ \quad (z_1\circ n)  (z_2\circ n) ,  \quad (z_1\circ z_2), \nonumber\\
(z_1\circ n')  (z_2\circ n), \quad (z_1\circ n')  (z_2\circ n'), \quad (z_1 \bullet  z_2).
\end{align}
The Casimir equation gives a set of six coupled differential equations which simplify using the ansatz \eqref{Gcal_11_Defect}. Firstly, we obtain that one of the equations can be dropped, since it is linearly dependent from the other ones (there are six differential equations for five functions). Secondly, three of the five remaining equations can be decoupled, so that only $\Wcal^{(1,1)}_0$ and $\Wcal_1$ are still coupled. 
Finally, the resulting set of equations can be solved exactly for the five functions.
In particular, when $\hl=0$  there are only four linearly independent conformal blocks labelled by the constants $w^{(\pf,\qf)}_0(0)$ (for $\pf,\qf=1,2$) which parametrize  the leading behaviour of $\Wcal^{(\pf,\qf)}_0$ for $\hr=0$ according to \eqref{Def_W_Defect},
\ba
\begin{array}{lcllcl}
\vspace{0.1 cm}
\Wcal^{(1,1)}_0(\hr)&=& \ w^{(1,1)}_0(0) \ \hD^{-2}  \ \hr \partial_\hr   \hr \partial_\hr   f(\hr) \ ,
&
\qquad
 \Wcal^{(2,1)}_0(\hr)&=& \ w^{(2,1)}_0(0) \ \hD^{-1}  \ \hr \partial_\hr   f(\hr) \ ,
\\
\vspace{0.1 cm}
 \Wcal^{(1,2)}_0(\hr)&=&  \ w^{(1,2)}_0(0) \ \hD^{-1}  \ \hr \partial_\hr   f(\hr) \ ,
 &
\qquad
\Wcal^{(2,2)}_0(\hr)&=& w^{(2,2)}_0(0)    \ f(\hr)~,
\\
\vspace{0.1 cm}
\Wcal_1(\hr)&=& -2 \, \hr \, w^{(1,1)}_0(0)  \ \hD^{-1}   g(\hr)~,
&&&
\end{array}
\ea
where 
\begin{equation}
\begin{split}
f(\hr) &= \  \hr^{\hD } \, _2F_1\left(p/2,\hD ;-p/2+\hD +1;\hr^2\right)~, \\
g(\hr) &=
\ \hr^{\hD} \, _2F_1\left(p/2+1,\hD+1;-p/2+\hD+1;\hr^2\right)~.
\end{split}
\end{equation}
The extra constant $w_1(0)$, associated to the leading behaviour $\Wcal_1=\hr^{\hD } (w_1(0)+O(\hr))$, does not appear since it is forced to vanish by consistency with the Casimir equation.
Similarly when  $\hl=1$ the Casimir equation forces $w^{(\pf,\qf)}_0(0)=0$,  therefore we obtain a single conformal block labeled by the constant $w_1(0)$,
\ba
\begin{split}
\Wcal^{(1,1)}_0(\hr)=& w_1(0) \frac{2 p \ \hr}{ (-1 + p - \hD)} g(\hr) 
\\
\Wcal_1(\hr)=&w_1(0)  \bigg[ \frac{ (p-\hD ) \left((\hD -1) \hr^2-(\hD +1)\right)}
{\hD  \left(\hr^2-1\right) (-\hD +p-1)}
\hr
  f(\hr) \nonumber \\
  & \ \ \ \ \ +
  \frac{p \left(\hr^2+1\right) }{\hD  \left(\hr^2-1\right) (-\hD +p-1)}
  \, _2F_1\left(\frac{p+2}{2},\hD ;-\frac{p}{2}+\hD +1;\hr^2\right)  \bigg] \ ,
  \end{split}
\ea
and all the other functions vanish.

There is another conformal block, for the exchange of a defect primary in the transverse representation $(s,1)$, 
\ba
\label{Gcal_11s1_Defect}
\Gcal_{\hD, \hl=0, (s,1)} &=&  \Wcal_0(\hr) {\bf P}^{q}_{s,1}(n , z_1 ; n' , z_2)
\ .
\ea
It easy to check that the transverse Casimir is automatically satisfied and that the parallel Casimir fixes the form of $\Wcal_0(\hr)$ to be equal to the scalar one of \cite{Lauria:2017wav} (see also eq. \eqref{D_CB_scalar}).

In appendix \ref{FinalRes} we present explicit expressions for the previous conformal blocks in the bulk-to-bulk basis \eqref{DEF_CB_Defect_Spin_Generic}.

%%%%%%%%%%%%%%%%%%%

%%%%%%%%%%%%%%%%%%%

%%%%%%%%%%%%%%%%%%%%%%%%%%%%%
\section{Zamolodchikov recurrence relation - defect channel}
\label{SpinningZam}
%%%%%%%%%%%%%%%%%%%%%%%%%%%%%
In this appendix, we show explicit examples of the Zamolodchikov recurrence relation for the defect conformal blocks explained in section \ref{BulkZamolodchikovSpinning}. 
We only focus on  the case of external operators in a traceless and symmetric representation, but the generalization to more complicated $SO(d)$ representations can be obtained following \cite{recrel}. 

Before we go to the examples, let us stress that most of the results presented in \cite{recrel} and reviewed in \cite{Lauria:2017wav} still apply to the defect case.
In particular it is convenient to define defect primary states as
\be 
|\hD,\hl ,s \,;z \rangle \equiv z_{a_1} \dots z_{a_\hl} \hO_{\hD \, \hl \,s}^{a_1 \dots a_\hl}(0)|0\rangle \ ,
\ee
 where the orthogonal indices are left implicit. We then define the following descendant states
\be
| \hD_A,\hl_A, s \, ; z \rangle = \hat \Dcal_A  |\hD,\hl,s \,;z \rangle \, ,
\label{hDA}
\ee
with $A=T,n$ and types $T=\I, \II, \III$ with $n=1,2 \dots$. The operators $\hat \Dcal_A$ are the same of \cite{recrel} after we replace $d \rightarrow p$, $l\rightarrow \hl$ and the scalar product with the parallel one $\pdot$ (for a flat defect) 
\be
\label{OA_def}
\begin{array}{ll}
\Dcal_{\I,n}|\hD,\hl,s\, ;z \rangle
&\equiv (z\pdot P)^n |\hD,l\, ;z \rangle \ ,
\\
\Dcal_{\II,n}|\hD,\hl,s\, ;z \rangle 
&\equiv
\frac{(D_z \pdot P)^n}{(2-p/2-\hl)_n(-\hl)_n} |\hD,\hl,s\, ;z \rangle \ ,
\\
\Dcal_{\III,n}|\hD,\hl,s\, ;z \rangle 
&
\equiv  \Vcal_0 \pdot \Vcal_1 \pdot\dots \pdot  \Vcal_{n-1}
|\hD,\hl,s\, ;z \rangle \ , 
\end{array}
\ee
where $P^\m$ is the generator of  translations and 
\be  \label{Vcalj}
\Vcal_j\equiv P\pdot P -2 \frac{ (P\pdot z) (P\pdot D_z)}{ (p/2+\hl+j-1) (p/2+\hl-j-2)}  \ .
\ee
The descendant states in \eqref{hDA} become primaries when $\hD=\hD_A^{\star}$ with
\be
\label{Dstar}
\begin{array}{l c l l}
\hD^\star_{\I, n} &\equiv& 1-\hl-n   & \qquad\qquad n=1,2,\dots \ , \\
\hD^\star_{\II, n}&\equiv& \hl+p-1-n  & \qquad\qquad n=1,2,\dots, \hl \ , \\
\hD^\star_{\III, n}&\equiv& \frac{p}{2}-n & \qquad\qquad n=1,2, \dots \ .
\end{array}
\ee
Finally, the inverse norm of the primary descendants have residues $\hat Q_A$ predicted by 
\begin{align}
\label{QA}
\begin{split}
&\hat Q_{\I,n}= -\frac{n }{2^n (n!)^2} \ , \\
&\hat Q_{\II,n} =  -\frac{n (-\hl)_n}{(-2)^n (n!)^2 (p+\hl-n-2)_n} \frac{(p/2+ \hl- n-1)}{ (p/2+\hl-1)} \ , \\
&\hat Q_{\III,n} =-\frac{n }{(-16)^n (n!)^2 (p/2-n-1)_{2 n}} \frac{(p/2+\hl-n-1)}{ (p/2+\hl+n-1)} \ .
\end{split}
\end{align}

The states also transform under transverse rotations, but the latter commute with the conformal transformations on the defect. Hence, the operator $\hat \Dcal_A$ in \eqref{hDA} is diagonal in the transverse spin $s$, and consequently so is the recurrence relation \eqref{D_CB_RecRel_Spinning}.
%%%%%%%%%%%%%%%%%
\subsection{Examples}
\label{subsec:examples_zam_defect}
%%%%%%%%%%%%%%%%%
\paragraph{Vector-scalar\\}
We consider the two-point function of a vector $\Ocal_1$ and a scalar $\Ocal_2$.
As we showed in table \eqref{TAB_2pt_Spin_Example_def},
in this case there are two partial waves  $\hG_{\hO}^{(1)}$ and $\hG_{\hO}^{(2)}$ which are both associated to the exchange of an operator $\hO$ labelled by the conformal dimension $\hD$ and the transverse spin $s$ (the only allowed parallel spin  is $\hl=0$, so we will drop this label).
We associate the label $\pf$ of $\hG_{\hO}^{(\pf)}$ to the tensor structures $Q^{(\pf)}$ defined in \eqref{btodefect2pt} 
\be
\label{JhO0s}
Q^{(1)}=\Vp{1,12} (K^2_{1})^{s} \, ,\qquad  Q^{(2)}=Y^2_{1,1}(K^2_{1})^{s-1} \, .
\ee
We also expand each partial wave $\hG_{\hO}^{(\pf)}$ in conformal blocks $\hg_{\hO}^{(\pf),k}$ as defined in \eqref{DEF_CB_Defect_Spin_Generic}. The label $k$ is associated to the choice of two-point function tensor structures $Q_k$ defined in  \eqref{newStructEmi},
\begin{align}\label{JbulkStructures}
Q_1=\Vp{1}\, , \qquad  Q_2=\Vt{1} \ .
\end{align}
 From table \eqref{TableDefRes} we see that the only type of poles which is allowed is the type \III (this is in fact the only type that does not change the parallel spin of the blocks at the residue). 
 Formula \eqref{D_CB_RecRel_Spinning} therefore reduces to
\begin{align}
\begin{split}
\label{D_CB_RecRel_scalar-vector}
\hat h_{\hD s}^{(\pf),k}(\hr,\heta)&=\hat h_{\infty s}^{(\pf),k}(\hr,\heta)+\sum_{n=1}^{\infty}  \sum_{\pf'=1,2} \hr^{2 n} \,  \frac{(R_{\III, n})_{\pf \pf'}}{\hD-\frac{p}{2}+n } \, 
\hat h_{\frac{p}{2}+n \,  s}^{(\pf'),k} (\hr,\heta) \ ,
\end{split}
\end{align}
where $\pf=1,2$ and $k=1,2$.
To compute the matrix $R_{\III, n}$ we use the prescription $(R_A)_{\pf \pf'}=(M^{(L)}_A)_{\pf \pf '} \hat Q_A M_A^{(R)}$ of formula  \eqref{R=MQM_general_defect}. The coefficient $\hat Q_A$ was defined in the beginning of this appendix, while $ M^{(R)}_{\III,n}$ is the same as the ones of the scalar blocks. To compute $M^{(L)}$ we follow  the recipe \eqref{Mdef}. We write the two bulk-defect tensor structures \eqref{JhO0s} in the Poincar\'e section
\ba
\begin{split}
\label{JhO0s_poincare}
\langle \Ocal_1(y,z_1) \hO_{\hD s}(x,w_2)  \rangle^{(1)} &= -\frac{{\left(\frac{w_2\tdot y}{\sqrt{y\tdot y}}\right)^s \left[2 (y\tdot y) (x \pdot z_1)+(y\tdot z_1) (y\tdot y-x\pdot x)\right]}{}}{(y\tdot y)^{\frac{\hD-\D_1+1}{2}} (y\tdot y+x\pdot x)^{\D_1+1}} \ ,
\\ 
\langle \Ocal_1(y,z_1) \hO_{\hD s}(x,w_2)   \rangle^{(2)} &= \frac{{\left(\frac{w_2\tdot y}{\sqrt{y\tdot y}}\right)^{s-1} \left[(y\tdot y) (w_2\tdot z_1)-(w_2\tdot y) (y\tdot z_1)\right]}}{(y\tdot y)^{\frac{\hD-\D_1}{2}+1} (y\tdot y+x\pdot x)^{\D_1}} \ .
\end{split}
\ea 
Here, without loss of generality, we placed the bulk primary $\Ocal_1$ at the origin in the parallel space, so that $y$ only has transverse components.

We then use the definition \eqref{hDA} of the primary descendants $\hO_{\D_A \hl_A \, s} = \hat \Dcal_A \hO_{\hD_A^\star \, \hl \, s}$ to obtain the following equation which defines the matrix $M_{\III,n}^{(L)}$,
\be
(\partial_x \pdot \partial_x)^n
\,
\langle \Ocal_1(y,z_1) \hO_{\hD_A^\star s}(x,w_2)  \rangle^{(\pf)} \equiv \sum_{\pf'=1}^{2} \left(M_{\III,n}^{(L)}\right)_{\pf \pf'} \langle \Ocal_1(y,z_1) \hO_{\hD_A s}(x,w_2)  \rangle^{(\pf')} \ .
\ee
Here, the differential operator $\hat \Dcal_{\III, n}$ reduces to $(\partial_x \pdot \partial_x)^n$ because $\hl=0$.
The result is
\be
\label{MIII_J}
\left(M_{\III,n}^{(L)}\right)_{\pf \pf'}=
(-4)^n \left(\frac{p}{2}-n\right)_{2 n}
\
\left(
\begin{array}{cc}
\frac{p+2 n}{p-2 n} & 0 \\
 0 & 1
\end{array} 
\right)_{\pf \pf'}
\ .
\ee
As mentioned in subsection \ref{SpinZamDef}, the matrices $R_A$ are expected to be very sparse since we can only couple conformal blocks which have exactly the same orthogonal part.
In the present case the two OPE tensor structures \eqref{JhO0s} cannot couple since  they have a different orthogonal structures, namely $(K^2_{1})^{s}\neq Y^2_{1,1}(K^2_{1})^{s-1}$.
We conclude that the recurrence relations \eqref{D_CB_RecRel_scalar-vector} for  $\pf=1,2$ are decoupled.

From the Casimir equation of the defect channel, it is easy to compute the large $\hD$ limit of the conformal blocks. The boundary conditions are provided by the defect OPE limit $\hr \rightarrow 0$.  The result is 
\begin{align}
\hat h_{\infty s}^{(1),1}(\hr,\heta)=&\left(1-\hat{r}^2\right)^{-\frac{p}{2}-1}\sqrt{(\hr^2+1)(\hr^2-2 \heta  \hr+1)} \ \ \Ccal_{s}(\heta)\, ,\nonumber\\
\hat h_{\infty s}^{(2),2}(\hr,\heta)=& -\left(1-\hat{r}^2\right)^{-\frac{p}{2}} \sqrt{1-\frac{2 \heta  \hr}{\hr^2+1}} \ \ s^{-1}\,\heta \; \Ccal'_{s}(\heta) \, ,
\end{align}
where $\Ccal_{s}$ is defined in \eqref{Ccals} and $\hat h_{\infty s}^{(1),2}(\hr,\heta)=\hat h_{\infty s}^{(2),1}(\hr,\heta)=0$. We  conclude that $\hat h_{\hD s}^{(1),2}=\hat h_{\hD s}^{(2),1}=0$, thus \eqref{D_CB_RecRel_scalar-vector} becomes a simple set of two decoupled recurrence relation for $\hat h_{\hD s}^{(1),1}$ and $\hat h_{\hD s}^{(2),2}$. One can use \eqref{D_CB_RecRel_scalar-vector} to compute the conformal blocks efficiently in a radial expansion. However a closed form solution for such blocks exist, as we show for example in \eqref{ScalVectExpr}.
%%%%%%%%%%%%%%%%%
\paragraph{Vector-vector\\}
%\label{vector-vector-zam}
%%%%%%%%%%%%%%%%%
We now study the bulk two-point function of two vectors $\Ocal_1$ and $\Ocal_2$. As shown in table \eqref{TAB_2pt_Spin_Example_def} this case is more involved, since there are six different partial waves for the exchange of operators $\hO$ in three possible representations labelled by $\hD$, the parallel spin $\hl$ and the transverse spin $s$. For $\hl=0$ there are four partial waves, $\hG_{\hD 0 s}^{(\pf,\qf)}$ with $\pf,\qf=1,2$, labelled by the same choice of OPE tensor structures $Q^{(\pf)}$ as defined in \eqref{JhO0s}. When $\hl=1$  there is a single OPE tensor structure available, $(\Hp{12}) (K^2_{1})^{s}$, which gives rise to the seed block $\hG_{\hD 1 s}$.
Finally we can build another seed block $\hG_{\hD 0 (s,1)}$ when the exchanged operator is in the mixed symmetric representation $(s,1)$ of the transverse spin. This case is  completely decoupled from the previous ones since the exchanged operator lives in a different transverse representation,  therefore we will consider it separately in the end of this section. 

Using \eqref{DEF_CB_Defect_Spin_Generic} we write the six partial waves $\hG_{\hO}^{(\pf,\qf)}$ in terms of conformal blocks $\hg_{\hO}^{(\pf,\qf),k}$.
The index $k$ is associated to the tensor structure $Q_{k}$ defined in formula \eqref{JJbulkStructures}.
 From table \eqref{TableDefRes} we see that in this case, beside infinitely many poles of type \III, there are two new allowed poles: the $A=(\I,1)$ for $\hG_{\hD 0 s}^{(\pf,\qf)}$ and the $A=(\II,1)$ for $\hG_{\hD 1 s}$.
Equation \eqref{D_CB_RecRel_Spinning} can be therefore written as the following set of recurrence relations
\begin{align}
\label{D_CB_RecRel_vec_vec}
\begin{split}
\hat h_{\hD 0 s}^{(\pf,\qf),k}(\hr,\heta) &= \hat h_{\infty 0 s}^{(\pf,\qf),k}(\hr,\heta)+\sum_{n=1}^{\infty}  \sum_{\pf',\qf'=1}^{2} \hr^{2n}  \frac{(R_{\III,n})_{\pf \pf'\qf \qf'}}{\hD-\frac{p}{2}+n} 
\hat h_{\frac{p}{2}+n \, 0 \, s}^{(\pf',\qf'),k} (\hr,\heta)  \\
&\qquad \qquad \ \ \  + \hr  \ \frac{(R_{\I,1})_{\pf \qf}}{\hD} \ \hat h_{1 \,1 \, s}^{k} (\hr,\heta) \ ,
\\
\hat h_{\hD 1 s}^{k}(\hr,\heta) &= \hat h_{\infty 1 s}^{k}(\hr,\heta)+\sum_{n=1}^{\infty} \hr^{2n}  \frac{(R_{\III,n})}{\hD-\frac{p}{2}+n} 
\hat h_{\frac{p}{2}+n \, 1 \, s}^{k} (\hr,\heta)  \\
&\qquad \qquad \ \   + \hr  \sum_{\pf,\qf=1}^{2} \ \frac{(R_{\II,1})_{\pf \qf}}{\hD-p+1} \ \hat h_{p \, 0 \, s}^{(\pf,\qf),k} (\hr,\heta) \ .
\end{split}
\end{align}
The coefficients $(R_{\III,n})_{\pf \pf'\qf \qf'}=(M^{(L)}_{\III,n})_{\pf \pf'} \hat Q_{\III,n} (M^{(L)}_{\III,n})_{\qf \qf'}$ are obtained using $M^{(L)}=M^{(R)}$ equal to the matrix computed in \eqref{MIII_J}. 
The coefficient $(R_{\I,1})_{\pf \qf}=(M_{\I,1})_{\pf} \hat Q_{\I,1} (M_{\I,1})_{\qf}$ is fixed in terms of 
\be
\label{MI_JJ}
(M_{\I,n})_{1}=-2 \ , \qquad\qquad (M_{\I,n})_{2}=0\, .
\ee
In the recurrence relation for $\hat h_{\hD 1 s}^{k}$ the residue of the type \III can be computed as $(R_{\III,n})=(M_{\III,n})^2  \hat Q_{\III,n}$, where 
\be
M_{\III,n}=\frac{(-4)^{n} (p+2 n) \left(-n+\frac{p}{2}-1\right)_{2 n}}{p-2 n} \ .
\ee
Finally the matrix $(R_{\II,1})_{\pf \qf}=(M_{\II,1})_{\pf}  \hat Q_{\II,1} (M_{\II,1})_{\qf}$ is obtained from
\be
\label{MII_JJ}
(M_{\II,1})_{1}=-2p \ , \qquad (M_{\II,1})_{2}=0 \ .
\ee
Since the second component in \eqref{MI_JJ} and \eqref{MII_JJ} vanishes, we find that $\hat h_{\hD 0 s}^{(1,1),k}$ is the only conformal block that couples to $\hat h_{\hD 1 s}^{k}$. This fact has a simple explanation: $\hG_{\hD 0 s}^{(1,1)}$ and $\hG_{\hD 1 s}$ are the only partial waves with the same orthogonal part in the OPE tensor structure, namely $(K^2_{1})^{s}$. 
Moreover, since \eqref{MIII_J} is diagonal we obtain that all the $\hat h_{\hD 0 s}^{(\pf,\qf),k}$ in \eqref{D_CB_RecRel_vec_vec}  are decoupled from each other. 
 The pole structure of formula \eqref{D_CB_RecRel_vec_vec} is now explained.

The last missing ingredient is the large delta behaviour of the conformal blocks. Solving the Casimir equation at large $\hD$ with initial conditions fixed by the leading OPE we find
\begin{align}
\hat{h}^{(1,1),1}_{\infty 0 s}(\hr,\heta)=& \left(\hr^2+1\right) \left(1-\hr^2\right)^{-\frac{p}{2}-2} \left(\hr^2-2 \heta  \hr+1\right) \Ccal_{s}(\heta),\nonumber\\
{}&{}\nonumber\\
\hat{h}^{(1,2),2}_{\infty 0 s}(\hr,\heta)=&  \hat{h}^{(2,1),3}_{\infty 0 s}(\hr,\heta)=-s^{-1}{\heta  \left(1-\hr^2\right)^{-\frac{p}{2}-1} \left(\hr^2-2 \heta  \hr+1\right) \Ccal'_{s}(\heta)},\nonumber\\
{}&{}\nonumber\\
\hat{h}^{(2,2),4}_{\infty 0 s}(\hr,\heta)=& \frac{\heta  \left(1-\hr^2\right)^{-\frac{p}{2}} \left(\hr^2-2 \heta  \hr+1\right) }{\left(\hr^2+1\right) s^2}\left(\Ccal'_{s}(\heta)+\heta \, \Ccal''_{s}(\heta)\right),\nonumber\\
\hat{h}^{(2,2),6}_{\infty 0 s}(\hr,\heta)=& s^{-2}{ \left(1-\hr^2\right)^{-\frac{p}{2}} \Ccal'_{s}(\heta)}{} ,\nonumber\\
{}&{}\nonumber\\
\hat{h}_{\infty 1 s}^{1}(\hr,\heta)=& 2 \hr \left(1-\hr^2\right)^{-\frac{p}{2}-2} \left(\hr^2-2 \heta  \hr+1\right) \Ccal_s(\heta),\nonumber\\
\hat{h}_{\infty 1 s}^{5}(\hr,\heta)=& {\left(1-\hr^2\right)^{-\frac{p}{2}} \Ccal_s(\heta)}{ },
\end{align}
where all the other functions are zero. We thus find that for most $k$ the functions \eqref{D_CB_RecRel_vec_vec}  are strictly zero. 

We can now move to the partial wave $\hG_{\hD 0 (s,1)}$. This is actually a trivial case, since the associated function $\hat h^k_{\hD 0 (s,1)}$ has the same recurrence relation as the scalar defect block. Indeed, all the dependence on the external spin is absorbed in the transverse piece of the partial wave. The only difference is the large $\hD$ behaviour, which has to be replaced by
\ba
\hat h^4_{\infty 0 (s,1)}(\hr,\heta)&=& \frac{2^{-s} s!\,  (2-q) q }{(s+1) (q+s-3) \left(\frac{q}{2}-1\right)_s}\frac{\heta\, \left(1-\hr^2\right)^{-\frac{p}{2}} \left(\hr^2-2 \heta  \hr+1\right) }{\left(\hr^2+1\right)}C_{s-2}^{\left(\frac{q}{2}+1\right)}(\heta ),\nonumber\\
\hat h^6_{\infty 0 (s,1)}(\hr,\heta)&=&{\frac{(q-2) 2^{-s} s! \left(1-\hr^2\right)^{-\frac{p}{2}}}{(s+1) (q+s-3) \left(\frac{q}{2}-1\right)_s}}\left[\left(\heta ^2-1\right) q C_{s-2}^{\left(\frac{q}{2}+1\right)}(\heta )+\heta  (q-2) C_{s-1}^{\left(\frac{q}{2}\right)}(\heta )\right].\nonumber\\
\ea

\paragraph{Seed blocks\\}
In this paragraph, we show that all the seed blocks in even dimensions have a finite number of poles in $\hD$. To see this, it is sufficient to check that only a finite number of poles of type \III contribute to the Zamolodchikov expansion \eqref{D_CB_RecRel_Spinning}, since type \I and \II poles are always finitely many (see the comment to table \eqref{TableDefRes}). 

We focus on the case of external operators $\Ocal_i$ in a traceless and symmetric representation of spin $l_i$. A seed partial wave appears when $l_1=l_2=\hl+s_2$ where $\hl$ and $s=(s_1,s_2)$ are respectively the parallel and transverse spin of the exchanged operator $\hO$. In this case the partial wave has a unique OPE tensor structure proportional to $(\Hp{12})^\hl$ (times the transverse part), which in the Poincar\'e section gives
\begin{equation}
\label{Qseed}
\langle \Ocal_1(y,z_1) \hO_{\hD \hl s}(x,z)  \rangle = b_{1 \hO} \frac{(z_1 \pdot z -  \frac{2(z\pdot x)(z_1\pdot x+y\tdot z_1)}{(y\tdot y+x\pdot x)})^\hl}{(y\tdot y)^{\frac{\D_1-\hD}{2}} (y\tdot y+x\pdot x)^{\hD}} \times (\text{transverse}).
\end{equation}
In equation \eqref{Qseed}, the position vector $y$ only has transverse components.
We omitted the transverse tensor structures since they are unimportant for this argument. One can easily check that 
\be
\hat \Dcal_{\III, n} \langle \Ocal_1(y,z_1) \hO_{\frac{p}{2}-n \, \hl \,  s}(x,z)  \rangle = M_{\III,n}^{\text{seed}} \langle \Ocal_1(y,z_1) \hO_{\frac{p}{2}+n \, \hl \, s}(x,z)  \rangle
\ee
for any  $\hl\geq 0$ and $n \geq 1$. The coefficient  $M_{\III,n}^{\text{seed}} $ reads
\begin{equation}\label{type3onmax}
M_{\III,n}^{\text{seed}}=\frac{(-4)^{n}  (2 \hl+2n+p-2) \left(-n+\frac{p}{2}-1\right)_{2 n}}{2 \hl-2 n+p-2} \ .
\end{equation}
When the dimension $p$ of the defect is an even number, the right hand side of \eqref{type3onmax} vanishes for all $n\geq p/2-1$ (beside the case $\hl=0$, where the zero is at $n\geq p/2$). Therefore, for even $p$,  any defect seed block has a finite number of poles in $\hD$ in the recurrence relation \eqref{D_CB_RecRel_Spinning}. This is a clear indication of the simplicity of the defect conformal blocks in even dimensions, which we expect  to reduce to rational functions  of $\hr$ multiplied by $\hr^\hD$ as explained in section \ref{Seed_Projectors}.

%%%%%%%%%%%%%%%%%%%

%%%%%%%%%%%%%%%%%%%%%%%%%%%%
\section{Spinning differential operators -  defect channel }
\label{App_Defect_Spinning_Diff_Ops}
%%%%%%%%%%%%%%%%%%%%%%%%%%%%

%%%%%%%%%%%%%%%%%%%%%%%
\subsection{Recurrence relation}
\label{app:recrelDiff}
%%%%%%%%%%%%%%%%%%%%%%%
In this appendix, we check that the differential operators \eqref{hD_explicit} are correct and can be used to generate all the bulk-defect tensor structures of $\langle \Ocal \hO \rangle$ (and consequently all the conformal blocks).
We consider defect operators $\hO$ with  traceless and symmetric parallel spin $\hl$ and transverse spin $s=(s_1,s_2)$.  The bulk operator $\Ocal$ has conformal dimension $\D$ and traceless and symmetric  spin $l$.

The list of the relevant tensor structures was presented in equations \eqref{btodefect2pt} and \eqref{BuildBlocksBulkToDef}. In this appendix we will denote them as follows 
\be
\label{BD_BuildingBlocks}
[\D,\hD,l,\hl,m_{11},n_{1},m_{12},n_2,s_2]
\equiv 
\frac{ \left(\Hp{12}\right)^{\hl} \  (\Hp{11})^{m_{11}} \ (S^2_1)^{s_2}  \ (\Vp{1,12})^{n_{1}}\ (Y_{1,1}^2)^{m_{12}} \  (K^2_{1})^{n_2}
}{(-2 P_1 \gbullet P_2)^{\hD} (P_1 \wbullet P_1)^{\frac{\D-\hD}{2}}} 
\ .
\ee
We remind that the integers $m_{ij}$ and $n_i$ need to satisfy the condition \eqref{condition_BD_integers} which depends on $l$, $s_1$ and $s_2$ .
The form of the operators $\hat{D}^\star_{i}$ can be fixed as in \eqref{diffOP_def1} by requiring that their action closes on the span of \eqref{BD_BuildingBlocks}.
In other words, we require their action on any of the \eqref{BD_BuildingBlocks} to be expressed as a linear combination of the \eqref{BD_BuildingBlocks}.
Using \eqref{diffOP_def1}, we explicitly find
\be
\label{recurrenceD}
\begin{array}{lll}
\hat{D}^\pdot_{1} [\D+1,\hD,l,\hl,m_{11},n_{1},m_{12},n_2,s_2] 
&=& 
-(\hl+n_1+\hD)  [\D,\hD,l+1,\hl,m_{11},n_{1}+1,m_{12},n_2,s_2]  \\
&&-n_1  [\D,\hD,l,\hl,m_{11}+1,n_{1}-1,m_{12},n_2,s_2] 
\ ,
\\
\\
\hat{D}^\tdot_{1} [\D+1,\hD,l,\hl,m_{11},n_{1},m_{12},n_2,s_2] 
&=& 
-n_2 [\D,\hD,l+1,\hl,m_{11},n_{1},m_{12}+1,n_2-1,s_2]  \\
&&+ m_{12}  [\D,\hD,l+1,\hl,m_{11}+1,n_{1},m_{12}-1,n_2+1,s_2] 
\ ,
\\
\\
\Hp{11} \ \ \ [\D,\hD,l,\hl,m_{11},n_{1},m_{12},n_2,s_2] 
&=&
 [\D,\hD,l+2,\hl,m_{11}+1,n_{1},m_{12},n_2,s_2]
 \ .
 \end{array}
\ee

From these equations it is clear that one can use the spinning operators to generate recursively all the building blocks \eqref{BD_BuildingBlocks}, if one knows all the bulk-defect seeds
\be
\label{Seed_Building_Block}
[\D,\hD,\hl,\hl,0,0,0,s_1,s_2] \, .
\ee
Notice that in the seeds \eqref{Seed_Building_Block} the bulk and the defect operators have the same spin $\hl=l$, as required by \eqref{condition_BD_integers}.
 For example, the scalar conformal block \eqref{D_CB_scalar} is associated to the bulk-defect seed $[\D,\hD,0,0,0,0,0,s_1,0]$. 
 
We can finally write the bulk-defect building blocks in the differential basis as follows
\be
\label{diffopbasis}
 \{ \D,\hD,l,\hl,m_{11},n_{1},m_{12},n_2,s_2 \} \equiv 
 (\Hp{11})^{m_{11}}(\hat{D}^\pdot_{1})^{n_{1}} (\hat{D}^\tdot_{1})^{m_{12}} \Sigma_1^{n_1+m_{12}} [\D,\hD,\hl,\hl,0,0,0,s_1,s_2] 
 \ ,
\ee
where $\Sigma^k$ implements the shift $\D \rightarrow \D+k$. Again the integers $m_{ij}$ and $n_i$ satisfy the condition \eqref{condition_BD_integers}. 
The basis \eqref{diffopbasis} and the basis \eqref{BD_BuildingBlocks} are related by a linear change of basis which can be obtained from \eqref{recurrenceD}.

%%%%%%%%%%%%%%%%%%%%%%%%%%%%%%%%%
\subsection{Examples}\label{ExamplesDiffDefBlocks}
%%%%%%%%%%%%%%%%%%%%%%%%%%%%%%%%%
In this subsection we exemplify how to compute  defect channel  blocks  by acting with the spinning differential operators \eqref{hD_explicit} on seed blocks. We will focus on  simple examples for which there is a unique seed: the scalar one. In the following we denote its partial wave as
\be
\label{Ghoscalar}
 \hG_{\hO}= \frac{\hg_{\hO}(\hr,\heta)}{(P_1 \tdot P_1 )^{\frac{\D_1}{2}}(P_2 \tdot P_2 )^{\frac{\D_2}{2}}} \ ,
\ee
where the exchanged operator $\hO$, has conformal dimension $\hD$, parallel spin $\hl=0$ and transverse spin $s$. The associated conformal block is \cite{Lauria:2017wav,Billo:2016cpy}
\begin{align}
\label{D_CB_scalar}
\hg_{\hO}(\hr,\heta)=\hr^\hD {}_2 F_1\left(\frac{p}{2},\hD,\hD-\frac{p}{2}+1; \hr^2\right)\Ccal_s(\heta )\, , 
\end{align}
where $\Ccal_s$ is defined in \eqref{Ccals}.

%%%%%%%%%%%%%%%%%%%%%%%%%%%%%%%%%
\paragraph{Vector-scalar\\}
%%%%%%%%%%%%%%%%%%%%%%%%%%%%%%%%%
In the case of one external vector (say the first operator), we have two independent conformal partial waves $\hG_{\hOcal}^{(\pf)}$, $\pf=1,2$ associated to the exchange of $\hl=0$ defect primaries. These can be obtained from the scalar conformal partial wave \eqref{Ghoscalar} as follows
\begin{align}
\hG_{\hOcal}^{(1)}=-\hat{\Delta}^{-1}\hat{D}^\gbullet_{1} \Sigma^{1,0} \hG_{\hOcal} \, ,
\qquad 
\hG_{\hOcal}^{(2)}=s^{-1}\hat{D}^\wbullet_{1} \Sigma^{1,0}  \hG_{\hOcal} \, ,
\end{align}
where  the normalization is fixed to recover blocks in the OPE basis \eqref{BD_BuildingBlocks}.
Each of the $\hG_{\hOcal}^{(\pf)}$ can be decomposed as in \eqref{DEF_CB_Defect_Spin_Generic}
with $Q_s$ given by \eqref{JbulkStructures}. We find
\begin{align}
\begin{split}
\label{ScalVectExpr}
\hg^{(1),1}_{\hOcal}(\hat{r},\hat{\eta})=& \hD^{-1}\frac{ \hat{r}}{1-\hat{r} ^2}\sqrt{\left(\hat{r} ^2+1\right) \left(-2 \hat{\eta}  \hat{r} +\hat{r} ^2+1\right)} \partial_{\hat{r}}\hg_{\hOcal}(\hat{r} ,\hat{\eta} )~, \\
\hg^{(2),2}_{\hOcal}(\hat{r},\hat{\eta})=& -s^{-1} \hat{\eta}  \sqrt{1-\frac{2 \hat{\eta}  \hat{r} }{\hat{r} ^2+1}} \partial_{\hat{\eta}} \hg_{\hOcal}(\hat{r} ,\hat{\eta} )~, \\
\end{split}
\end{align}
where $\hg_{\hOcal}$ is defined in \eqref{D_CB_scalar}.

%%%%%%%%%%%%%%%%%%%
\paragraph{Vector-vector\\}
%%%%%%%%%%%%%%%%%%%
There are six independent bulk-to-defect conformal partial waves associated to the exchange of a defect primary $\hOcal$. The four of them $\hG_{\hOcal}^{(\pf,\qf)}$ with $\hl=0$ (and arbitrary transverse spin $s$) can be obtained by applying \eqref{hD_explicit}  to the scalar conformal partial wave $\hG_{\hOcal}$ as follows 
\be
\label{VcalVectExpr}
\begin{array}{rlrl}
\hG_{\hOcal}^{(1,1)}&=\hD^{-2}\hat{D}^\gbullet_{1}\hat{D}^\gbullet_{2} \Sigma^{1,1}\hG_{\hOcal}~, 
& \qquad
\hG_{\hOcal}^{(1,2)}&=({\hD s})^{-1}\hat{D}^\gbullet_{1}\hat{D}^\wbullet_{2} \Sigma^{1,1}\hG_{\hOcal}~, \\
\hG_{\hOcal}^{(2,1)}&=({\hD s})^{-1} \hat{D}^\wbullet_{1}\hat{D}^\gbullet_{2} \Sigma^{1,1}\hG_{\hOcal}~, 
& \qquad
\hG_{\hOcal}^{(2,2)}&=s^{-2}\hat{D}^\wbullet_{1}\hat{D}^\wbullet_{2} \Sigma^{1,1}\hG_{\hOcal}~,
\end{array}
\ee
where we fixed the normalization consistently with \eqref{BD_BuildingBlocks}. 
The remaining two conformal partial waves are seeds and are obtained by different methods (see the appendix \ref{app:recrelDiff} and subsection \ref{Seed_Projectors}). Their explicit expression is reported in appendix \ref{FinalRes}.

%%%%%%%%%%%%%%%%%%%%
\paragraph{Two currents\\}
%\label{conservDefOPE}
%%%%%%%%%%%%%%%%%%%%
As reported in the third line of table \eqref{TAB_2pt_Spin_Example_def}, there is a total of six conformal partial waves  in the defect OPE of a two-point function of vector operators. 
In the following we shall consider the case of the two-point function of conserved currents (which have protected dimension $\Delta=d-1$ and  satisfy the conservation equation $(\partial_P \cdot D_Z) \Ocal_{\Delta,1}(Z,P)=0$).

In order to understand the constraint of conservation it is convenient to classify the possible bulk-defect tensor structures \eqref{btodefect2pt}.
When the defect operator has parallel spin $\hl=0$ and transverse spin $s$ there are two independent tensor structures. 
\begin{align}
\quad \langle \Ocal_{\Delta,1}(Z_1,P_1)\hOcal_{\hD,0,s}(P_2,W_2) \rangle = \frac{
b_{\Ocal \hO}^{(1)}(K^2_{1})^{s-1}(Y_{1,1}^2)+b_{\Ocal \hO}^{(2)}(K^2_{1})^{s}\Vp{1,12}
}{(-2 P_1\gbullet P_2)^{\hD}(P_1\wbullet P_1)^{\frac{\Delta-\hD}{2} }}.
\label{app:scalVec2}
\end{align}
It is easy to see that the expression \eqref{app:scalVec2} only satisfies conservation when the OPE coefficients are related as follows
\begin{align}
\label{conservationJJdef}
(q+s-2)b_{\Ocal \hO}^{(1)}=-{ (\hD-p)}b_{\Ocal \hO}^{(2)}\ , \quad (s>0) \ .
\end{align}
When $s=0$ the coefficient $ b_{\Ocal \hO}^{(1)}$ is absent and conservation implies that the correlation function vanishes, unless $\hD=p$.\footnote{The presence of such a defect operator denotes the breaking of the global symmetry associated to the conserved current by the defect, see \emph{e.g.} \cite{Bianchi:2018scb,Bianchi:2018zpb}.}
The conserved structure corresponding to  \eqref{conservationJJdef} can be generated also in the differential basis \eqref{diffopbasis}
\begin{align}\label{consDiffOp}
\hat{D}^{J}_1=\hat{D}_1^\gbullet+\frac{(p-\hD)}{(q+s-2)}\hat{D}_1^\wbullet.
\end{align}
Therefore, \eqref{consDiffOp} can be used to build the single conserved block  $\hG^{JJ}_{\hD,\hl=0,s}$ as follows  \begin{align}\label{conservedDefBlocks}
 \hG^{JJ}_{\hD,\hl=0,s}&=\hG^{(1,1)}_{\hD,\hl=0,s}+\frac{ (p-\hD)}{(q+s-2)}\hG^{(1,2)}_{\hD,\hl=0,s}+\nonumber\\
 &+\frac{ (p-\hD)}{ (q+s-2)}\hG^{(2,1)}_{\hD,\hl=0,s}+\left(\frac{p-\hD}{q+s-2} \right)^2\hG^{(2,2)}_{\hD,\hl=0,s}.
 \end{align} 
 
As we have shown in table \eqref{TAB_2pt_Spin_Example_def}, we can also build the two seed conformal blocks $\hG_{\hD,\hl=1,s}$ and $\hG_{\hD,\hl=0,(s,1)}$. Since they are seed blocks, they are automatically conserved, as we argued in  subsection \ref{Seed_Projectors}. As a check of this statement one can consider the bulk-defect structure associated to the block  $\hG_{\hD,\hl=1,s}$,
\ba
\label{app:scalVec1}
\quad \langle \Ocal_{\Delta,1}(Z_1,P_1)\hOcal_{\hD,\hl=1,s}(P_2,Z_2,W_2) \rangle = \frac{
b_{\Ocal \hO}
(K^2_{1})^{s-1}\Hp{12}}{(-2 P_1\gbullet P_2)^{\hD}(P_1\wbullet P_1)^{\frac{\Delta-\hD}{2} }} \ .
\ea
It is easy to see that  \eqref{app:scalVec1} is conserved. Ultimately, this is a trivial consequence of the fact that the operator $(\partial_\m \Ocal^\m_{\Delta,1})$ is a scalar primary, which cannot couple to a defect primary with $\hl=1$.

In sum, when the external operators are two conserved currents, there is a total of three conformal blocks: $\hG^{JJ}_{\hD,\hl=0,s}$, $\hG_{\hD,\hl=1,s}$ and $\hG_{\hD,\hl=0,(s,1)}$.

%%%%%%%%%%%%%%%%%%%

%%%%%%%%%%%%%%%%%%%%%%%%%%%%%%%%
\section{The explicit vector-vector blocks in the defect channel}\label{FinalRes}
%%%%%%%%%%%%%%%%%%%%%%%%%%%%%%%%
In this appendix we collect the results for the defect blocks for external vector operators, which are relevant to the examples presented in section \ref{sec:example}. These blocks are computed in a closed form with various techniques. In subsection \ref{Seed_Projectors} it is explained how to obtain all the seed blocks as projectors and how to get the most generic block by acting on a seed  with differential operators (as exemplified in appendix \ref{ExamplesDiffDefBlocks}). In appendix \ref{App_Examples_Defect_CB_Casimir_Expansion} it is shown how to obtain the same blocks by directly solving the Casimir equation taking advantage of a suitable ansatz. Finally, in appendix \ref{subsec:examples_zam_defect} an explicit recurrence relation for the radial expansion of the blocks was derived (in this case however the resummation of the series was not attempted). All the techniques give the same result.

Even if the blocks are already written in three different ways throughout the paper we decided, for the sake of clarity, to report them here in their most transparent form, as function of the cross ratios which multiply the basis of bulk-bulk tensor structures, following the definition \eqref{DEF_CB_Defect_Spin_Generic}. In particular, for two external vectors, a conformal partial wave is fixed in terms of five functions $g_{\Ocal}^{(\pf,\qf),k}(\hr,\heta)$ ($k=1,\dots 5$) which multiply the basis of $Q_k$ defined in \eqref{JJbulkStructures}.
The blocks associated to the exchange of a $\hl=0$ defect primary with spin $s$ can be computed for example using \eqref{VcalVectExpr},
\begin{align}
\hg_{\hD0s}^{(1,1),1}(\hr,\heta)=&\frac{1}{\hD^{2}}\,\frac{ \hr  \left(-2 \heta  \hr +\hr ^2+1\right)}{\left(\hr ^2-1\right)^3}\left[\hr  \left(\hr ^4-1\right) \partial_{\hr}^2\hg_{\hD,s}(\hr,\heta)+\left(\hr ^4-4 \hr ^2-1\right) \partial_{\hr}\hg_{\hD,s}(\hr,\heta)\right],\nonumber\\
\hg_{\hD0s}^{(1,1),5}(\hr,\heta)=&-\frac{2}{\hD^{2}}\,\frac{ \hr ^2 }{\hr ^2-1}\partial_{\hr}\hg_{\hD,s}(\hr,\heta),\nonumber\\
{}&{}\nonumber\\
\hg_{\hD0s}^{(1,2),2}(\hr,\heta)=&\hg_{\hD0s}^{(2,1),3}(\hr,\heta)=-\frac{1}{\hD s}\,\frac{ \heta  \hr  \left(2 \heta  \hr -\hr ^2-1\right) }{\hr ^2-1}\partial_{\heta}\partial_{\hr}\hg_{\hD,s}(\hr,\heta),\nonumber\\
{}&{}\nonumber\\
\hg_{\hD0s}^{(2,2),4}(\hr,\heta)=&-\frac{1}{s^2}\,\frac{ \heta  \left(2 \heta  \hr -\hr ^2-1\right) }{\hr ^2+1}\left[\partial_{\heta}\hg_{\hD,s}(\hr,\heta)+\heta  \partial_{\heta}^2\hg_{\hD,s}(\hr,\heta)\right],\nonumber\\
\hg_{\hD0s}^{(2,2),6}(\hr,\heta)=& s^{-2} \partial_{\heta}\hg_{\hD,s}(\hr,\heta),
\end{align}
where $\hg_{\hD,s}(\hr,\heta)$ is defined in \eqref{D_CB_scalar}.
The seed block associated to the exchange of a $\hl=0$ defect primary with mixed symmetry $(s,1)$ has only $Q_4$ and $Q_6$ components in the basis \eqref{JJbulkStructures}:
\be
\begin{array}{ll}\label{seedsBlocksMixed}
\hg_{\hD0(s,1)}^4(\hr,\heta)=& \frac{2^{-s} s!\,q\,(2-q)}{(s+1) (q+s-3) \left(\frac{q}{2}-1\right)_s} \,\frac{\heta   \left(\hr^2-2 \heta  \hr+1\right) }{\hr^2+1} C_{s-2}^{(\frac{q}{2}+1)}(\heta ) \Fcal_{0,0}(\hr),\\
\hg_{\hD0(s,1)}^6(\hr,\heta)=& \frac{2^{-s} s!(q-2)}{(s+1) (q+s-3) \left(\frac{q}{2}-1\right)_s}\left[\left(\heta ^2-1\right) q C_{s-2}^{(\frac{q}{2}+1)}(\heta )+\heta  (q-2) C_{s-1}^{(\frac{q}{2})}(\heta )\right] \Fcal_{0,0}(\hr) \ ,
\end{array}
\ee
where we introduced the auxiliary function
\begin{equation}
\Fcal_{\alpha,\beta}(\hr)
\equiv \hr^{\hD } \ _2F_1\left(\frac{p}{2}+\alpha,\hD+\beta ;\hD -\frac{p}{2}+1; \hr ^2\right) \, .
\end{equation}
The seed block associated to the exchange of a $\hl=1$ defect primary with symmetric and traceless transverse spin $s$ has only $Q_1$ and $Q_5$ components in the basis \eqref{JJbulkStructures}:
\begin{align}\label{seedsBlocksVector}
\hg_{\hD1s}^1(\hr,\heta)&= \frac{2 p}{(\hD+1-p)} \frac{\left(\hr^2+1\right) \hr \left(\hr^2-2 \heta  \hr+1\right)}{\left(\hr^2-1\right)^2 }\Ccal_s(\heta )\Fcal_{1,1}(\hr)+\nonumber\\
&+\frac{2 \hr\left(\hr^2-2 \heta  \hr+1\right) }{\hD \left(\hr^2-1\right)^3 (-\hD+p-1)}\Ccal_s(\heta )\times\nonumber\\
&\times\left[(p-\hD) \left(-\hD+(\hD-1) \hr^2-1\right) \Fcal_{0,0}(\hr)+p \left(\hr^2+1\right) \Fcal_{1,0}(\hr)\right],\nonumber\\
\hg_{\hD1s}^5(\hr,\heta)&=\frac{\Ccal_s(\heta ) }{\hD  \left(\hr^2-1\right) (-\hD+p-1)}\times\nonumber\\
&\times \left[(p-\hD) \left(-\hD+(\hD-1) \hr^2-1\right) \Fcal_{0,0}(\hr)+p \left(\hr^2+1\right) \Fcal_{1,0}(\hr)\right].
\end{align}

%%%%%%%%%%%%%%%%%%%

\bibliographystyle{./utphys}
\bibliography{./spinning_defect_submission}

\end{document}